\documentclass[11pt, a4paper, oneside]{book}

\usepackage[table,xcdraw, dvipsnames]{xcolor}

\usepackage[T1]{fontenc} 
\usepackage[utf8]{inputenc}
\usepackage{listings} 
\usepackage[spanish, es-tabla]{babel} 
\usepackage{url} 
\usepackage{graphics,graphicx, float} 
\usepackage[gen]{eurosym} 
\usepackage{enumerate}
\usepackage{hyperref}
\usepackage{graphicx}
\usepackage{tabularx}
\usepackage{booktabs}
\usepackage{csquotes}
\usepackage{emptypage}
\usepackage{float}
\usepackage{tikz}
\usepackage{framed}
\usepackage{eurosym}

\usepackage[toc,page]{appendix}
\addto\captionsspanish{

}

\usepackage{enumitem}
\usepackage{pifont}

\usepackage{fancyhdr} 
\usepackage[backend=biber, style=numeric, sorting=ynt]{biblatex}
\usepackage[ruled,vlined,spanish,onelanguage]{algorithm2e}

\usepackage{tcolorbox}
\usepackage{lipsum}
\usepackage{tocloft}
\usepackage{etoolbox}

\usepackage{subfig}

\usepackage[table]{xcolor}
\usepackage{collcell}
\usepackage{hhline}
\usepackage{pgf}
\usepackage{multirow}

\def\colorModel{rgb} 

\newcommand\ColCell[1]{
  \pgfmathparse{#1<50?1:0}  
    \ifnum\pgfmathresult=0\relax\color{white}\fi
    \pgfmathsetmacro\compA{#1/12}   
    \pgfmathsetmacro\compB{1-#1/12} 
    \pgfmathsetmacro\compC{1}        
  \edef\x{\noexpand\centering\noexpand\cellcolor[\colorModel]{\compA,\compB,\compC}}\x #1
  } 
\newcolumntype{E}{>{\collectcell\ColCell}m{0.4cm}<{\endcollectcell}}  
\newcommand*\rot{\rotatebox{90}}

\usepackage[acronym, toc, nogroupskip]{glossaries}
\makeglossaries
\newglossaryentry{latex}
{
    name=latex,
    description={Is a mark up language specially suited for scientific documents}
}

\newglossaryentry{maths}
{
    name=mathematics,
    description={Mathematics is what mathematicians do}
}

\newglossaryentry{formula}
{
    name=formula,
    description={A mathematical expression}
}

\newglossaryentry{caracteristica}{
    name=característica,
    description={Cada una de las variables...}
}

\newacronym{ad}{AD}{Alzheimer's disease}
\newacronym{mci}{MCI}{Mild cognitive impairment}
\newacronym{cn}{CN}{Cognitively normal}
\newacronym{cnn}{CNN}{Convolutional neural network}
\newacronym{sgd}{SGD}{Stochastic gradient descendent}
\newacronym{mri}{MRI}{Magnetic resonance imaging}
\newacronym{pet}{PET}{Positron emission tomography}
\newacronym{dsm}{DSM-V}{Diagnóstico y manual estadístico de los trastornos mentales}
\newacronym{mmse}{MMSE}{Mini-Mental State Examination}
\newacronym{lr}{LR}{Learning rate}
\newacronym{ct}{CT}{Computed tomography}

\hypersetup{
	colorlinks=true,	
	linkcolor=black,	
	urlcolor=cyan		
}

\begin{document}

    \frontmatter

	\begin{titlepage}
\newlength{\centeroffset}
\setlength{\centeroffset}{-0.5\oddsidemargin}
\addtolength{\centeroffset}{0.5\evensidemargin}
\thispagestyle{empty}

\noindent\hspace*{\centeroffset}\begin{minipage}{\textwidth}

\centering
\includegraphics[width=0.9\textwidth]{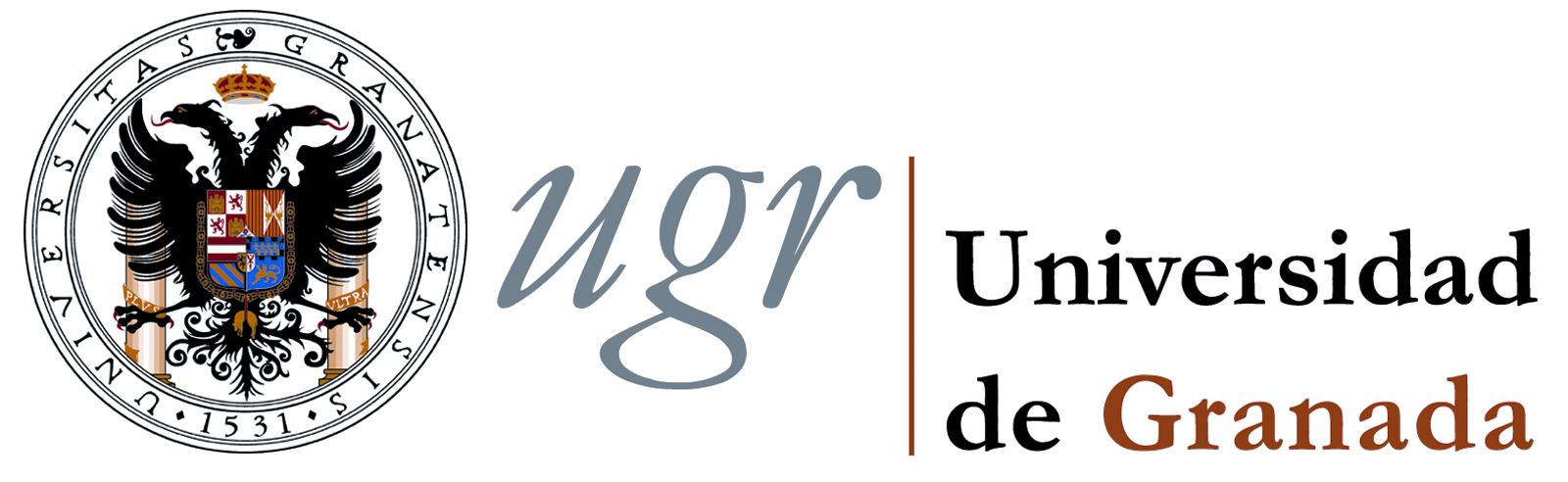}\\[1.4cm]

\textsc{ \Large TRABAJO FIN DE GRADO\\[0.2cm]}
\textsc{ GRADO EN INGENIERIA INFORMATICA}\\[1cm]

{\Large\bfseries Aplicación de redes neuronales convolucionales profundas al diagnóstico asistido de la enfermedad de Alzheimer \\}
\noindent\rule[-1ex]{\textwidth}{3pt}\\[3.5ex]
\end{minipage}

\vspace{2.5cm}
\noindent\hspace*{\centeroffset}
\begin{minipage}{\textwidth}
\centering

\textbf{Autor}\\ {Ángel de la Vega Jiménez}\\[2.5ex]
\textbf{Director}\\ {Fermín Segovia Román}\\[2cm]
\includegraphics[width=0.3\textwidth]{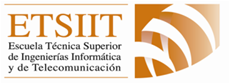}\\[0.1cm]
\textsc{Escuela Técnica Superior de Ingenierías Informática y de Telecomunicación}\\
\textsc{---}\\
Granada, Septiembre de 2021
\end{minipage}
\end{titlepage}

	\thispagestyle{empty}

\begin{center}
{\large\bfseries Aplicación de redes convolucionales profundas al diagnóstico asistido de la enfermedad de Alzheimer}\\ 
\end{center}
\begin{center}
Ángel de la Vega Jiménez\\
\end{center}

\vspace{0.5cm}
\noindent{\textbf{Palabras clave}: \textit{redes neuronales convolucionales, enfermedad de Alzheimer, clasificación, resonancia magnética, tomografía por emisión de positrones, transferencia de aprendizaje, aumento de datos, ADNI, Tensorflow}}
\vspace{0.7cm}

\noindent{\textbf{Resumen}\\
	
En la actualidad, el diagnóstico de la enfermedad de Alzheimer es un proceso complejo y propenso a errores. Una mejora de este diagnóstico podría permitir una detección más temprana de la enfermedad y mejorar la calidad de vida de los pacientes y sus familiares. 

Para este trabajo, utilizaremos 249 imágenes cerebrales de dos modalidades: PET y MRI, tomadas de la base de datos ADNI, y etiquetadas en tres clases según el grado de desarrollo de la enfermedad de Alzheimer.

Proponemos el desarrollo de una red neuronal convolucional para llevar a cabo la clasificación de estas imágenes, durante el cual estudiaremos la profundidad adecuada de las redes para este problema, la importancia del preprocesado de las imágenes médicas, el uso de las técnicas de transferencia de aprendizaje y de aumento de datos como herramientas para reducir los efectos del problema que supone tener pocos datos, y el uso simultáneo de múltiples modalidades de imagen médica.

Planteamos también la aplicación de un método de evaluación que garantiza un buen grado de repetibilidad de los resultados aun utilizando un conjunto de datos de reducido tamaño. Siguiendo este método de evaluación, nuestro mejor modelo final, que hace uso de la transferencia de aprendizaje con datos de COVID-19, consigue una exactitud del 68\%. Por otra parte, en un conjunto de test independiente, este mismo modelo consigue un 70\% de exactitud, un resultado prometedor dado el pequeño tamaño de nuestro conjunto de datos.

Concluimos además, que el aumento de la profundidad de las redes ayuda en este problema, que el preprocesado de las imágenes es un proceso fundamental para abarcar este tipo de problemas médicos, y que el uso de la técnica de aumento de datos y el uso de redes preentrenadas con imágenes de otras enfermedades pueden aportar mejoras notables.

\cleardoublepage
\thispagestyle{empty}

\begin{center}
	{\large\bfseries Application of deep convolutional networks to assisted diagnosis of Alzheimer's disease}\\
\end{center}
\begin{center}
	Ángel de la Vega Jiménez\\
\end{center}}
\vspace{0.5cm}
\vspace{0.7cm}

\noindent{\textbf{Abstract}\\

Currently, the diagnosis of Alzheimer's disease is a complex and error-prone process. Improving this diagnosis could allow earlier detection of the disease and improve the quality of life of patients and their families. 

For this work, we will use 249 brain images from two modalities: PET and MRI, taken from the ADNI database, and labelled into three classes according to the degree of development of Alzheimer's disease.

We propose the development of a convolutional neural network to perform the classification of these images, during which, we will study the appropriate depth of the networks for this problem, the importance of pre-processing medical images, the use of transfer learning and data augmentation techniques as tools to reduce the effects of the problem of having too little data, and the simultaneous use of multiple medical imaging modalities.

We also propose the application of an evaluation method that guarantees a good degree of repeatability of the results even when using a small dataset. Following this evaluation method, our best final model, which makes use of transfer learning with COVID-19 data, achieves an accuracy d
68\%. In addition, in an independent test set, this same model achieves 70\% accuracy, a promising result given the small size of our dataset.

We further conclude that augmenting the depth of the networks helps with this problem, that image pre-processing is a fundamental process to address this type of medical problem, and that the use of data augmentation and the use of pre-trained networks with images of other diseases can provide significant improvements.

\cleardoublepage

\thispagestyle{empty}

\noindent\rule[-1ex]{\textwidth}{2pt}\\[4.5ex]

Yo, \textbf{Ángel de la Vega Jiménez}, alumno del Grado de Ingeniería Informática de la \textbf{Escuela Técnica superior de Ingeniería Informática y Telecomunicaciones de la Universidad de Granada}, con DNI 76645067M, autorizo la ubicación de la siguiente copia de mi Trabajo Fin de Grado en la biblioteca del centro para que pueda ser consultada por las personas que lo deseen.

\vspace{3cm}

\begin{figure}[H]
    \includegraphics[width=3.5cm]{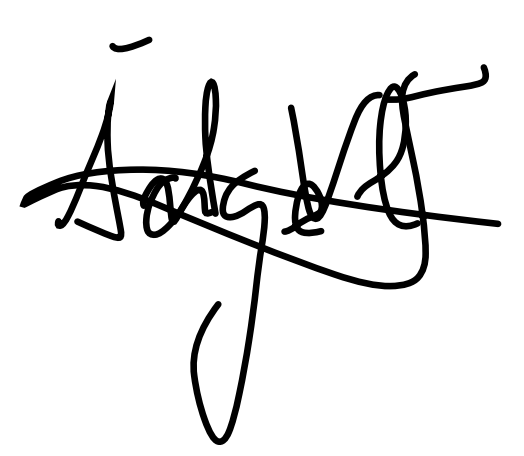}
\end{figure}
\noindent Fdo: Ángel de la Vega Jiménez

\vspace{4cm}

\hspace*{\fill} Granada a 3 de septiembre de 2021.

}

\chapter*{Agradecimientos}

Gracias a mi familia, amigos, y a todos los que me han apoyado durante estos últimos siete meses.

Gracias a todos los profesores, por darme gran parte de los conocimientos necesarios para desarrollar este trabajo, y sobre todo, por darme las bases para poder seguir aprendiendo. Y gracias a mi tutor, Fermín, por toda la ayuda y dudas resueltas.

	\newpage
	\tableofcontents

	\newpage
	\listoffigures

	\listoftables 

    \printglossaries

    \mainmatter

	\part{Introducción}
	\chapter{Motivación}

\section{Enfermedad de Alzheimer}
\label{enfermedad_alzheimer}
La enfermedad de Alzheimer es el tipo más común de demencia \footnote{Demencia: pérdida suficientemente grave de las funciones mentales de una persona como para que su vida diaria se vea afectada (pérdidas de memoria, problemas en el habla...) \cite{demencia_que_es}.}, representando entre un 60 y un 80\% de los casos \cite{what_is_alzheimer}.

Se trata de una enfermedad que avanza progresivamente a lo largo de varios años. Primero afecta a las zonas que controlan el lenguaje, el pensamiento y la memoria. Con el tiempo, evoluciona hacia una pérdida completa de la capacidad para interactuar con el entorno, hasta que finalmente se van perdiendo las funciones biológicas, lo que implica la muerte de la persona. 

Este tipo de demencia suele aparecer en la mayoría de los casos por encima de los 60 años, aumentando el riesgo conforme la persona envejece.

No existe ningún tipo de cura \cite{treatments_2021}, pero los tratamientos para el Alzheimer pueden enlentecer el avance de los síntomas por un tiempo limitado y mejorar la calidad de vida tanto de los enfermos como de sus cuidadores. Entre estos tipos de tratamientos se encuentra el donepezilo, un fármaco aprobado por la Agencia Europea del Medicamento para tratar todas las etapas de la patología.

\section{Una enfermedad difícil de diagnosticar}

En la actualidad, el diagnóstico de la enfermedad de Alzheimer se basa fundamentalmente en la clínica (síntomas) del paciente y en su tiempo de evolución, así como en test neuropsicológicos, siendo las técnicas de neuroimagen (resonancia magnética, PET, SPECT) un apoyo para este diagnóstico \cite{compendio}. Se trata de una enfermedad difícil de diagnosticar, especialmente en etapas tempranas debido, en parte, a la similitud con la sintomatología de otros tipos de demencia (frontotemporal, vascular, por cuerpos de Lewy...), e incluso con trastornos psicológicos producidos por depresión \cite{pmid28868066, depresion_demencia_1}. Esta última dificultad, puede llevar en ocasiones al falso diagnóstico de la enfermedad, o en el lado opuesto, a un rechazo de la misma en fases tempranas, o simplemente, a una gran incertidumbre ante la incapacidad de poder dar un diagnóstico claro.

La enfermedad de Alzheimer es incurable, pero a pesar de ello, un diagnóstico temprano puede hacer que la calidad de vida del paciente sea mejor durante el transcurso de la misma, y no menos importante, permitirá a las personas cercanas una adaptación más progresiva a todo lo que esto conlleva. Por otro lado, evitar el falso diagnóstico, o la incertidumbre en el diagnóstico en las etapas tempranas, puede reducir enormemente el estrés que esto produce tanto en el paciente (según su estado de consciencia) como en las familias.

\section{¿Puede ayudar el aprendizaje automático?}

Desde 2012, cuando el grupo de la universidad de Toronto liderado por Alex Krizhevsky logró, por medio del uso de redes convolucionales profundas, superar holgadamente a todos los enfoques existentes para la clasificación de imágenes naturales \cite{NIPS2012_4824}, las redes neuronales convolucionales profundas se han convertido en el estándar para multitud de problemas perceptuales (clasificación de imágenes, reconocimiento del habla, traducción, conducción autónoma...), llegando incluso a superar a la percepción humana en algunos casos \cite{DBLP:journals/corr/HeZR015, alphago}.

Siguiendo esta tendencia, en el ámbito de la medicina, las redes convolucionales profundas están siendo aplicadas a multitud de problemas (detección de tumores, anomalías cardiacas, neumonía, etc)\cite{Sarvamangala2021, LUNDERVOLD2019102}, y están a la cabeza de importantes competiciones en este ámbito, como el reto HVSMR 2016 sobre segmentación de resonancia magnética cardiovascular \cite{hvsmr} o el reto RSNA sobre la detección de neumonía en radiografías \cite{rsna}. 

Todo lo anterior nos hace pensar que para el diagnóstico precoz de la enfermedad de Alzheimer, es probable que la aplicación de redes neuronales profundas sobre imágenes médicas tenga un gran potencial. Además, este diagnóstico sería realizado únicamente por medio de imágenes cerebrales, disminuyendo la necesidad del diagnóstico clínico por parte del médico, y aumentando el rendimiento en el diagnóstico en consecuencia.

\chapter{Objetivos}
\label{objetivos}

Ya hemos visto que en el diagnóstico de la enfermedad de Alzheimer, actualmente, las imágenes cerebrales no tienen una utilidad real por sí solas, sino que son un apoyo al diagnóstico.

A lo largo de este trabajo, estudiaremos la posibilidad de realizar un diagnóstico de esta enfermedad basado exclusivamente en imágenes cerebrales, usando para ello redes neuronales convolucionales. Más concretamente, partiremos de un  conjunto de datos formado por imágenes cerebrales 3D (\acrshort{mri} y \acrshort{pet}) de distintos pacientes, etiquetadas en tres clases posibles: \begin{itemize}
\itemsep0em
    \item \textbf{\acrshort{ad}} (\acrlong{ad}): son aquellas imágenes cerebrales pertenecientes a pacientes que presentan la enfermedad de Alzheimer.
    \item \textbf{\acrshort{mci}} (\acrlong{mci}): son aquellas imágenes cerebrales de pacientes con un deterioro cognitivo leve, y que podrían sufrir en el futuro la enfermedad de Alzheimer.
    \item \textbf{\acrshort{cn}} (\acrlong{cn}): son las pertenecientes a pacientes con un estado cognitivo normal.
\end{itemize}

Usando estos datos, trataremos de crear una red neuronal convolucional que sea capaz de tomar como entrada una nueva imagen cerebral (nunca antes vista) y clasificarla en una de las tres clases de forma correcta.

\noindent Nuestros \textbf{objetivos principales} serán los dos siguientes:\begin{enumerate}[label=\textbf{\arabic*}.]
    \item Implementar una red neuronal convolucional para la clasificación de imágenes cerebrales entre las clases \acrshort{mci}, \acrshort{ad} y \acrshort{cn}.
    \item Estudiar cuál es la \textbf{profundidad} adecuada de las redes neuronales convolucionales para resolver este tipo de problemas.
\end{enumerate}

Además de estos objetivos iniciales, tras estudiar en profundidad las redes convolucionales y hacer una amplia revisión de estudios relacionados (que en el capítulo \ref{estudios_relacionados} describiremos más en detalle), surgen \textbf{nuevos objetivos} por los siguientes motivos:\begin{enumerate}[label=\textbf{\arabic*}.]
    \setcounter{enumi}{2}
    
    \item Dados los excelentes resultados que se consiguen mediante el uso de transferencia de aprendizaje en el caso de clasificación de imágenes cuando se poseen pocos datos (como será nuestro caso), planteamos como objetivo \textit{utilizar conjuntos de datos de otras enfermedades para la aplicación de la técnica de transferencia de aprendizaje} (esta técnica se explicará en la sección \ref{preentrenamiento}). 
    
    \item La técnica de aumento de datos (sección \ref{data_augmentation}) también suele dar buenos resultados, pero no hemos encontrado estudios que hagan uso de esta técnica sobre imágenes 3D para el diagnóstico de la enfermedad de Alzheimer, por lo que planteamos \textit{estudiar el uso de la técnica de aumento de datos sobre imágenes 3D para la clasificación de imágenes médicas}.

    \item Tampoco hemos encontrado ningún estudio que haga uso de imágenes de dos modalidades de imagen al mismo tiempo para realizar el diagnóstico, por lo que planteamos el objetivo de \textit{estudiar el uso simultáneo de imágenes cerebrales de dos modalidades, MRI y PET, para la mejora del diagnóstico}.
    
    \item En algunos de los artículos revisados, existe un sesgo optimista en la evaluación de los modelos, que hace que los resultados dados no transmitan bien cómo sería el desempeño en condiciones reales. Además, en ocasiones son resultados que pueden variar en función de elementos aleatorios. Planteamos en consecuencia el siguiente objetivo: \textit{realizar una fase de experimentación y evaluación de modelos que asegure, en la medida de lo posible, la reproducibilidad de los resultados, y que estime la bondad de los modelos en condiciones reales}.
    
    \item En muchas ocasiones, no se dan detalles sobre el motivo de elección de las métricas de evaluación, o se usan métricas que pueden dar lugar a confusiones. Esto nos obliga a plantearnos el siguiente objetivo: \textit{valorar distintas métricas de error para la evaluación de los modelos y seleccionar la más adecuada al problema dado}.

    \item Casi siempre se afirma que el preprocesado de las imágenes médicas es necesario para conseguir buenos resultados, pero dadas las características de las redes convolucionales profundas (sección \ref{extrayendo_caracteristicas}), creemos de especial interés \textit{estudiar la necesidad del preprocesado de las imágenes médicas para la aplicación de técnicas de aprendizaje automático}.

\end{enumerate}

Como último objetivo, nos gustaría \textit{lograr una implementación eficiente y escalable, por medio del uso de funciones avanzadas de Tensorflow y las unidades de procesamiento tensorial (TPU)}.

\chapter{Planificación y presupuesto}
\section{Planificación}

\newenvironment{formal}{
\def\FrameCommand{{\color{YellowOrange}\vrule width 2pt}\hspace{2pt}}
\MakeFramed{\advance\hsize-\width}
\vspace{2pt}\noindent\hspace{-7pt}\vspace{3pt}
}{\vspace{3pt}\endMakeFramed}

En esta sección, veremos la planificación que hemos seguido, en la cual podemos distinguir ocho tareas principales: \begin{itemize}
    \itemsep0em
    \item \textbf{Estudio e investigación}: toma de contacto con todos los conceptos que no conocíamos y que eran necesarios para la realización de este trabajo, además de la revisión del estado del arte y la investigación realizada para diseñar los experimentos.
    \item \textbf{Preprocesado y carga}: todo lo relacionado con el preprocesado de las imágenes y la carga de los datos para el entrenamiento de las redes.
    \item \textbf{Estudio de la profundidad}: una tarea amplia que engloba la experimentación con numerosas arquitecturas con distintas profundidades.
    \item \textbf{Aumento de datos}: el desarrollo de todas las funciones necesarias para poder aplicar la técnica de aumento de datos sobre imágenes 3D.
    \item \textbf{Imágenes crudas}: experimentos llevados a cabo haciendo uso de imágenes sin preprocesar.
    \item \textbf{Transfer learning}: la aplicación de la técnica de transferencia de aprendizaje haciendo uso de datos de pacientes con COVID-19.
    \item \textbf{Dos entradas}: implementación y experimentación con redes convolucionales que toman como entrada dos imágenes de distintas modalidades al mismo tiempo.
    \item \textbf{Memoria}: redacción de la memoria.
\end{itemize}

\begin{figure}[H]
    \centering
    \includegraphics[width=12cm]{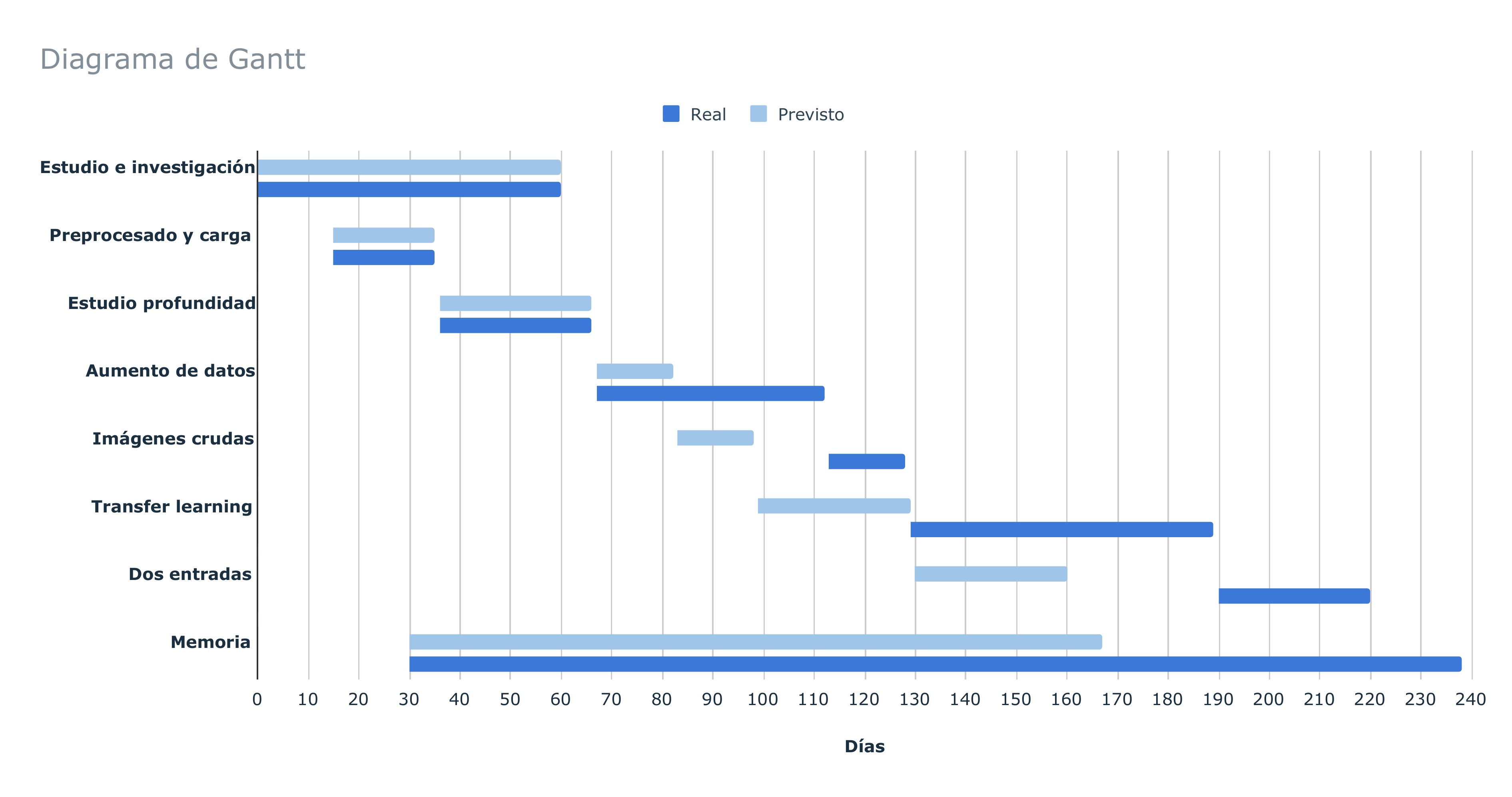}
    \caption[Diagrama de Gantt.]{Diagrama de Gantt. El día cero se corresponde con el día 4 de enero de 2021, y el día 238, con el 30 de agosto.}
    \label{fig:diagrama_gantt}
\end{figure}

En la figura \ref{fig:diagrama_gantt}, vemos como quedan repartidas en el tiempo estas tareas.

\subsection{Problemas}

En el diagrama de Gantt (figura \ref{fig:diagrama_gantt}) se puede apreciar que el tiempo real que hemos empleado no se corresponde totalmente con el que habíamos previsto en un principio debido a varios problemas: \begin{itemize}
    \itemsep0em 
    \item En la tarea de aumento de datos, contábamos con que existirían una serie de funciones ya implementadas en bibliotecas conocidas, pero esto no fue así, por lo que hubo que implementar mucho más de lo esperado, lo que causó un retraso de unos 10 días, y además, el tiempo de cómputo de los experimentos de esta tarea fue más largo de lo esperado, causando un segundo retraso de 20 días aproximadamente.
    \item En la tarea de transfer learning, contábamos con que existirían ciertas redes neuronales ya implementadas, pero de nuevo no fue así, y su implementación causó un retraso de 15 días. Por otro lado, algunos problemas con la disponibilidad del hardware causaron un retraso de otros 15 días aproximadamente.
    \item La redacción de la memoria también se ha visto retrasada, ya que su finalización requería haber terminado todos los experimentos.
\end{itemize}

Aunque la finalización del proyecto en un principio estaba prevista para el 20 de junio aproximadamente, todos estos problemas han atrasado su terminación hasta el 30 de agosto. 

\subsection{Tablero de Trello}

Además de esta planificación genérica dividida en ocho tareas principales, hemos utilizado un tablero de Trello para tener una organización a más bajo nivel a medida que íbamos avanzando dentro de estas tareas generales. Estos tableros nos permiten crear distintas listas, y dentro de cada lista podemos añadir tantas tarjetas como queramos. En nuestro caso, hemos visto de gran utilidad crear las siguientes listas para definir los distintos estados de nuestras tareas:\begin{itemize}
    \itemsep0em
    \item \textbf{Por hacer:} en esta lista se introducen aquellas tareas que hay que hacer, pero en las que no se ha comenzado a trabajar aún. Además, hemos procurado ordenarlas por orden de prioridad.
    \item \textbf{En proceso:} aquellas tareas que están siendo realizadas en un momento concreto.
    \item \textbf{Bloqueado:} en ocasiones surgen problemas no esperados. En esta lista introducimos las tareas que se han visto bloqueadas por alguno de estos problemas, que aunque pueden verse retrasadas, no queremos olvidar.
    \item \textbf{Terminado:} tareas completamente terminadas, y que salvo excepciones muy concretas, no volverán a sufrir modificaciones.
    \item \textbf{Preguntar:} en esta lista se almacenan todas aquellas dudas que quisiéramos preguntar. Normalmente, las dudas de esta lista se resolvían cada viernes, después de la clase de teoría de la asignatura de Robótica Industrial, ya que el profesor era nuestro tutor.
\end{itemize}

Durante toda la semana, conforme íbamos trabajando en el proyecto, íbamos moviendo las tarjetas a la lista correspondiente según su estado. Y luego, cada sábado, volvíamos a ajustar el tablero, archivando las tareas finalizadas, e introduciendo nuevas tareas previstas para la semana siguiente.

\begin{figure}[H]
    \centering
    \includegraphics[width=\textwidth]{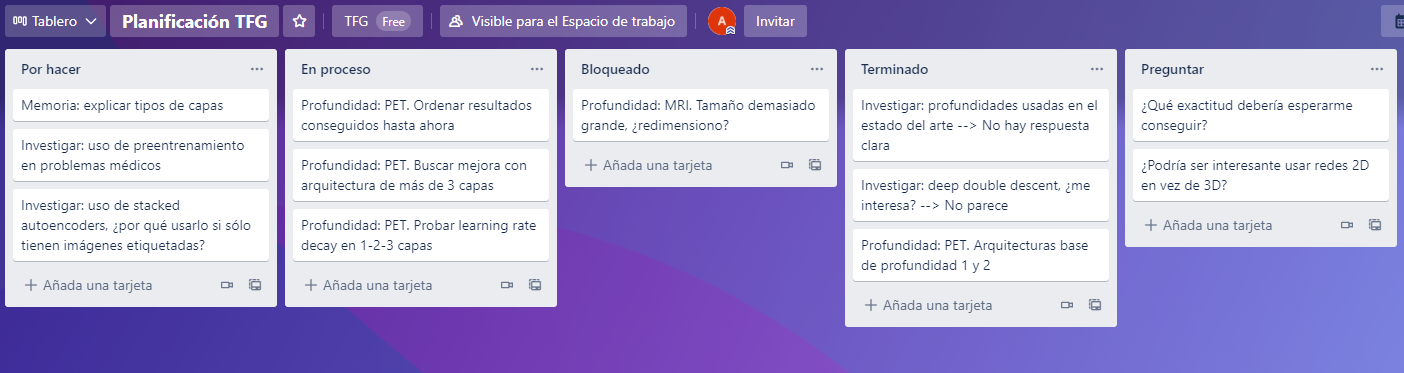}
    \caption[Tablero de Trello]{Tablero de Trello durante la etapa del estudio de la profundidad.}
    \label{fig:tablero_trello}
\end{figure}

\section{Presupuesto}

En esta sección veremos los distintos recursos utilizados para la realización de este proyecto y su coste asociado.

\section{Recursos materiales}

En la tabla \ref{recursos_materiales} se muestran los recursos materiales más importantes para el desarrollo de este trabajo.

\begin{table}[H]
\begin{tabular}{l|l|l|l|l}
Recurso           & Coste      & Coste UGR & Meses de uso & Coste \\ \hline
Ordenador         & Gratuito * & Gratuito                         & 8            & 0 \euro   \\
Suscripción Colab & 9.99\euro/mes     & 9.99\euro/mes                           & 8            & 79.92 \euro  \\
Drive (1TB) **    & 9.99\euro/mes  & Gratuito                         & 8            & 0 \euro   \\
Kaggle            & Gratuito   & Gratuito                         & 8            & 0 \euro   \\
Overleaf          & Gratuito   & Gratuito                         & 8            & 0 \euro    \\
Matlab            & 119 \euro      & Gratuito                         & 2            & 0 \euro    \\
SPM12             & Gratuito   & Gratuito                         & 2            & 0 \euro    \\
Mricron           & Gratuito   & Gratuito                         & 6            & 0 \euro   
\end{tabular}
\caption{Recursos materiales utilizados}
\label{recursos_materiales}
\end{table}

\noindent * no necesitamos que sea un ordenador potente, nos vale con que tenga conexión a internet. Lo consideramos gratuito ya que en general, cualquier persona que se proponga realizar un proyecto de estas características, tendrá acceso a un ordenador (sea cual sea) antes de comenzar el proyecto.

\noindent ** Nos habría bastado con 200GB, pero las tarifas de almacenamiento de Google Drive dan el salto desde 100GB (que no nos sirve), hasta 1TB.

Los recursos materiales, han tenido un coste total de 79.92\euro, que sería de 278.84\euro~ en el caso de un usuario que no tenga acceso a una licencia universitaria de Matlab, ni a un almacenamiento gratuito en Google Drive (al que tenemos acceso con nuestra cuenta de la UGR).

\section{Recursos humanos}

En la tabla \ref{horas_dedicadas} se muestra de forma aproximada el tiempo dedicado a cada una de las tareas generales de este trabajo.

Podemos ver que hay tareas que aún habiendo consumido más días de trabajo, han requerido el mismo número, o incluso menos horas de trabajo que otras tareas realizadas en menos días. Esto se debe a dos motivos: por un lado, hemos intentado contar solo aquellas horas de trabajo propiamente dicho, y no hemos contado las horas de ``espera'' cuando ejecutábamos experimentos, por lo que las fases con experimentos pesados ocupan muchos días (debido a la espera). Por otro lado, no hemos dispuesto del mismo número de horas al día durante todos los meses, eso también hace que algunas tareas se hayan prolongado por más días que otras.

\begin{table}[H]
\centering
\begin{tabular}{l|l|l}
Recurso                 & Días & Horas \\ \hline
Estudio e investigación & 60   & 90   \\
Preprocesado y carga    & 20   & 30    \\
Estudio profundidad     & 30   & 60    \\
Aumento de datos        & 60   & 60    \\
Imágenes crudas         & 15   & 20  \\
Transfer learning       & 60   & 75   \\
Dos entradas            & 30   & 60   \\
Memoria                 & 213  & 240  
\end{tabular}
\caption{Horas aproximadas dedicadas a cada fase}
\label{horas_dedicadas}
\end{table}

En total, hemos dedicado unas \textbf{635 horas}. Teniendo en cuenta que el salario medio de un ingeniero informático con poca experiencia laboral se encuentra entorno a 24880\euro~ brutos anuales \cite{salario_junior}, y teniendo en cuenta que el total de horas trabajadas en un año (suponiendo jornada de 8 horas y 30 días de vacaciones) son unas 1800 horas, el salario bruto por hora sería de unos 14\euro/hora.

Multiplicando por el número de horas empleadas, el coste de recursos humanos para este proyecto sería de unos 8890\euro~ brutos.

\subsection{Coste total}

Sumando el coste de los recurso materiales y los recursos humanos, el coste total del proyecto sería de unos \textbf{9000\euro}.
	
	\part{Fundamentos teóricos}
    \chapter{Aprendizaje automático}

  



\begin{tcolorbox}[
  colback=SkyBlue!5!white,
  colframe=SkyBlue!75!black,
  title={Sumario}]
  
En este capítulo presentaremos los fundamentos teóricos básicos del aprendizaje automático, que se utilizarán durante el resto del trabajo. Estos son los principales temas que se abarcarán:

\begin{itemize}
\itemsep 0em 
    \item Idea intuitiva de qué es el aprendizaje automático.
    \item En qué consiste el problema de clasificación.
    \item Qué es la función de pérdida y en qué consiste el entrenamiento de un modelo, basando la explicación en el modelo de regresión logística, elegido por ser un modelo simple que nos permitirá entender correctamente los conceptos importantes.
    \item Idea de los conceptos de error de generalización, \textit{overfitting} y regularización. 
    \item Evaluación de la bondad de un modelo de aprendizaje automático.
\end{itemize}

\end{tcolorbox}

Si enseñamos a una persona cualquiera una fotografía de un avión, es muy probable que sepa decirnos que en esa fotografía aparece un avión. Luego podríamos preguntarle qué algoritmo utiliza para poder distinguir un avión, pero no obtendríamos una respuesta demasiado satisfactoria. Esto es porque no es algo que aprendamos estudiando la definición matemática avión, sino que se aprende viendo ejemplos de aviones. Dicho de otra forma, \textit{aprendemos de los datos} \cite{abu-mostafa_2012}.

Supongamos ahora que queremos que dada una fotografía, un computador diga ``sí'' en el caso de que aparezca un avión y ``no'' en caso contrario. Si tomáramos el problema desde el punto de vista de la programación clásica tendríamos que escribir un programa que tomando como entrada una fotografía, ejecutara una serie de reglas (algoritmo) tal que acabaran dando como salida la respuesta que deseamos. Como podemos imaginar, escribir este programa no es factible.

Sin embargo podemos tomar un paradigma de programación distinto, el del aprendizaje automático. En este caso escribiríamos un programa de aprendizaje automático, que tomaría como entradas ejemplos de fotos etiquetadas según contengan un avión o no (esto podríamos conseguirlo de forma sencilla) y este \textit{aprendería} una función tal que al recibir como entrada una nueva foto, obtiene como salida la respuesta deseada.

\begin{figure}[H]
  \centering

\tikzset{every picture/.style={line width=0.75pt}} 

\begin{tikzpicture}[x=0.75pt,y=0.75pt,yscale=-1,xscale=1]

\draw   (154,46) -- (224,46) -- (224,86) -- (154,86) -- cycle ;
\draw    (107,54) -- (152,54) ;
\draw [shift={(154,54)}, rotate = 180] [color={rgb, 255:red, 0; green, 0; blue, 0 }  ][line width=0.75]    (10.93,-3.29) .. controls (6.95,-1.4) and (3.31,-0.3) .. (0,0) .. controls (3.31,0.3) and (6.95,1.4) .. (10.93,3.29)   ;
\draw    (107,77) -- (152,77) ;
\draw [shift={(154,77)}, rotate = 180] [color={rgb, 255:red, 0; green, 0; blue, 0 }  ][line width=0.75]    (10.93,-3.29) .. controls (6.95,-1.4) and (3.31,-0.3) .. (0,0) .. controls (3.31,0.3) and (6.95,1.4) .. (10.93,3.29)   ;
\draw    (225,66) -- (270,66) ;
\draw [shift={(272,66)}, rotate = 180] [color={rgb, 255:red, 0; green, 0; blue, 0 }  ][line width=0.75]    (10.93,-3.29) .. controls (6.95,-1.4) and (3.31,-0.3) .. (0,0) .. controls (3.31,0.3) and (6.95,1.4) .. (10.93,3.29)   ;
\draw   (154,119) -- (224,119) -- (224,159) -- (154,159) -- cycle ;
\draw    (107,127) -- (152,127) ;
\draw [shift={(154,127)}, rotate = 180] [color={rgb, 255:red, 0; green, 0; blue, 0 }  ][line width=0.75]    (10.93,-3.29) .. controls (6.95,-1.4) and (3.31,-0.3) .. (0,0) .. controls (3.31,0.3) and (6.95,1.4) .. (10.93,3.29)   ;
\draw    (107,150) -- (152,150) ;
\draw [shift={(154,150)}, rotate = 180] [color={rgb, 255:red, 0; green, 0; blue, 0 }  ][line width=0.75]    (10.93,-3.29) .. controls (6.95,-1.4) and (3.31,-0.3) .. (0,0) .. controls (3.31,0.3) and (6.95,1.4) .. (10.93,3.29)   ;
\draw    (225,139) -- (270,139) ;
\draw [shift={(272,139)}, rotate = 180] [color={rgb, 255:red, 0; green, 0; blue, 0 }  ][line width=0.75]    (10.93,-3.29) .. controls (6.95,-1.4) and (3.31,-0.3) .. (0,0) .. controls (3.31,0.3) and (6.95,1.4) .. (10.93,3.29)   ;

\draw (60,48) node [anchor=north west][inner sep=0.75pt]  [font=\scriptsize] [align=left] {Entradas};
\draw (153.4,55.6) node [anchor=north west][inner sep=0.75pt]  [font=\scriptsize] [align=left] {\begin{minipage}[lt]{47pt}\setlength\topsep{0pt}
Programación
\begin{center}
clásica
\end{center}

\end{minipage}};
\draw (56,71) node [anchor=north west][inner sep=0.75pt]  [font=\scriptsize] [align=left] {Programa};
\draw (276.6,59.6) node [anchor=north west][inner sep=0.75pt]  [font=\scriptsize] [align=left] {Respuesta};
\draw (60,121) node [anchor=north west][inner sep=0.75pt]  [font=\scriptsize] [align=left] {Entradas};
\draw (155.4,126.6) node [anchor=north west][inner sep=0.75pt]  [font=\scriptsize] [align=left] {\begin{minipage}[lt]{47pt}\setlength\topsep{0pt}
Aprendizaje
\begin{center}
automático
\end{center}

\end{minipage}};
\draw (48,144) node [anchor=north west][inner sep=0.75pt]  [font=\scriptsize] [align=left] {Respuestas};
\draw (276.6,132.6) node [anchor=north west][inner sep=0.75pt]  [font=\scriptsize] [align=left] {Función};
\end{tikzpicture}
    \label{fig:paradigma_clasica_ml}
\label{antiguo_nuevo_paradigma}
\end{figure}

Un programa de aprendizaje automático se dice que es \textit{entrenado} en lugar de ser programado. El entrenamiento consiste en presentar a este algoritmo muchos ejemplos de una cierta tarea y ``ajustarlo'' de forma que dado un \textit{nuevo} ejemplo de la tarea, sea capaz de resolverla correctamente.

\section{El problema de clasificación}

Existen distintos problemas que se pueden resolver mediante aprendizaje automático (regresión, clustering, segmentación, detección...). Sin embargo, no sería de gran interés explicar estos tipos de problemas para nuestros propósitos, por lo que nos centraremos en el problema de clasificación.

El problema de clasificación entra dentro de un tipo de aprendizaje conocido como aprendizaje supervisado. Este tipo de aprendizaje consiste en aprender una función que asigne a cada dato de entrada una salida, basándose en un conjunto de datos de ejemplo para los que ya sabemos como debería ser la salida, es decir partimos de un conjunto de parejas entrada-salida \cite{russell_norvig_chang_2010}. 

En un problema de clasificación, intentamos aprender una función con salidas discretas. Es decir, estamos tratando de asignar a cada dato de entrada una categoría discreta (clase).

Un ejemplo de problema de clasificación es el visto en el apartado anterior: ``dada una imagen, etiquetarla como SI si es la foto de un avión y como NO en caso contrario''.

\subsection{Elementos de un problema}
\label{elementos_clasificacion}

 Vamos a formalizar las ideas anteriores: para plantear un problema de aprendizaje automático supervisado (en el que se encuentra el de clasificación) necesitamos los siguientes elementos:\begin{itemize}
     \item Un conjunto de ejemplos entrenamiento, que consiste en un conjunto de tuplas $\boldsymbol{D} = (x^{(1)}, y^{(1)}), (x^{(2)}, y^{(2)}), ..., (x^{(m)}, y^{(m)})$ donde cada $x^{(i)}$ es una entrada al problema, y cada $y^{(i)}$ la respuesta. En nuestro ejemplo anterior, cada tupla sería una imagen etiquetada, siendo cada $x^{(i)}$ la imagen, y $y^{(i)}$ su etiqueta (sí o no).
     
     \item Una función objetivo desconocida que asigna a cada entrada su salida correspondiente $\boldsymbol{f}:X \rightarrow Y$; $y^{(i)} = f(x^{(i)})$
     
     \textit{Asumimos} que esta $f$ \textit{existe}. En nuestro ejemplo, esta $f$ sería la función que dada una imagen, la clasifica según aparezca o no un avión en ella. Sabemos que $f$ existe porque una persona es capaz de clasificarla, aunque desconocemos cómo es esa $f$ (puede ser muy compleja).
     
     \item Un algoritmo de aprendizaje $\boldsymbol{A}$, que utiliza el conjunto de datos $D$ para elegir una función $g: X \rightarrow Y$ que aproxima a $f$ lo mejor posible.
     
     \item La función $g$ es elegida por el algoritmo de aprendizaje entre un conjunto de funciones candidatas, que llamamos conjunto de hipótesis $\boldsymbol{H}$. Por ejemplo, $H$ podría ser el conjunto de todas las funciones lineales.
     
     \item Para elegir la función $g$, el algoritmo de aprendizaje utiliza una función de pérdida $\boldsymbol{J}$, que sirve para saber el rendimiento de cada función candidata en la tarea a resolver.
     
 \end{itemize}
 
 Con estos elementos, podemos dar paso a la definición de aprendizaje automático dada por Tom Mitchell: \textit{``Se dice que un programa informático aprende de la experiencia E con respecto a algún tipo de tarea T y una medida de rendimiento P, si su rendimiento en la tarea T, medido por P, mejora con la experiencia E'' }\cite{mitchell_1997}.
 
 En esta definición, podríamos decir que la experiencia E se correspondería con el conjunto de datos $D$, la tarea T sería el problema concreto que queremos resolver (por ejemplo un problema de clasificación), y podríamos decir que P sería la función de pérdida $J$. El algoritmo de aprendizaje automático se entrenaría con $D$ para mejorar su rendimiento medido por $J$ en la tarea concreta.
 
 \section{Concepto de entrenamiento}
 
 Hasta ahora hemos dicho que para resolver un problema mediante aprendizaje automático vamos a presentar al algoritmo de aprendizaje una serie de ejemplos de la tarea a resolver, de forma que al ser \textit{entrenado} éste aprenda a resolverla para nuevos ejemplos, pero no hemos especificado en qué consiste este entrenamiento.
 
 A continuación, vamos a plantear todos los elementos necesarios para resolver un problema de clasificación sencillo mediante aprendizaje automático, para finalmente entender en qué consiste el entrenamiento.
 
 \subsection{Notación}
 
 Para familiarizarnos con la notación vamos a poner un ejemplo de tarea de clasificación sencilla: ``dado el peso y la longitud de un ratón, decir si éste es obeso o no lo es''. 

 En el conjunto de entrenamiento $D = (x^{(1)}, y^{(1)}), ..., (x^{(m)}, y^{(m)})$, cada $x^{(i)}$ será un \textit{vector columna} con las características de un determinado ratón (siendo $x_0^{(i)}$ la característica correspondiente al término independiente, siempre igual a 1 por convención, $x_1^{(i)}$ su peso y $x_2^{(i)}$ la longitud), y cada $y^{(i)} \in \{0, 1\}$, donde $0$ indica que el ratón no es obeso, y $1$ que sí lo es.
 
 En esta tarea concreta sólo tenemos dos características, el peso y la altura del ratón, pero podríamos tener un número cualquiera de características. En ese caso, cada característica $j$ de un determinado ejemplo $i$ se notaría como $x_j^{(i)}$.
 
 \subsection{Conjunto de hipótesis}
 
 \renewcommand{\thefootnote}{\roman{footnote}}
 
 Para abarcar un problema de clasificación como el del ejemplo anterior necesitamos primero elegir un conjunto de hipótesis $H$. En este caso utilizaremos $H = h_w = g(w_0x_0 +w_1x_1 + ... +w_nx_n)$ \footnote{$g$ denota la función sigmoide: $g(z) = \frac{1}{1 + e ^{-z}}$} $= g(w^Tx)$ \footnote{Notar que cuando escribimos $w$ sin subíndices, nos referimos a un vector columna de dimensión $(n+1)\times1$, con $n$ el número de características de cada ejemplo. $w^T$ denota el vector traspuesto de $w$.}, donde cada $w_j$ son los \textbf{parámetros} que queremos \textbf{aprender} (es decir, aún no están fijos). Por otro lado, $\boldsymbol{x}$ representa un dato de entrada al que aplicamos la función $h_w$, por lo que su valor es fijo.
 
 \renewcommand{\thefootnote}{\arabic{footnote}}

 \begin{figure}
     \centering
     \includegraphics[width=10cm]{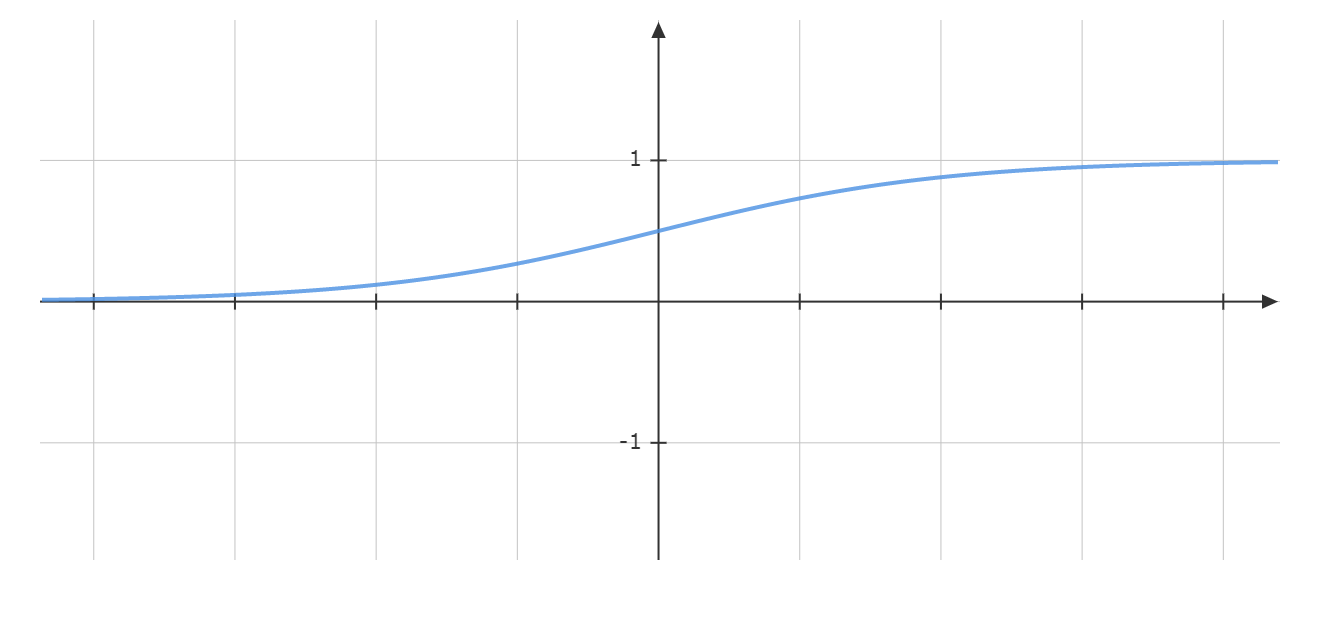}
     \caption[Función sigmoide]{Función sigmoide ($\frac{1}{1 + e ^{-x}}$). Esta función tiene la propiedad de que su salida está acotada entre $0$ y $1$, cruzando por $y=0.5$ cuando $x=0$}
 \end{figure}
 
Una vez que aprendamos los parámetros $w$, $h_w(x)$ nos dará la \textbf{probabilidad} de que nuestra salida para la entrada $x$ sea 1. Por ejemplo, $h_w(x)=0.7$ nos dice que hay una probabilidad del 70\% de que la salida sea 1. La probabilidad de que nuestra predicción sea 0 es simplemente el complemento de la probabilidad de que sea 1 (por ejemplo, si la probabilidad de que sea 1 es del 70\%, entonces la probabilidad de que sea 0 es del 30\%).

Sin embargo, esta función nos dará como resultado un número decimal entre $0$ y $1$, pero para un problema de clasificación queremos salidas discretas. Para conseguirlo, podemos transformar la salida de nuestra función de la siguiente forma: 
\[ \begin{cases} 
      1 & h_w(x) \geq 0.5 \\
      0 & h_w(x) < 0.5
   \end{cases}
\]

Este tipo modelo de clasificación simple que acabamos de presentar se conoce como \textbf{regresión logística}.

 \subsection{Función de pérdida}
 
 Hasta ahora tenemos un conjunto de hipótesis $H = h_w = g(\boldsymbol{w^Tx})$, donde cada $w_i$ son parámetros que no están fijos. 
 
 El siguiente paso es fijar estos parámetros de forma que nuestra función $h_w$ ``acierte'' clasificando los datos lo mejor posible. Para ello, tenemos que decidir una \textbf{función de pérdida $J$} que nos diga cómo es el desempeño de nuestra función $h_w$ en la tarea de clasificación según los valores de $w$ que elijamos. $J$ es la siguiente en este caso \footnote{Función usualmente usada para este tipo de problema, una de las propiedades interesantes es su concavidad.}:

\begin{equation}
    J(\boldsymbol{w}) = \frac{1}{m}\sum_{i=1}^{m}L(h_w(\boldsymbol{x^{(i)}}), y^{(i)})
\label{perdida_logistica}
\end{equation}

Donde $L$ se define como:

\[L(h_w(x), y) = - log(h_w(x)) \hspace{0.5em} si \hspace{0.5em} y = 1\]
\[L(h_w(x), y) = - log(1 - h_w(x)) \hspace{0.5em} si \hspace{0.5em} y = 0\]

Si nos fijamos, en el caso de que la respuesta correcta sea $1$ y nuestra función $h_w$ haya predicho $1$, entonces el coste será 0, y por contra, si $h_w$ predice $0$, el coste se aproxima a infinito. Lo mismo ocurre en el caso de que la respuesta correcta sea $0$.

En definitiva, tenemos una función de pérdida que arrojará un valor pequeño cuando los ejemplos se estén clasificando de forma correcta, y grande cuando haya un elevado número de errores en la clasificación.

 \subsection{Descenso de gradiente}
 
 Ya tenemos una función de pérdida que nos dice cómo de bien lo está haciendo nuestro algoritmo de clasificación. Como hemos visto, esta función tendrá como salida valores pequeños en el caso de clasificaciones correctas, y por tanto, está claro que queremos \textit{minimizar} su valor. Esta minimización es lo que usualmente conocemos como \textbf{entrenamiento del modelo}.
 
 Para minimizarla, primero podemos escribirla de la siguiente forma (se puede comprobar que es totalmente equivalente a la expresión \ref{perdida_logistica}): \[ J(w) = -\frac{1}{m}\sum_{i=1}^{m}[y^{(i)}log(h_w(x^{(i)}) + (1 - y^{(i)})log(1 - h_w(x^{(i)}))]\]

 Y calcular su gradiente respecto de los pesos $w_j$:
 
 \[ \frac{\partial}{\partial w_j}J(w) = \frac{1
 }{m}\sum_{i=1}^{m}(h_w(x^{(i)}) - y^{(i)})x_j^{(i)}\]
 
 Luego, aplicaremos el algoritmo de descenso de gradiente sobre esa función. Este algoritmo consiste en calcular en sucesivas iteraciones el gradiente de la función (respecto de los parámetros $w$) para obtener la dirección de la pendiente máxima, y actualizar los parámetros en la dirección opuesta de la pendiente, minimizando así el valor de la función.

\begin{algorithm}[H]
\label{alg:descenso_gradiente}
\SetAlgoLined
\KwResult{$w$}
 Inicializar $w$
 
 \For{$i = 0, 1, 2, ...$}{
    $w_j \gets w_j - \alpha \frac{\partial}{\partial w_j}J(w)$\;
  }
 
\caption{Descenso de gradiente}
\end{algorithm}

En el algoritmo anterior vemos que los pesos $w$ se inicializan al principio. Existen formas sofisticadas de inicializar estos pesos, pero queda fuera de los objetivos de este capítulo, por lo que podemos imaginar que la inicialización es aleatoria (es una inicialización perfectamente válida).

Respecto a la variable $\alpha$, se trata de la \textbf{tasa de aprendizaje} (usualmente notada como \acrshort{lr}, del inglés, learning rate), y se encarga de controlar la longitud de los ``pasos'' en cada actualización. Es un parámetro libre que hay que fijar, y por ahora, nos basta con saber que no debe ser ni demasiado grande, ya que el algoritmo no convergería, ni demasiado pequeño, ya que se necesitarían demasiadas iteraciones.

A pesar de que el algoritmo anterior es válido, no se utiliza en la realidad por la necesidad de calcular el gradiente en cada iteración utilizando para ello todo el conjunto de datos de entrenamiento. Si este conjunto fuese realmente grande, cada iteración de este algoritmo sería extremadamente costosa. 
 
En su lugar, se utiliza la variante llamada \textbf{gradiente descendente estocástico} (\acrshort{sgd}) \cite[p.~97]{abu-mostafa_2012}, que (de forma resumida) consiste en dividir el conjunto de datos de entrenamiento en pequeños grupos disjuntos\footnote{Estos pequeños grupos se conocen como \textbf{minibatches}.} que son usados para el cálculo del gradiente, y se itera múltiples veces sobre el conjunto de datos completo\footnote{Una iteración sobre el conjunto de datos completo, de minibatch en minibatch, se conoce como \textbf{época}.}. Además de que con este algoritmo cada actualización de los pesos es mucho más eficiente, se ha visto que en funciones no convexas usualmente consigue alcanzar mejores mínimos de la función que la variante anterior, evitando puntos de silla y mínimos locales \cite{masters2018revisiting}.
 
 \section{Error de generalización}
 
 Ya sabemos cómo ajustar los parámetros $w$ de nuestro modelo de forma que el error cometido en el conjunto de entrenamiento sea mínimo, pero debemos recordar que el objetivo final del modelo es que funcione ante nuevos datos \textit{nunca antes vistos} en el entrenamiento, esto es lo que conocemos como \textbf{generalización}.
 
 La pregunta es, ¿garantiza tener un bajo error dentro del conjunto de entrenamiento que vayamos a hacerlo bien con nuevos ejemplos?. La respuesta a esta pregunta no es trivial. No vamos a demostrarlo, pero para nuestros objetivos es suficiente con saber que si el conjunto de entrenamiento consiste en ejemplos \textbf{independiente e idénticamente distribuidos} tomados de una población con una distribución de probabilidad $P$, y además es \textit{suficientemente grande}, entonces una función que minimice el error dentro del conjunto de entrenamiento, es probable que cometa también un error bajo en otros datos nunca vistos (tomados de $P$).
 
\begin{figure}
     \centering
     \includegraphics[width=6cm]{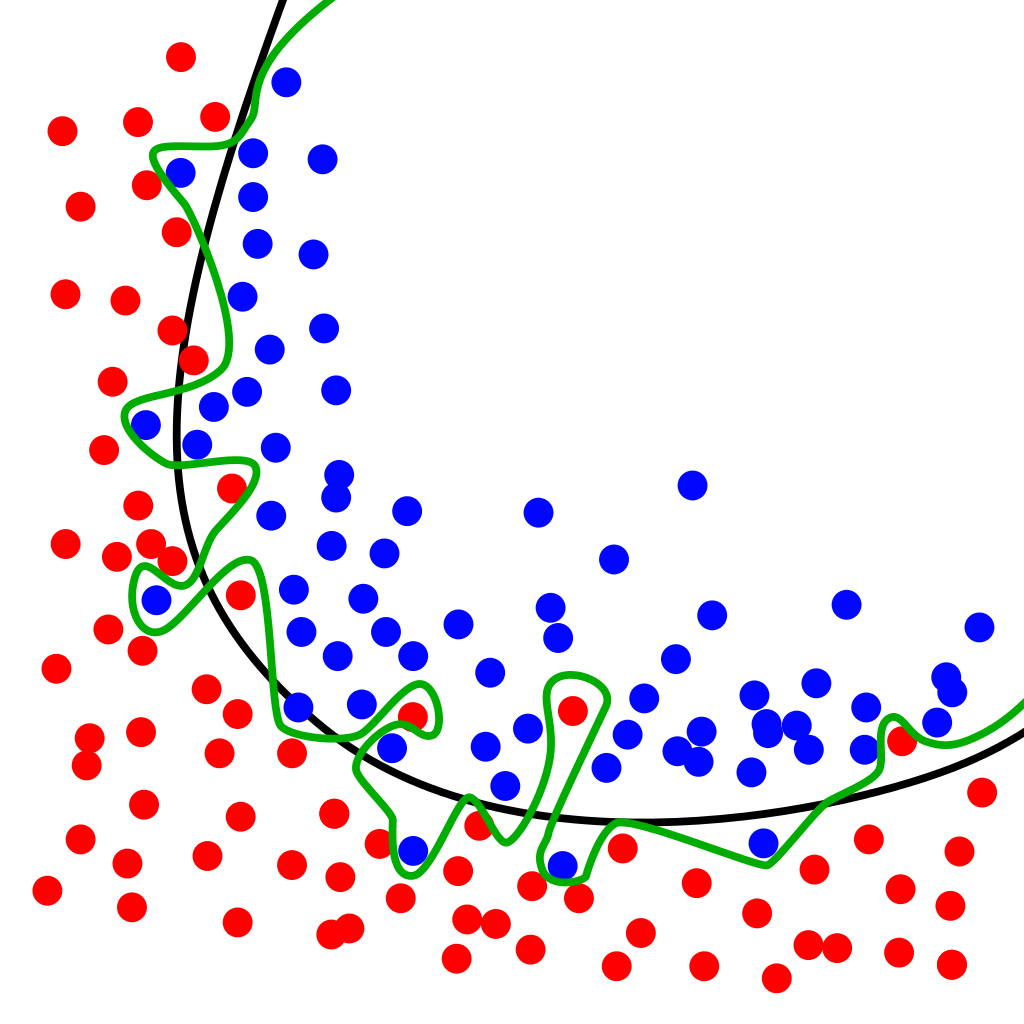}
     \caption[Overfitting]{Overfitting \cite{eswiki:129965975}.}
     \label{fig:overfitting}
 \end{figure}

 Sin embargo, hay que tener cuidado con la afirmación anterior. Supongamos la tarea de separar los puntos azules de los rojos de la figura \ref{fig:overfitting}.
 
 Usualmente, \textbf{existe ruido} dentro de los datos del problema (por ejemplo, etiquetas mal puestas), y además, existen ``peculiaridades'' dentro de los datos de entrenamiento. En la figura \ref{fig:overfitting}, la línea verde representa a un clasificador \textit{excesivamente complejo} que ha aprendido a clasificar perfectamente todos los datos de entrenamiento, incluso los puntos ruidosos y las peculiaridades del conjunto concreto. Podríamos decir que ``ha memorizado, pero no ha entendido la función subyacente que clasifica los datos'', por lo que no generalizará bien a nuevos datos nunca antes vistos.
 
 Cuando un modelo comete un error muy bajo dentro del conjunto de entrenamiento, pero alto cuando clasifica nuevos datos, decimos que existe \textbf{overfitting}, o que presenta una \textit{alta varianza} (ya que el error ante nuevos datos variará mucho dependiendo del conjunto de entrenamiento).
 
 Por otro lado tenemos la línea negra, que representa a un modelo que realmente ha aprendido de los datos: tiene un error bajo dentro del conjunto de entrenamiento, aunque comete algunos errores, pero al ser una función más simple funcionará bien ante nuevos datos. Podríamos decir que esta segunda función ha aprendido la función subyacente que clasifica los datos, y no se ha adaptado al ruido y peculiaridades de la muestra de entrenamiento.
 
 En el lado opuesto al overfitting, estaría el caso en que el clasificador fuera una línea recta. Está claro que con una línea recta sería imposible clasificar bien los datos (ni siquiera los de entrenamiento). En este caso, el error cometido sería alto tanto en el conjunto de datos de entrenamiento, como ante nuevos datos. Cuando esto ocurre decimos que el modelo sufre de \textbf{underfitting}, o que tiene un \textit{alto sesgo} (de algún modo podríamos verlo como si el modelo hubiera asumido que la función objetivo es más sencilla de lo que realmente es).

 \label{regularización}
 Falta introducir un último concepto importante, el de \textbf{regularización}: normalmente, cuando abarcamos un problema de aprendizaje automático, utilizamos  \textit{clases de funciones muy complejas}, con muchos parámetros $w$ a fijar. El problema de estas funciones tan complejas es que tienen la capacidad suficiente para memorizar completamente el conjunto de entrenamiento (especialmente si el conjunto es pequeño), es decir, son muy propensas al overfitting.
 
 Para reducir esta tendencia al overfitting, se pueden aplicar una serie de estrategias para reducir la complejidad de estas funciones y así mejorar su capacidad de generalización. La aplicación de estas estrategias es lo que conocemos como ``regularizar un modelo''.
 
Existen numerosas técnicas de regularización. Más adelante (sección \ref{seccion_regularizacion}), explicaremos algunas de las más utilizadas para regularizar las redes neuronales convolucionales.
 
\section{Evaluando un modelo}

Como hemos visto, el hecho de que un modelo tenga un error bajo dentro del conjunto de entrenamiento \textit{no nos garantiza} que vaya a tener un buen comportamiento ante nuevos datos, ya que puede sufrir de overfitting.

Además del overfitting, existen otros motivos por los que un modelo podría no funcionar bien ante nuevos datos: errores de implementación, toma incorrecta de los datos, etc...

\subsection{Entrenamiento - validación - test}
\label{train_validation_test}

Con el objetivo de estimar el error que cometerá nuestro modelo ante nuevos datos, lo que se hace usualmente es dividir el conjunto de datos inicial en tres subconjuntos: entrenamiento, validación y test.

El objetivo de cada uno de estos conjuntos es el siguiente:\begin{itemize}
    \itemsep0em
    \item \textbf{Entrenamiento} (training): este es el conjunto de datos que se utiliza para entrenar nuestro modelo. Hay que recordar que no podemos evaluar la bondad del modelo en este conjunto, ya que sería una evaluación enormemente sesgada.
    
    \item \textbf{Validación} (validation): este conjunto se utiliza para poder estimar la bondad del modelo ante datos nunca vistos y \textit{tomar decisiones} en base a ello. Por ejemplo, si vemos que el error cometido en training es bajo, pero en validación es alto, sabemos que estamos sufriendo de overfitting y tendremos que tomar alguna decisión para solucionarlo.
    
    Es importante darse cuenta de que por cada decisión que tomemos, estamos haciendo que el error cometido dentro de este conjunto sea menor, por lo que esta estimación del error estará progresivamente más sesgada, ya que en cierto modo, al tomar una decisión, estamos ``aprendiendo'' con los datos de validación.
    
    \item \textbf{Test}: este conjunto sirve exclusivamente para dar una estimación no sesgada del error que cometerá nuestro modelo. Es muy importante resaltar que este conjunto nunca será utilizado ni para el entrenamiento, ni para tomar ninguna decisión sobre el modelo.
    
\end{itemize}

Respecto al tamaño de estos conjuntos, no existe una regla fija. Lo importante en cualquier caso es que tanto validación como test tengan un tamaño suficiente para que sean representativos, teniendo en cuenta que cuanto mayor sea su tamaño, mejor será la estimación, pero menos datos tendremos para entrenar, y en consecuencia peor será el modelo obtenido.

\subsection{Validación cruzada k-fold}
\label{k_fold}

En muchas ocasiones, y sobre todo en el caso de conjuntos de datos pequeños, una partición del tipo ``train-validation-test'' no será suficiente, ya que el valor de la estimación que hagamos sobre el conjunto de validación, será muy dependiente de la partición concreta de los datos, por lo que no podremos comparar de forma ``fiable'' distintos modelos \cite[p.~122]{goodfellow2016deep}.

La validación cruzada k-fold es un procedimiento para obtener una estimación del desempeño de un modelo ante datos no anteriormente vistos, pero de forma \textbf{menos dependiente} del conjunto de validación concreto escogido. El procedimiento es el siguiente:\begin{enumerate}
\itemsep0em
    \item Mezclar el conjunto de datos aleatoriamente
    \item Dividir el conjunto de datos en k grupos disjuntos
    \item Para cada grupo:
    \begin{enumerate}
    \itemsep0em
        \item Tomar el grupo como conjunto de test
        \item Tomar el resto de grupos como conjunto de entrenamiento
        \item Entrenar el modelo con el conjunto de entrenamiento y evaluarlo en el conjunto reservado para test
        \item Guardar el resultado de la evaluación y descartar el modelo
    \end{enumerate}
    \item Obtener la bondad del modelo como la media de cada una de las evaluaciones.
\end{enumerate}

Existen algunas variaciones de este método que ofrecen una estimación más exactas, como \textbf{k-fold repetido} (para reducir la variabilidad), que consiste en aplicar k-fold $n$ veces y promediar los resultados, y \textbf{k-fold estratificado}, en el que se asegura que cada uno de los $k$ grupos tienen la misma distribución de clases. Estas variaciones suelen aplicarse cuando el conjunto de datos que poseemos es especialmente pequeño (como será nuestro caso).

\noindent\textbf{Nota:} aunque hagamos k-fold o k-fold repetido, siempre existirá un cierto sesgo optimista en la estimación, por lo que siempre se mantendrá un conjunto de \textbf{test} que no se utilizará hasta que hayamos seleccionado el mejor modelo final, con el objetivo de poder dar una estimación \textbf{no sesgada} del error del modelo.

\subsection{Métrica de error}

Además de las técnicas anteriores, es muy importante seleccionar una métrica de error que se ajuste a nuestro problema (exactitud, precisión, área por debajo de la curva ROC, etc...), y que nos indique ``cómo de bien'' se está resolviendo.

La selección de una métrica de error es muy dependiente del problema concreto, por lo que se explicará en el \textbf{apartado \ref{metrica_explicacion}},
cuando expliquemos en detalle nuestra propuesta para resolver el problema.

\begin{tcolorbox}[
  colback=Green!5!white,
  colframe=Green!75!black,
  title={Recapitulación}]

Tras este capítulo de introducción a los fundamentos del aprendizaje automático sabemos:  

\begin{itemize}
\small
\itemsep 0.1em 

    \item El problema de clasificación mediante aprendizaje automático consiste en \textbf{aprender} una función que clasifique nuevos datos a partir de un conjunto de ejemplos ya etiquetados.
    \item La \textbf{función de pérdida} nos dice cómo de bien está realizando su tarea un modelo de aprendizaje automático.
    \item El \textbf{entrenamiento} consiste en minimizar el valor de la función de pérdida (usualmente mediante descenso de gradiente).
    \item La \textbf{generalización} es la capacidad de un modelo de realizar correctamente la tarea ante datos nunca antes vistos.
    \item El \textbf{overfitting} ocurre cuando un modelo tiene un bajo error en el conjunto de entrenamiento, pero alto ante nuevos datos.
    \item El \textbf{underfitting} ocurre cuando un modelo tiene un alto error tanto en entrenamiento como en test.
    \item Para crear y \textbf{evaluar} un modelo utilizaremos tres conjuntos: training, validación y test. En el caso de conjuntos de datos pequeños, usaremos k-fold, stratified k-fold, e incluso repeated k-fold.
    
\end{itemize}

\end{tcolorbox}

    \chapter{Redes neuronales}
\label{redes_neuronales}

\begin{tcolorbox}[
  colback=SkyBlue!5!white,
  colframe=SkyBlue!75!black,
  title={Sumario}]

En este capítulo se explicarán muy brevemente algunas de las ideas esenciales tras las redes neuronales, que nos servirán como paso intermedio para poder entender las redes convoluciones. Entre otras cosas, explicaremos:
  
\begin{itemize}
\itemsep 0.1em 
    \item El problema de los modelos lineales como el visto en el capítulo anterior para la clasificación de datos no linealmente separables.
    \item El concepto de neurona artificial y red neuronal, así como una explicación intuitiva de por qué funcionan.
    \item Una idea intuitiva de cómo se entrena una red neuronal mediante los algoritmos de fordward-propagation y back-propagation.
    \item La necesidad de aumentar la profundidad de las redes. 
    \item El concepto de capacidad de una red neuronal y cómo influye en el error de generalización.
\end{itemize}

\end{tcolorbox}

\section{Datos no linealmente separables}

\begin{figure}[H]
    \centering
    \includegraphics[width=6cm]{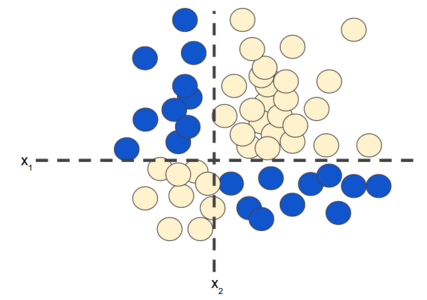}
    \caption[Datos no linealmente separables]{Datos no linealmente separables \cite{imagen_no_separable}.}
    \label{fig:non_linear}
\end{figure}

Supongamos el problema de separar los puntos amarillos de los puntos azules que aparecen en la figura  \ref{fig:non_linear}, e intentemos imaginar una línea recta que separe los puntos: podemos ver claramente que no es posible, ya que los datos no son linealmente separables en este caso.

 \renewcommand{\thefootnote}{\roman{footnote}}

El hecho de que los datos no sean linealmente separables quiere decir que no podemos predecir su etiqueta por medio de una recta definida por una ecuación de la forma $w_0 + w_1x_1 + w_2x_2 = 0$ \footnote{En el caso de un número de características mayor a 2, llamaremos hiperplano a cualquier ecuación de la forma $w_0 +w_1x_1 + ... +w_nx_n = 0$, siendo un hiperplano una extensión de la recta a mayor dimensionalidad. Recordar que con $\boldsymbol{w}$ nos referimos a los parámetros entrenables.}, y por tanto, nuestro modelo de regresión logística visto anteriormente no podría separarlos.

 \renewcommand{\thefootnote}{\arabic{footnote}}

Una posible opción para poder separar los datos sería crear nuevas características polinómicas combinando las características iniciales $x_1$ y $x_2$, de forma que los nuevos datos fueran separables por un hiperplano en una dimensión más alta, pero esta opción carece de interés para nuestros objetivos, por lo que no le prestaremos mayor antención.

La otra opción es utilizar modelos no lineales, como las redes neuronales.

\section{La neurona artificial}

\begin{figure}[H]
    \centering
    \includegraphics[width=5cm]{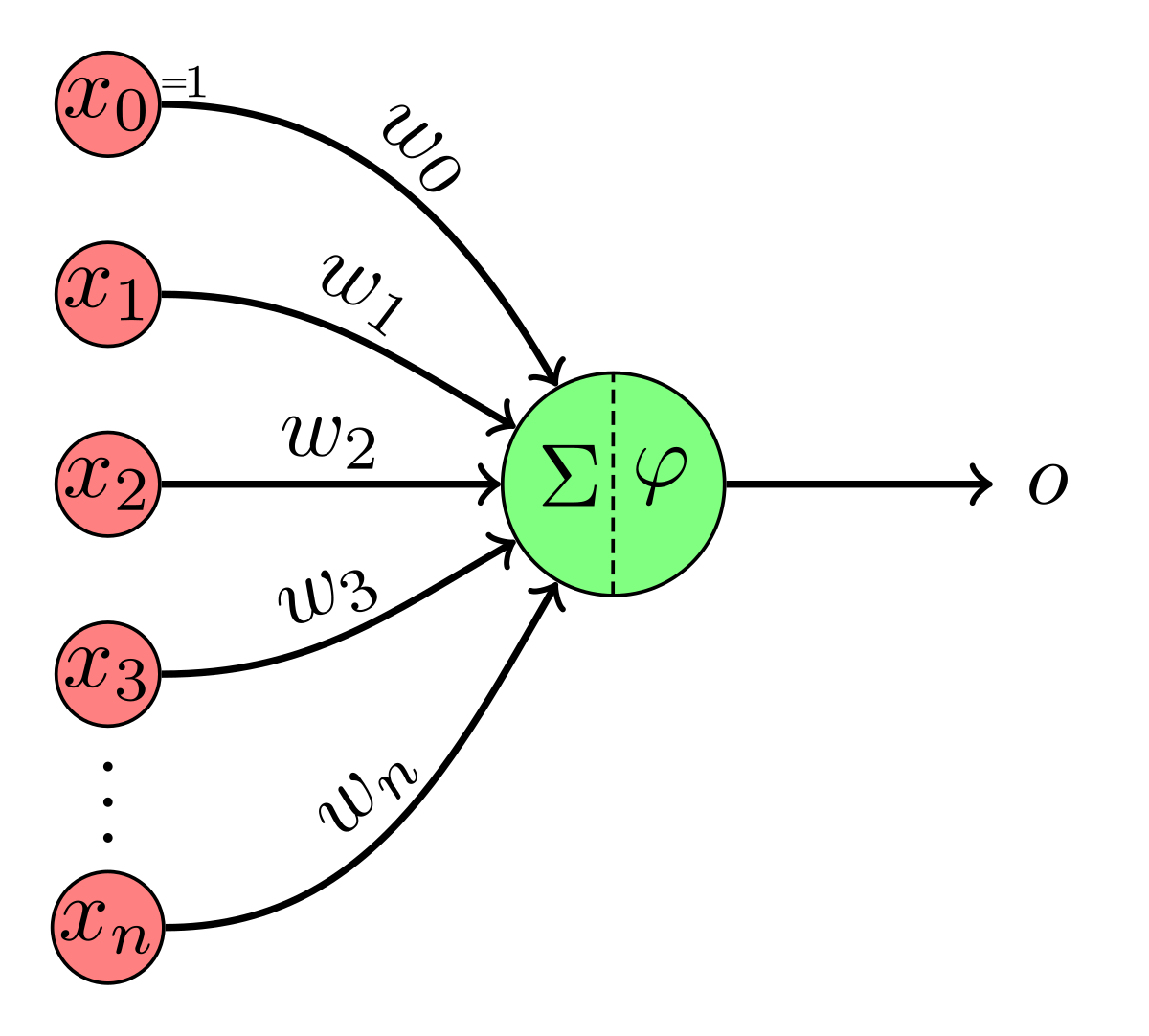}
    \caption[Neurona artificial]{Neurona artificial \cite{wiki:artificial_neuron}. Para que el grafo represente realmente al modelo de regresión logística, $\varphi$ debe ser la función sigmoide (no lineal), y la salida ($o$) se obtendría, como vimos, usando el 0.5 como umbral de clasificación.}
    \label{fig:artificial_neuron}
\end{figure}

Para entender como pueden ayudarnos las redes neuronales con problemas no linealmente separables, podemos primero representar nuestro modelo de regresión logística como un grafo, tal y como el que aparece en la figura \ref{fig:artificial_neuron}.

\noindent \textbf{Nota:} cuando hablemos de redes neuronales, la función $\varphi$ es una función no lineal y derivable conocida como \textit{función de activación}.

Podemos darnos cuenta de que el problema de separar los puntos de la figura \ref{fig:non_linear} no se puede resolver con una sola neurona, ya que podríamos decir que una sola neurona es equivalente al modelo de regresión logística, y como hemos dicho, este modelo sólo puede separar datos linealmente separables.

\section{Red neuronal}

Entonces, ¿cómo podemos conseguir un modelo capaz de resolver problemas no lineales?, la solución es conectar neuronas como las vista anteriormente de una forma estratégica, formando una red neuronal. Veamos un ejemplo: \begin{figure}[H]
    \centering
    \includegraphics[width=7cm]{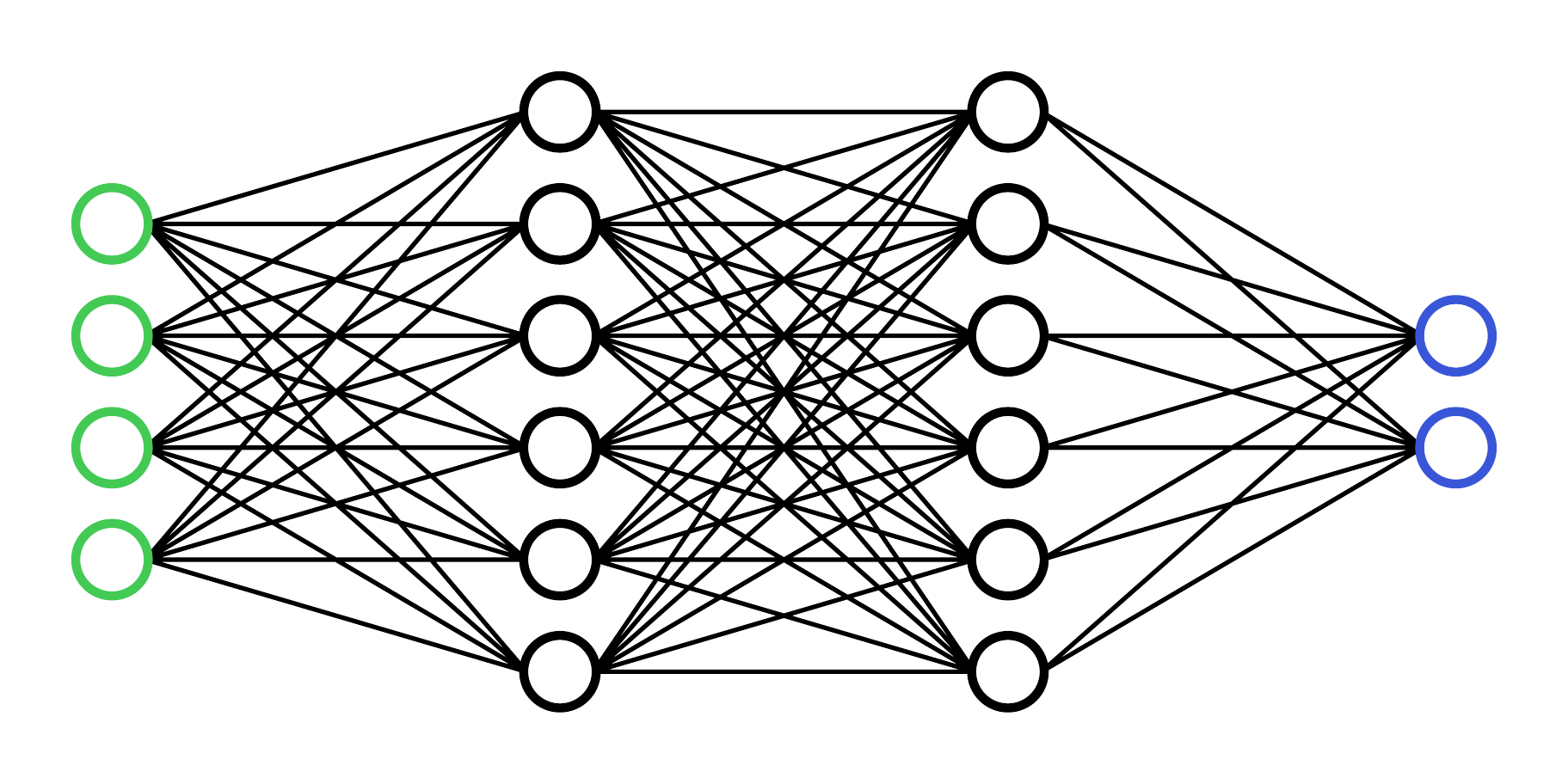}
    \caption[Red neuronal con dos capas ocultas.]{Red neuronal con dos capas ocultas \cite{zhou_2021}.}
    \label{fig:capas_ocultas}
\end{figure}

Vamos a ver los elementos que aparecen en el esquema de la figura anterior: \begin{itemize}
    \item Los nodos verdes de la izquierda representan las características de entrada ($x_1, ..., x_n$), y en conjunto, se conocen usualmente como la \textbf{capa de entrada}.
    \item Cada uno de los nodos en color negro representan neuronas tal y como la que se ha visto en la figura \ref{fig:artificial_neuron}, y como sabemos, cada una produce una determinada salida ante una entrada.
    
    \item Cada una de las filas verticales de nodos de color negro conectados a la capa inmediatamente anterior, se conoce como \textbf{capa oculta}.
    
    \item La última capa de neuronas, en azul, se conoce como \textbf{capa de salida}, y como su nombre indica, es la capa final que nos dará la salida obtenida por la red neuronal.
    
\end{itemize}

\subsection{¿Por qué funciona?}

Si nos fijamos en la primera capa oculta, cada una de las neuronas de esta capa recibirá como entrada los datos de entrada al problema, sin ninguna transformación. Una vez que cada una de las neuronas de esta primera capa ha calculado una cierta función sobre los datos de entrada, pasarán esta salida a las neuronas de la segunda capa oculta.

Entonces, la entrada a las neuronas de la segunda capa oculta es una transformación de los datos iniciales, y de nuevo, las neuronas de la segunda capa realizarán otra nueva transformación sobre su entrada.

Por último, las neuronas de la capa de salida recibirán como entrada la transformación no lineal de los datos de entrada producida por la ``acumulación'' de las transformaciones de las dos capas ocultas anteriores, y volverán a aplicar su función para obtener la salida final.

Este cálculo de \textbf{transformaciones no lineales} sobre los datos de entrada, es lo que permiten a las redes neuronales la clasificación de datos no linealmente separables.

\subsection{Funciones de activación}

Hemos visto entonces que las redes neuronales son capaces de calcular transformaciones no lineales (que pueden llegar a ser muy complejas) de los datos de entrada. Esto ocurre por el hecho de que apilar no-linealidades sobre no-linealidades da lugar a funciones cada vez más potentes. Pero esto sólo puede ocurrir gracias a que tenemos una \textbf{función no lineal $\boldsymbol{\varphi}$} a la salida de cada neurona, ya que por contra, cuando apilamos funciones lineales, la función resultante sigue siendo una función lineal, y no aportaríamos más ``potencia'' al modelo.

Existen distintas funciones de activación que suelen utilizarse en la práctica:\begin{itemize}
    \item Sigmoide: es la que vimos en el modelo de regresión logística.
    \item \textbf{ReLU} (rectified linear unit): es una función más simple de calcular que la sigmoide, $f(x) = max(0, x)$.  Usualmente, esta función aporta mejores resultados que la anterior. Esta superioridad suele relacionarse con el hecho de tener un rango de respuesta más amplio (la función sigmoide se satura en 1 ó 0 rápidamente, a ambos lados, mientras que ReLu sólo satura a la izquierda), que evita el problema del desvanecimiento del gradiente\footnote{Sin entrar en demasiados detalles, el problema de \textbf{desvanecimiento de gradiente}, ocurre cuando el gradiente de la función de pérdida respecto de los parámetros de las primeras capas se hace extremadamente pequeño. Esto hace que ni siquiera grandes cambios en el valor de los parámetros tengan un efecto notable en la salida de la red, lo que complica enormemente el entrenamiento.} durante el entrenamiento de las redes \cite{enwiki:1006243644, agarap2019deep}.
\end{itemize}

\subsection{Clasificación multiclase}

Si necesitáramos clasificar las entradas en $C$ clases diferentes mediante una red neuronal, lo primero que haríamos es representar cada clase como un vector de longitud $C$, en el que todos sus elementos son cero a excepción del elemento que ocupa la posición de la clase que queremos representar, que tomará como valor 1. Por ejemplo, la clase 1 se representaría con un vector de longitud $C$ de la forma $y = [1, 0, ..., 0]$.

En la capa de salida, colocaríamos $C$ neuronas que darán como salida una serie de valores $f_1, f_2, ..., f_c$, y luego, mediante la función \textbf{softmax} convertiríamos estos valores en probabilidades: \[ \hat{y} = softmax(f_1, ... , f_c) = (\frac{e^{f_1}}{\sum_{j=1}^{C}e^{f_j}}, ..., \frac{e^{f_C}}{\sum_{j=1}^{C}e^{f_j}})\]

De forma que si uno de los $f_j$ tiene un valor mucho mayor, su correspondiente valor de softmax será cercano a 1, mientra que el resto serán cercanos a 0.

La forma de conocer la clase final, será simplemente aplicando la función $max(softmax(f_1, ..., f_C))$.

Por último, como función de pérdida para el entrenamiento utilizaríamos la función de \textbf{entropía cruzada}\footnote{La entropía cruzada es ampliamente usada como función de pérdida para clasificación multiclase. Entender los detalles exactos de por qué se usa, queda fuera de los objetivos de este trabajo.}: \[ Loss(y, \hat{y}) = - \sum_{j=1}^{C}y_j \cdot log\hat{y}_j\]

\section{Entrenando una red neuronal}

Como vimos, entrenar un modelo de aprendizaje automático consistía en \textit{minimizar} el valor de una determinada función de pérdida dentro del conjunto de entrenamiento, y usualmente se minimizaba mediante el algoritmo de \textit{gradiente descendente estocástico o alguna variación de él.}

Este concepto no cambia para las redes neuronales, de nuevo necesitaremos una función de pérdida, que se calculará a partir de la salida (dada por la capa de salida) de la red neuronal, y aplicaremos el mismo algoritmo de minimización.

El único ``problema'', es que calcular el valor de la salida de una red neuronal es un procedimiento algo distinto al de calcular el valor de una función corriente, al igual que el cálculo del gradiente de la función de pérdida de una red neuronal. Para calcular estos dos elementos se usan dos algoritmos, que resumiremos muy brevemente, sin entrar en los detalles matemáticos: \begin{itemize}
    \item Algoritmo de propagación hacia adelante ó \textbf{forward-propagation}: se utiliza para calcular el valor de salida de la red neuronal, y consiste simplemente en ir propagando hacia adelante (desde los datos de entrada hasta la capa de salida), el cálculo de las neuronas de la capa inmediatamente anterior.
    
    \item Algoritmo de propagación hacia atrás ó \textbf{back-propagation}: se trata de un algoritmo relativamente complejo, pero en esencia, consiste en calcular el gradiente de la función de pérdida, partiendo del valor de la función en la capa de salida, e ir aplicando la regla de la cadena para ir ``propagando'' el gradiente hasta la primera capa oculta \cite{rumelhart1986learning}.
\end{itemize}

\subsection{Optimizadores}
 
 Actualmente, en el ámbito de las redes, el algoritmo \acrshort{sgd} no se utiliza realmente. En su lugar, se utilizan otros algoritmos mejorados como \textbf{Adam} \cite{kingma2014adam} ó \textbf{RMSProp} (propuesto por el profesor Geoff Hinton en sus clases).
 
 No es nuestro objetivo explicar estos optimizadores, pero la idea importante a conocer es que en general consiguen una optimización más rápida, además de lograr mejores mínimos en caso de funciones no convexas.

\section{¿Cómo afecta el número de capas?}
\label{numero_capas}

El teorema de aproximación universal \cite{hornik1989multilayer} nos dice que una red neuronal con una única capa oculta, y con el suficiente número de neuronas en dicha capa, podría aproximar a \textbf{cualquier} función continua en un subconjunto cerrado y acotado de $R^n$. 


En la práctica, este teorema quiere decir que sea cual sea la función que estemos tratando de aprender, sabemos que una red neuronal con una sola capa oculta y el número suficiente de neuronas será capaz de \textit{representar} a esa función. Sin embargo, no tenemos una garantía de que el algoritmo usado para el entrenamiento sea capaz de \textit{aprender} dicha función, el aprendizaje podría fallar por dos razones principales:\begin{itemize}
    \item El algoritmo de optimización utilizado para el entrenamiento (\acrshort{sgd}, por ejemplo), podría no ser capaz de encontrar el valor de los parámetros que correspondan con la función deseada.
    \item El algoritmo de optimización podría elegir la función incorrecta debido al overfitting.
\end{itemize}

Por otro lado, el teorema de aproximación universal dice que existe una red lo suficientemente grande para conseguir aproximar cualquier función que queramos, pero no especifica cuál debería ser el tamaño de esa red. En 1993 se dieron algunos límites superiores sobre el tamaño necesario en una red neuronal de una sóla capa oculta para aproximar ciertas clases de funciones, y lamentablemente, demostraron que hay casos en que el número de neuronas necesario debería ser exponencial en términos del número de dimensiones de la entrada \cite{barron1993universal}.

Por tanto, una red neuronal de una sola capa podría en teoría aproximar cualquier función que queramos, pero la capa podría ser de un tamaño inmanejable, o el algoritmo de aprendizaje podría fallar para aprender y generalizar correctamente.

Más tarde, se demostró que para algunos tipos de funciones, el \textbf{aumentar la profundidad} de las redes (número de capas ocultas), hace que el número de neuronas necesario sea extremadamente menor \cite{montufar2014number}. 

\begin{figure}
    \centering
    \includegraphics[width=10cm]{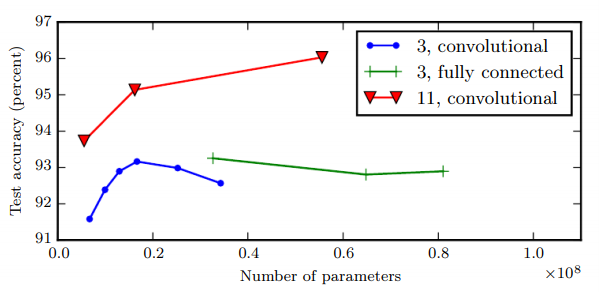}
    \caption[Efecto de la profundidad de las redes]{Parte del experimento de Ian Goodfellow sobre la clasificación de dígitos manuscritos mediante redes neuronales profundas \cite{goodfellow2013multi}. Figura tomada del libro Deep Learning \cite[p.~203]{goodfellow2016deep}.}
    \label{fig:experimento_profundidad_goodfellow}
\end{figure}

En general, se ha visto que \textbf{las redes neuronales profundas funcionan mejor}. Por ejemplo, en el experimento de la figura \ref{fig:experimento_profundidad_goodfellow}, se muestra que ante el mismo número de parámetros, una red más profunda obtiene un rendimiento notablemente superior, y no sólo eso, también se enseña que las redes más profundas, pueden soportar un número de parámetros mucho mayor sin sufrir overfitting.

Un detalle a tener en cuenta, es que cuando decidamos utilizar una red neuronal profunda, estamos asumiendo que la función que estamos tratando de aproximar consiste en la composición de muchas funciones simples (cada capa aplica una función simple) \cite[p.~201]{goodfellow2016deep}.

\section{Capacidad: overfitting y underfitting}
\label{capacidad}

Antes de pasar a las redes convolucionales vamos a introducir el concepto de \textbf{\textit{capacidad}} de una red. En aprendizaje automático, se le llama \textit{capacidad} de un modelo a la amplitud de la clase de funciones que éste puede representar. En el caso de una red neuronal, la capacidad vendrá determinada principalmente por dos aspectos: el número de capas y el número de parámetros.

Un modelo con un mayor número de parámetros tiene mayor capacidad de \textit{memorización} \cite[p.~97]{chollet2017deep}, y por tanto puede aprender fácilmente una correspondencia entre los datos de entrenamiento y sus etiquetas, pero podría ser una correspondencia sin ningún poder de generalización\footnote{De forma intuitiva, si una red tiene muchos parámetros, es capaz de memorizar un elevado número de patrones irrelevantes en los datos de entrenamiento, sin embargo si tiene menos parámetros, tendrá que ``esforzarse'' en enfocarse únicamente en los patrones importantes, dándole una mayor opción de que generalice bien.}. Dicho de otro modo, una red con demasiada capacidad será una red tendiente al \textit{overfitting}.

En el lado contrario, una red con una capacidad demasiado reducida, no será capaz de memorizar si quiera los patrones importantes dentro de los datos, por lo que sufrirá de \textit{underfitting}. Por tanto, habrá que buscar un equilibrio entre estos dos extremos.

Hay que destacar que aunque tanto el aumento de capas como el aumento del número de parámetros aumentan la capacidad de un modelo, como se dijo en la sección anterior, en general, las redes con más capas funcionan mejor.

\begin{tcolorbox}[
  colback=Green!5!white,
  colframe=Green!75!black,
  title={Recapitulación}]

Las ideas clave a recordar sobre este capítulo son:  

\begin{itemize}
\small
\itemsep 0.1em 
    \item Las redes neuronales pueden realizar transformaciones \textbf{no lineales} complejas de los datos, por medio de sucesivas transformaciones más simples.
    \item En general, \textbf{las redes con un mayor número de capas funcionan mejor} para muchos problemas, aunque el motivo no está del todo claro.
    \item La \textbf{capacidad} de un modelo se define como la amplitud de la clase de funciones que puede representar, y se controla por medio del número de parámetros y del número de capas.
    \item Una red con demasiada capacidad tenderá al \textbf{overfitting}, mientras que una con una capacidad baja sufrirá de \textbf{underfitting}.
\end{itemize}

\end{tcolorbox}

    \chapter{Clasificación de imágenes: CNN}
\label{capitulo_cnn}

\begin{tcolorbox}[
  colback=SkyBlue!5!white,
  colframe=SkyBlue!75!black,
  title={Sumario}]

En este último capítulo sobre los fundamentos teóricos, estudiaremos el uso de redes convolucionales profundas para la clasificación de imágenes. En concreto veremos: 

\begin{itemize}
\itemsep 0em 
    \item Por qué las redes neuronales convencionales no son el mejor modelo para la clasificación de imágenes.
    \item Qué es la operación de convolución y una red convolucional.
    \item Cuáles son las principales capas que existen en una red convolucional.
    \item Algunas técnicas de regularización.
    \item Algunas arquitecturas de redes convolucionales modernas.
    \item El motivo de la tendencia general a hacer las redes más profundas
    \item El concepto de transferencia de aprendizaje (transfer learning).
\end{itemize}

\end{tcolorbox}

Las redes neuronales convolucionales (\acrshort{cnn}) son similares a las redes neuronales ``normales'' vistas en el capítulo anterior: están formadas por capas de neuronas que tienen parámetros entrenables. Cada una de estas neuronas recibe una entrada desde la capa anterior, realiza el producto de la entrada por unos pesos (los parámetros) y aplica posteriormente una función no lineal (sigmoide, ReLU,...). En la última capa de la red, seguiremos teniendo una salida que expresa por ejemplo, la clase a la que pertenece una determinada imagen, y tendremos también una función de pérdida que ``compara'' la salida de la red con la salida esperada para saber cómo de bien se está realizando la tarea.

El principal cambio respecto a las redes convencionales es el hecho de que la arquitectura de las redes convolucionales \textit{asume que las entradas de la red serán imágenes} (en las que existe dependencia local de los píxeles), lo que permitirá un nuevo concepto de neurona, que reducirá enormemente el número de parámetros entrenables, aunque ahora se verá en detalle.

\section{Problema: clasificación de imágenes}

Supongamos que queremos resolver un problema de clasificación de imágenes por medio de redes neuronales convencionales.

Vamos a suponer que las imágenes utilizadas son pequeñas, con un tamaño de 32x32x3, es decir, (32 de ancho, 32 de alto y 3 canales de color). En este caso, si tomáramos cada píxel de la imagen como una característica, una neurona en la primera capa oculta de una red neuronal convencional como las del apartado anterior tendría $32*32*3 = 3072$ parámetros.

Si en lugar de imágenes pequeñas, tuviéramos una imagen de 5 megapíxeles, cada neurona tendría 15 millones de parámetros entrenables. Aunque en la práctica no se usan resoluciones tan altas, podemos ver que el número de parámetros entrenables sería enormemente grande si enfocamos el problema con redes normales.

El problema de este número tan elevado de parámetros es que la red tendría una enorme \textbf{\textit{capacidad}}, lo que causaría un grave problema de \textit{overfitting}. Para solucionar este problema del elevado número de parámetros, las redes neuronales convolucionales utilizarán otro tipo de operación: la \textbf{convolución}.

\section{Convolución sobre imágenes}

\noindent\textbf{Notación:} de ahora en adelante, llamaremos máscara ó kernel a una matriz de números de tamaño $k \times k$.

Aunque la operación de convolución tiene un significado más amplio, para nuestro objetivo diremos que se trata de un tipo ``especial'' de operación que se realiza sobre imágenes. Se trata de una operación que calcula una función del vecindario local de cada píxel de una imagen \cite[p.~111]{szeliski2010computer}, y viene definida por un filtro o máscara que nos dice cómo combinar los valores de este vecindario. Cuando se aplica esta operación a una imagen, podríamos decir que \textit{extraemos parte de la información que hay en ella}, aunque luego detallaremos esta idea.

\subsection{Operación de convolución}
\label{convolucion_2d}

Para formalizar la idea de convolución, supongamos una imagen $I$ (una matriz bidimensional de números) y una máscara $F$ de dimensiones $k \times k$, con $k$ más pequeño que las dimensiones de la imagen. La operación de convolución de la máscara $F$ sobre la imagen $I$ se define como (siendo $G$ la matriz resultado): \[ G(i, j) = (K * I)(i,j) = \sum_{u=0}^{k} \sum_{v=0}^{k} I(i - u, j - v)F(u, v)\]

Sin embargo, cuando hablamos de convolución en el ámbito de las redes neuronales convolucionales, usualmente la operación que se utiliza es la correlación cruzada (misma operación, dándole la vuelta a la máscara): \[ G(i, j) = \sum_{u=0}^{k} \sum_{v=0}^{k} I(i + u, j + v)F(u, v)\]

\begin{figure}[H]
    \centering
    \includegraphics[width=8cm]{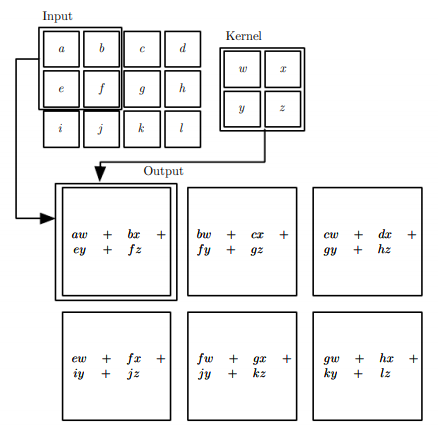}
    \caption[Operación de correlación]{Operación de correlación \cite[p.~334]{goodfellow2016deep}. Consiste en ``deslizar'' una máscara (Kernel en la figura) sobre la imagen (Input), de forma que en cada paso, se realiza la suma ponderada de los píxeles de la imagen por los pesos contenidos en la máscara, y se almacena el resultado en una matriz resultado (Output). Vemos como la esquina superior izquierda de la salida se genera por la aplicación de la máscara sobre los 4 píxeles de la esquina superior izquierda de la imagen.}
    \label{fig:operacion_correlacion}
\end{figure}

En la figura \ref{fig:operacion_correlacion} vemos la operación de correlación de forma visual. También se aprecia en la figura, que al aplicar una convolución (o correlación), la matriz de salida queda restringida a aquellas posiciones en las que la máscara ``encaja'' perfectamente, por lo que obtenemos una salida de menor tamaño. Si quisiéramos mantener el tamaño de salida, usualmente se introducen píxeles de relleno en la imagen para compensar esta pérdida (padding).

En el ámbito de las redes convolucionales, la salida $G$ de la operación se conoce como \textit{mapa de características}, y tendrá un valor alto en aquellas localizaciones donde la máscara tenga una alta \textit{similitud} con la porción de la imagen a la que se aplica \cite{stanford_image_filtering}, y bajo cuando la similitud sea baja.

\subsection{Convolución como extractor de características}

Algo realmente interesante de la operación anterior, es que dependiendo del filtro (o máscara) utilizado, podremos extraer distinto tipo de información de una imagen.

\begin{figure}[H]
    \centering
    \subfloat[Imagen original \cite{lena}]{%
        \includegraphics[width=0.4\textwidth]{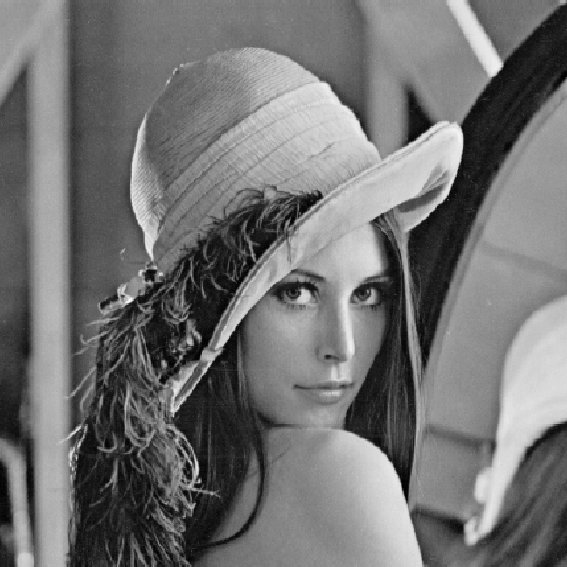}%
        \label{fig:a}%
        }%
    \hfill%
    \subfloat[Tras aplicar el filtro]{%
        \includegraphics[width=0.4\textwidth]{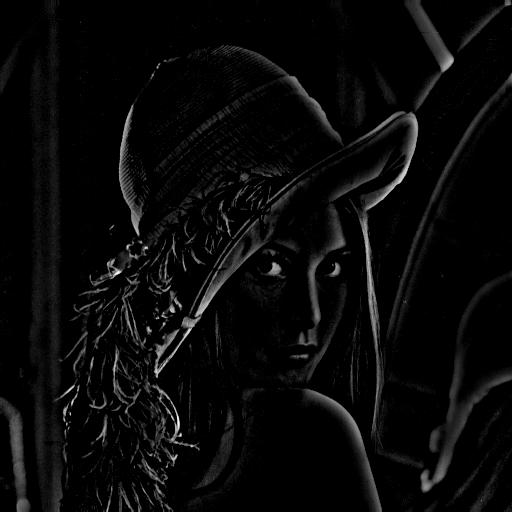}%
        \label{fig:b}%
        }%
    \caption[Filtro de laplaciana]{Filtro de laplaciana. En la imagen de la derecha, un valor más próximo al blanco representa un valor más alto (fuerte respuesta del filtro).}
    \label{fig:lena}
\end{figure}

Por ejemplo, vamos a partir del siguiente filtro, y vamos a realizar la operación de convolución del filtro por una imagen. \footnote{Se trata de un filtro de Laplaciana (segunda derivada), aunque no es un detalle vital para nuestra explicación.} El resultado se muestra en la figura \ref{fig:lena}. \[\begin{bmatrix}
0 & 1 & 0\\
1 & -4 & 1\\
0 & 1 & 0
\end{bmatrix}\]

Cuando un determinado filtro se ``desliza'' sobre la imagen, este producirá una respuesta alta sobre algún cierto tipo de patrón local de la imagen, generando un mapa de características que resume la presencia de ese tipo de patrón. Por ejemplo, el filtro anterior está diseñado de forma que genera un valor alto al pasar sobre los \textbf{bordes}, como se puede apreciar en la figura \ref{fig:lena}. Esta respuesta ante un cierto tipo de patrones podríamos entenderla como la \textit{extracción} de un cierto tipo de característica, como la extracción de bordes en este caso. Además, la modificación de los valores del filtro permitirá extraer distinto tipo de información \cite[p.~107]{forsyth2011computer}.

\section{Red convolucional}

Conociendo la operación de convolución y las redes neuronales convencionales, es sencillo entender en qué consiste una red convolucional.

En las redes ``normales'', cada capa estaba formada por varias neuronas, y cada una de ellas estaba conectada por medio de un peso (parámetro) a cada una de las salidas de la capa anterior. Ahora, en lugar de estas neuronas, \textbf{en cada capa tendremos un determinado de número \textit{filtros}}, y cada número contenido en cada filtro, será un \textit{parámetro entrenable}.

La operación que se realizará en cada capa, será una convolución de la entrada a dicha capa con cada uno de los filtros de la capa, y la salida será la concatenación de las salidas obtenidas al convolucionar la entrada con cada uno de los filtros. Luego, al igual que en las redes convencionales, se le aplicará a la salida una función \textit{no lineal derivable} (sigmoide, ReLU, u otra).

Por otro lado, dijimos que las redes neuronales, en esencia, lo que consiguen es aprender los parámetros necesarios, de forma que la red implemente una función no lineal (como una composición de las funciones de cada capa), que transforma los datos de entrada hasta una salida, que puede ser, por ejemplo, tan simple como una etiqueta.

Las redes convolucionales, por medio del aprendizaje de los parámetros de los filtros, aprenderán una serie de transformaciones sobre las imágenes, que extraerán progresivamente la información relevante para resolver un determinado problema (clasificar una imagen, por ejemplo).


\subsection{Capas}

A continuación veremos algunos de los tipos principales de capas con los que se construye una red convolucional. Más adelante, veremos otros tipos muy usados.

\subsubsection{Capa convolucional}

La capa convolucional será la fundamental en las redes convolucionales, ya que es la que contiene la inmensa mayoría de parámetros entrenables del modelo, es decir, son las capas que \textit{aprenden}.

\noindent\textbf{Nota:} hasta ahora, habíamos dicho que un filtro era una matriz de tamaño $k \times k$ que se podía convolucionar con una matriz de un cierto tamaño $n \times m$. Sin embargo, usualmente tenemos una tercera dimensión que conocemos como el número de canales o \textit{profundidad} (3 canales en el caso de imágenes RGB, por ejemplo). La forma de aplicar la convolución de un filtro $ k \times k \times c$ con una matrix $n \times m \times c$ se mantiene idéntica, sin más que ``deslizar'' el filtro sobre las dimensiones $m$ y $n$, y multiplicando punto a punto los $k * k * c$ parámetros del filtro por los valores de la posición correspondiente de la matriz. La salida de aplicar una convolución de un filtro sobre una matriz, \textbf{siempre} será una matriz de dimensiones $m' \times n' \times \boldsymbol{1}$, con $m' = m - k + 1, n' = n - k + 1$

Los parámetros entrenables de una capa convolucional consisten en un conjunto de filtros, cada uno con el mismo número de canales que su entrada. Cada uno de estos filtros, cuando es ``convolucionado'' con la entrada, produce una salida bidimensional. Luego, las salidas de cada convolución se concatenan formando la salida final de la capa, por ejemplo: si en una determina capa tenemos 6 filtros, la salida de la capa sería de dimensiones $m' \times n' \times 6$.

Intuitivamente, durante el entrenamiento, la red aprenderá los parámetros necesarios para que los filtros consigan extraer información relevante de la entrada.

Una capa convolucional se puede especificar mediante los siguientes \textbf{hiperparámetros} principales \footnote{Llamamos hiperparámetros a aquellos parámetros de una arquitectura que \textit{no} son entrenables.}:\begin{itemize}
    \item \textbf{Profundidad:} corresponde al número de filtros que queremos utilizar, cada uno de los cuales puede aprender algo diferente de la entrada. 
    \item \textbf{Anchura:} especifica el tamaño de cada uno de los filtros. Notar que sólo habría que especificar las dos primeras dimensiones, ya que la tercera viene dada por la profundidad de la entrada.
    \item \textbf{Stride:} cuando vimos la operación de convolución, en cada paso se deslizaba el filtro en una única posición, si saltáramos las posiciones de 2 en 2 o más, diríamos que estamos usando un stride de 2, 3...
    \item \textbf{Relleno (padding):} como dijimos, en ocasiones podríamos desear introducir ``relleno'' sobre la entrada para evitar la disminución de tamaño que produce la convolución.
    \item \textbf{Activación:} función de activación (no lineal) que se aplica a cada uno de los valores de la salida (punto a punto).
\end{itemize}
    
\begin{formal}
{Nota sobre el \bf número de parámetros}

Tomando como entrada una imagen RGB de resolución $1920\times1080$ tendríamos $1920 * 1080 * 3 = 6220800$ píxeles, que es igual al número de parámetros que tendría una sola neurona convencional en la primera capa (en el caso de usar redes neuronales convencionales).

Por contra, usando un filtro de tamaño $3 \times 3 \times 3$ (tamaño usualmente utilizado), tan sólo tendríamos 27 parámetros.

\end{formal}

\subsubsection{Capa de pooling}

Si únicamente utilizáramos capas convolucionales, la imagen inicial iría reduciendo su tamaño espacial progresivamente, pero esta reducción sería excesivamente lenta, requiriendo un número quizás demasiado elevado de capas, que implicaría un número elevado de parámetros.

\begin{figure}[H]
    \centering
    \includegraphics[width=7cm]{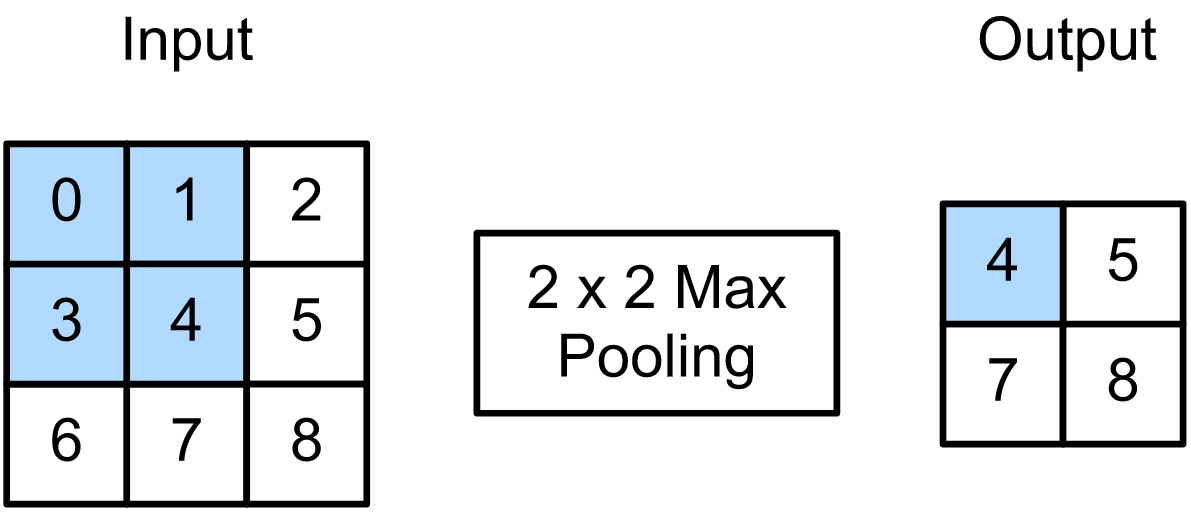}
    \caption[Pooling]{Max pooling \cite[sección 6.5]{zhang2021dive}. Consiste en deslizar una ventana cuadrada de un cierto tamaño (normalmente $2 \times 2$) y dar como salida el máximo de los valores contenidos en la ventana. Existen otras variaciones como \textit{average pooling}, donde se calcula el promedio de valores en la ventana, ó \textit{global average pooling}, donde la ventana tiene el mismo tamaño que la entrada, y la salida es el promedio de todos los valores. En el caso de una entrada con múltiples canales, la operación se aplica de forma independiente sobre cada canal, dando una salida con el mismo número de canales que la entrada.}
    \label{fig:max_pooling}
\end{figure}

Es común insertar una capa de ``pooling'' entre capas convolucionales consecutivas. Su función es la de reducir progresivamente la dimensión espacial de la representación para reducir el número de parámetros entrenables en la red, reduciendo el costo computacional, y controlando el overfitting. La operación realizada por este tipo de capa queda descrita en la figura \ref{fig:max_pooling}.

\subsubsection{Capa totalmente conectada}

Una capa totalmente conectada es el tipo de capa que vimos en las redes convencionales. Están formadas por un conjunto de neuronas, donde cada una de ellas está ``unida'' a todas las salidas de la capa de entrada.

En las \acrshort{cnn}, generalmente, estas capas se sitúan como las últimas capas, después de las capas convolucionales, y actúan como clasificador final. Podríamos decir que las capas convolucionales se encargan de extraer las características útiles de las imágenes, y las últimas capas totalmente conectadas utilizan estas características (resumen de la imagen) para realizar la clasificación final.

\subsubsection{Batch normalization}

Aunque no se ha dicho previamente, en muchas ocasiones la estandarización de las características de entrada a un modelo (hacer que la media del valor de una determinada característica sea 0, y su varianza 1) hace que el proceso de minimización (aprendizaje) sea más rápido.

Por otro lado, el entrenamiento de las redes neuronales se hace más difícil a medida que la profundidad aumenta. Conseguir que una red profunda aprenda en un tiempo razonable puede llegar a ser difícil.

La técnica de normalización a nivel de minibatch (Batch normalization \cite{ioffe2015batch}) podríamos decir que es una evolución de la estandarización que se hace sobre un conjunto de datos. Es una técnica muy popular y efectiva para \textbf{acelerar la convergencia} en redes profundas y funciona de la siguiente forma: en cada iteración del entrenamiento (es decir para cada minibatch), primero se normalizan las entradas restando su media y dividiendo por la desviación típica. Luego, se aplica un coeficiente de escalado y otro de desplazamiento (que tienen que ser aprendidos). Además, durante el entrenamiento se calculan una media y varianza en movimiento (que ``simularán'' la media y varianza del conjunto de datos completo), que se usarán en tiempo de inferencia para realizar la normalización.

\noindent\textbf{Nota:} a lo largo de este trabajo, es probable que en algún momento nos refiramos al parámetro \textbf{momento} (momentum) de las capas Batch normalization. No entraremos en muchos detalles, pero de forma intuitiva, este parámetro ajusta la importancia que se le da a la media y varianza en movimiento respecto a la media y varianza del minibatch, y se usa en la actualización de estas medias en movimiento.

\subsection{Regularización}
\label{seccion_regularizacion}

En el apartado \ref{regularización} hablamos del término \textbf{regularización}, que en esencia, consistía en reducir la complejidad de un modelo con el fin de que la clase de funciones capaz de representar se vea reducida, es decir, reducir ``la potencia'' del modelo. Esto se hacía con el fin de reducir el \textit{overfitting}.

Ahora presentaremos algunas de las técnicas más utilizadas en las \acrshort{cnn} con el fin de regularizar. En algunos casos, no entraremos en los detalles exactos de por qué algunas de estas técnicas funcionan, sino que daremos una idea intuitiva.

\subsubsection{Capa dropout}

Este tipo de capa puede ser introducida tanto antes de una capa convolucional, como antes de una totalmente conectada.

Este tipo de capa tiene la función de ``apagar'' (dar valor cero) la salida de algunas neuronas de forma \textbf{aleatoria}. En el artículo original donde se propuso la idea \cite{srivastava2014dropout}, los autores defienden que el overfitting de las redes neuronales se caracteriza por un estado en el que cada capa depende de un patrón muy específico de activaciones de la capa previa, a lo que denominan \textit{coadaptación}. Al ``apagar'' ciertas neuronas de la capa anterior, la técnica de dropout obliga a que las neuronas aprendan sin depender únicamente de ciertas neuronas, rompiendo así esta coadaptación.

\begin{figure}[H]
    \centering
    \includegraphics[width=7cm]{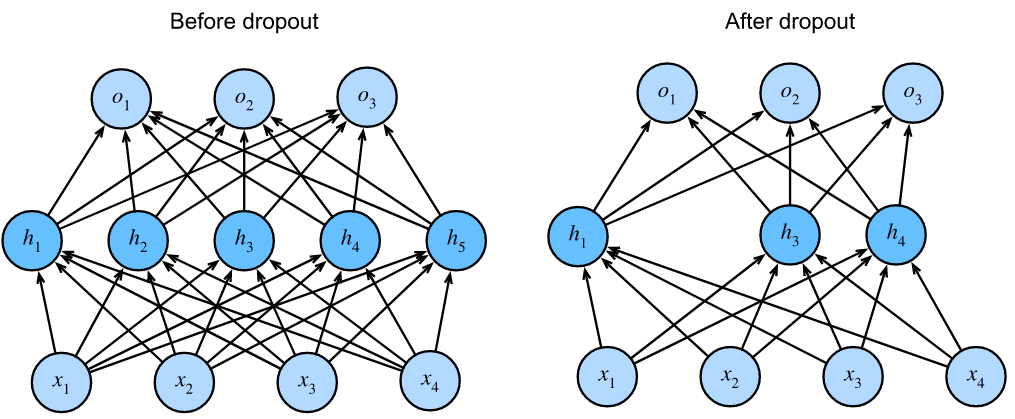}
    \caption[Dropout]{Capa de dropout \cite[sección 4.6]{zhang2021dive}. Tras aplicar una capa de dropout sobre la última capa ($x_1, ..., x_4$), se han desactivado ciertas neuronas aleatorias de la capa anterior ($h_2, h_5$), de forma que las neuronas de esta última capa se ven obligadas a aprender con las neuronas que no han sido desactivadas.}
    \label{fig:dropout}
\end{figure}

Más allá del por qué de su funcionamiento, lo cierto es que en general, añadir alguna capa de dropout en las últimas capas totalmente conectadas de una red, es capaz de reducir el overfitting.

\subsubsection{Regularización L2}

Para explicar esta técnica, podemos partir de la asunción de que la complejidad de una determinada hipótesis $h$ se puede medir por el tamaño de los coeficientes (parámetros) usados para representarla.

Partiendo de ahí, esta técnica consiste en sumar a la función de pérdida del modelo un nuevo término que penalizará el tamaño de los coeficientes, de forma que estos se mantengan pequeños. Dado el vector de parámetros $w$ de un modelo, de forma simplificada, lo que haríamos sería sumar a la función de pérdida el cuadrado de la norma L2 del vector: $\| w \| ^2 = w_1^2 + ... + w_n^2$.

Para ser más exactos, en lugar de sumar directamente $\| w \|^2$, el término que se suma es de la forma $\lambda \cdot \| w \|^2$, siendo $\lambda$ un parámetro que especifica la fuerza de la penalización aplicada. Un valor de $\lambda$ igual a cero, sería equivalente a no aplicar regularización.

\subsubsection{Early stopping}

Cuando entrenamos una red convolucional (o cualquier otro modelo), el algoritmo de aprendizaje itera múltiples veces sobre el conjunto de datos de entrenamiento. Cada una de las veces que se itera sobre el conjunto de datos completo se conoce como \textbf{época}.

Si el modelo entrenado tiene la suficiente capacidad, a medida que avanzan las épocas del entrenamiento, el error dentro del conjunto de entrenamiento irá disminuyendo progresivamente, hasta al final llegar a valores cercanos al cero. Sin embargo, como sabemos, es muy probable que no vaya ocurriendo lo mismo si lo evaluamos ante datos no vistos en el entrenamiento debido al overfitting.

\begin{figure}[H]
    \centering
    \includegraphics[width=6cm]{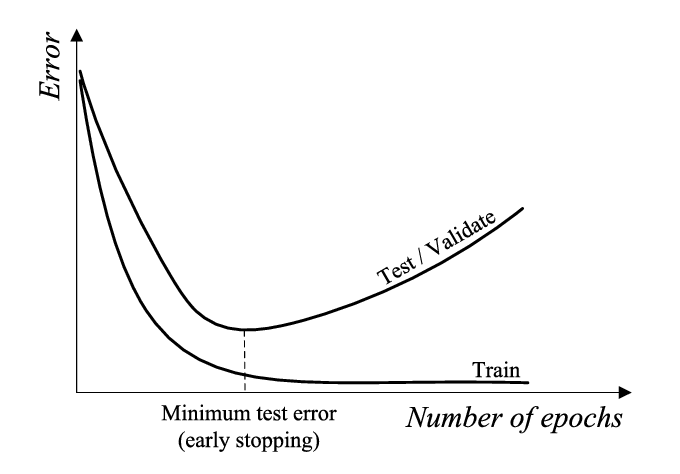}
    \caption[Curva de aprendizaje]{Curva de aprendizaje \cite{early_stopping_img}.}
    \label{fig:early_stopping}
\end{figure}

Para aplicar la técnica de early stopping, se reserva un pequeño conjunto de validación (10-20\% de los datos habitualmente), y se va comprobando, durante el entrenamiento, cómo va progresando el error cometido tanto en ese conjunto, como en el de entrenamiento. Usualmente, la variación de ambos errores se representa en una gráfica conocida como \textbf{curva de aprendizaje}, como la de la figura \ref{fig:early_stopping}
\label{curva_aprendizaje}

La técnica consiste, simplemente, en \textbf{detener el entrenamiento} del modelo en aquella época que minimice el error en el conjunto de validación, antes de que el overfitting haga que el error de generalización aumente.

\subsubsection{Data augmentation}
\label{data_augmentation}

Tener un conjunto de datos \textbf{grande} es un requisito esencial para la buena generalización de las redes neuronales profundas, y esto se hace especialmente importante en el caso de la \textit{visión por computador}.

La técnica de \textit{data augmentation} (aumento de datos en español) consiste en generar imágenes similares a las imágenes presentes en los datos de entrenamiento. Esta generación se realiza mediante una serie de \textbf{modificaciones aleatorias} de las imágenes de entrenamiento: giros, cambios de intensidad y color, zoom, etc...

Aunque esto consigue un efecto de aumentar el tamaño del conjunto de datos, hay que tener en cuenta que no será tan efectivo como conseguir nuevos datos reales, ya que aunque sean distintas, siempre existirá una alta correlación entre las imágenes ``aumentadas'' y las originales.

Además del efecto de aumentar el tamaño del conjunto de datos, esta técnica puede estar motivada por el hecho de que estos cambios aleatorios pueden permitir a los modelos depender menos de ciertos atributos concretos, mejorando así su capacidad de generalización.

\section{Algunas redes}

Ahora que tenemos todas las piezas para construir una \acrshort{cnn}, podemos estudiar aunque sea muy brevemente algunas de las arquitecturas clásicas, que nos pueden servir de inspiración para posteriormente construir una arquitectura propia.

\subsection[LeNet-5]{LeNet-5 \texorpdfstring{\cite{lenet5}}{}}

LeNet-5 fue una de las primeras \acrshort{cnn} que captaron la atención de un gran público gracias a su rendimiento en tareas de visión por computador, más concretamente, se utilizó para la clasificación de dígitos manuscritos (con un conjunto de datos relativamente pequeño: 600000 ejemplos en training y 10000 en test). 

\begin{figure}[H]
    \centering
    \includegraphics[width=12cm]{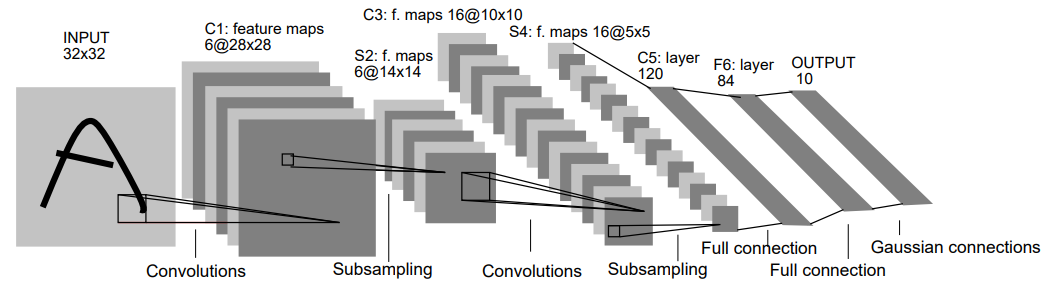}
    \caption[LeNet-5]{LeNet-5: resumen de la arquitectura \cite{lenet5}.}
    \label{fig:lenet_5}
\end{figure}

En la figura \ref{fig:lenet_5} vemos un resumen de la arquitectura, donde los ``feature maps'' son las salidas de cada capa convolucional (y entrada de la siguiente). Estas son algunas de las características a destacar:\begin{itemize}
    \itemsep0em
    \item \textbf{Tamaño de los filtros}: se utilizan filtros de tamaño ($5 \times 5$) en todas las capas convolucionales.
    \item \textbf{Número de filtros}: se utilizan también pocos filtros en cada capa convolucional (6 en la primera y 16 en la segunda).
    \item \textbf{Función de activación}: aunque no viene especificado en la figura \ref{fig:lenet_5}, después de cada capa se utiliza como función de activación la \textit{sigmoide}.
    \item \textbf{Pooling}: tampoco se especifica, pero para reducir la resolución (subsampling) se usan capas de ``average pooling''.
    \item \textbf{Número de capas}: pocas, a penas dos capas convolucionales y tres totalmente conectadas.
\end{itemize}

\subsection[AlexNet]{AlexNet \texorpdfstring{\cite{NIPS2012_4824}}{}}

Esta red de 8 capas hizo definitivamente populares a las redes neuronales convolucionales, ganando el reto ``Imagenet Large Scale Visual Recognition Challenge 2012'', con un conjunto de datos de más de \textbf{1 millón} de ejemplos, con 1000 ejemplos de cada una de las 1000 categorías. 

\begin{figure}[H]
    \centering
    \includegraphics[width=6cm]{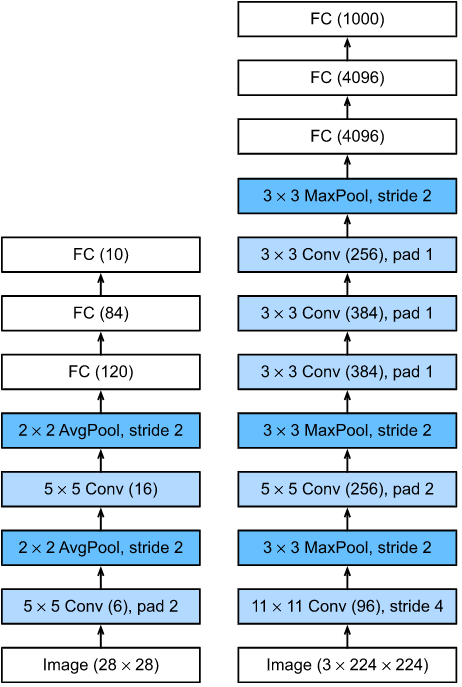}
    \caption[AlexNet]{LeNet (izquierda) vs AlexNet (derecha) \cite[sección 7.1]{zhang2021dive}.}
    \label{fig:lenet_vs_alexnet}
\end{figure}

\noindent Veamos los principales \textbf{cambios frente a LeNet5}: \begin{itemize}
    \itemsep0em
    \item Es \textbf{más profunda} (8 capas).
    \item Se usan filtros de distinto tamaño. Al principio, se utilizan filtros de tamaño $11\times11$ por el hecho de que las imágenes de ImageNet son más grandes, y los objetos ocupan más espacio. Los autores defienden que usando filtros de $11\times11$ podrán captar mejor los objetos presentes en la imagen. Luego, el tamaño se reduce a $5 \times 5$ y finalmente a $3 \times 3$.
    \item El número de filtros utilizados es enormemente mayor: 96 en la primera capa convolucional, 256 en la segunda...
    \item Respecto a la función de activación, se utiliza \textbf{ReLU} en lugar de la sigmoide. A partir de esta arquitectura, la mayoría las desarrolladas posteriormente han seguido usando ReLU.
    \item En lugar de average pooling, se usa \textbf{max pooling}. 
    \item Aunque no se especifica en el esquema, se utiliza un \textbf{dropout} de 0.5 en las dos últimas capas ocultas (LeNet sólo usaba regularización L2).
    \item En el entrenamiento, se hizo uso del \textbf{aumento de datos}.
\end{itemize}

\subsection[ResNet]{ResNet \texorpdfstring{\cite{he2016deep}}{}}

\label{ResNets}

Progresivamente las redes convolucionales se fueron haciendo más profundas, con la idea de que a mayor número de capas, mayor es la clase de funciones que una red puede representar, pero, ¿es necesariamente una red con más capas más ``expresiva'' que una con menos capas?: lo cierto es que no necesariamente.

Supongamos una función objetivo $f$ que estamos tratando de encontrar (o al menos, aproximar).

Supongamos ahora, que una arquitectura con un determinado número de capas es capaz de representar una determinada clase de funciones $H$, y supongamos que mediante el entrenamiento en un conjunto de datos, esta red es capaz de encontrar una función $f^*_{H}$ que aproxima a $f$.

Podría parecer lógico que si tenemos otra arquitectura más potente (con más capas), que es capaz de representar una clase de funciones $H'$, podríamos esperar entonces que esta arquitectura podría encontrar una función $f^*_{H'}$ ``mejor'' que $f^*_{H}$. El problema es que si $H \not \subseteq H'$, no hay garantía de que esto ocurra, de hecho, $f^*_{H'}$ podría ser peor.

Por tanto, lo ideal sería que cuando creamos una arquitectura con una clase de funciones más amplia, las clases de funciones más simples estén contenidas en ellas, garantizando así una ganancia en expresividad. De aquí surge la idea de ResNet.

Si cuando añadimos una capa a un modelo, pudiéramos entrenarla de forma que aprendiera la \textbf{función identidad} $f(x) = x$, entonces este nuevo modelo con una capa más sería estrictamente más expresivo que el modelo original, ya que podría aprender una mejor solución gracias a esa nueva capa, pero si una mayor profundidad no aportara nada, simplemente podría mantener la función identidad.

\subsubsection{El bloque residual}

\begin{figure}[H]
    \centering
    \includegraphics[width=8cm]{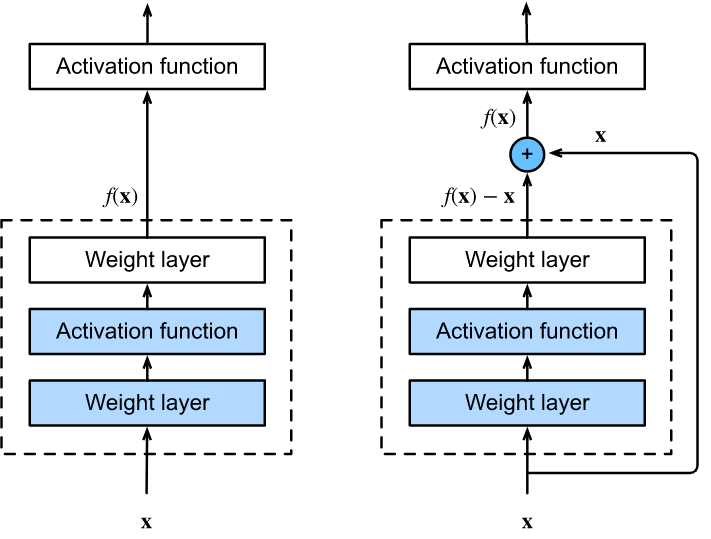}
    \caption[Bloque residual]{Un bloque convencional (izda) y el bloque residual que usa ResNet (dcha) \cite[sección 7.6]{zhang2021dive}. La entrada se denota por $\boldsymbol{x}$. La operación $+$ dentro del círculo azul indica una suma punto a punto de matrices. La flecha entre la entrada y la salida del bloque (llamada conexión residual), indica que la entrada se copia directamente para ser sumada a la salida del bloque.
    }
    \label{fig:bloque_residual}
\end{figure}

Para conseguir la función identidad, surge la idea del bloque residual, que tiene la estructura que aparece en la figura \ref{fig:bloque_residual}. 

Ahora, suponiendo que la función objetivo que queremos obtener mediante el aprendizaje es $f(x)$, el bloque convencional de la izquierda en la figura \ref{fig:bloque_residual} tendría que aprender directamente la función $f(x)$, mientras que el bloque residual tendría que aprender la diferencia (residuo) entre $f(x)$ y la entrada $x$. 

Aunque pueda parecer que no, esto tiene una ventaja, y es que si ahora suponemos que la función que se quiere aprender es la identidad, $f(x) = x$, el bloque convencional tendría que aprender la función identidad por medio de la composición de varias capas no lineales, mientras que el bloque residual sólo tendría que dar valor cero a todos los pesos, lo cual es más sencillo \cite{he2016deep}.

\textbf{En definitiva}, la idea importante que subyace a las redes residuales, es que la función identidad puede ser aprendida fácilmente por las distintas capas, por lo que añadir nuevas capas, como poco, hará que la arquitectura sea tan expresiva como una red menos profunda. Gracias a esta idea, se han desarrollado con éxito redes residuales con más de \textbf{cien capas}.

\section{Tendencia a más profundidad}
\label{tendencia_profundidad}

\begin{figure}[H]
    \centering
    \includegraphics[width=12cm]{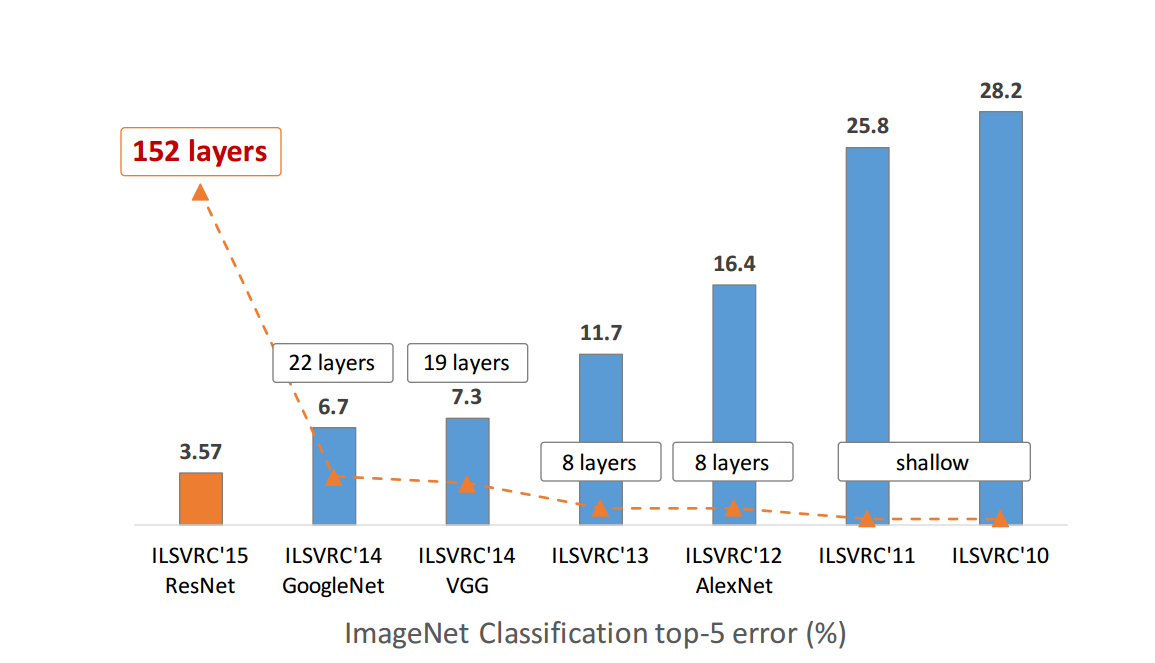}
    \caption[Revolución de la profundidad]{Imagen tomada de la presentación que hizo el autor principal del artículo sobre ResNet \cite{kaiminghe_revolucion_profundidad}.}
    \label{fig:depth_revolution}
\end{figure}

Como hemos visto en los ejemplos anteriores, a lo largo de los años ha habido una tendencia a hacer las redes más profundas. En la figura \ref{fig:depth_revolution} se muestra un resumen de la profundidad de las arquitecturas ganadoras del reto ILSVRC sobre clasificación de imágenes entre los años 2010 y 2015. Vemos como desde AlexNet, la profundidad de las redes se ha incrementado de forma extrema, llegando a \textbf{152} capas en 2015.

\subsection{¿Por qué más profundas?}

\label{por_que_mas_profundas}

Aunque parezca sorprendente dada la popularidad del aprendizaje profundo, lo cierto es que \textbf{no existe una respuesta del todo clara} sobre el por qué las redes más profundas podrían tener tan buen comportamiento (esto no es siempre cierto, teorema de No free lunch).

Sin embargo, existe un consenso sobre algunos de los motivos que pueden explicar este hecho. Dos de ellos se discutieron en la sección \ref{numero_capas}, y resumidamente eran: \begin{itemize}
\itemsep0em
    \item Quizás las redes neuronales poco profundas necesitan más neuronas para tener la misma capacidad que una red más profundas.
    \item Quizás, las redes menos profundas son más difíciles de entrenar con los algoritmos de aprendizaje existentes (quizás por tener más mínimos locales).
\end{itemize}

Un tercer motivo es el que vamos a explicar a continuación.

\subsubsection{Extrayendo características}
\label{extrayendo_caracteristicas}

Hace años, antes de 2012, el problema de clasificación de imágenes se enfocaba de un modo muy distinto. El proceso general era el siguiente: (i) Extraer ``a mano'' una serie de características interesantes para la clasificación de las imágenes, de forma que las imágenes queden \textit{resumidas} en unas pocas características y (ii) Usar estas características con un clasificador clásico, como regresión logística. Podríamos decir que aquí la complejidad estaba en decidir qué característica era útil extraer de las imágenes, y cómo extraerla.

A partir de 2012, todo cambió, y es que un grupo de investigadores, incluyendo algunos como Andrew Ng, Yann Lecun, Geoff Hinton y Alex Krizhevsky (de ahí el nombre AlexNet) pensaron que las características a extraer debían ser también aprendidas (en lugar de extraerse a mano). Además, pensaban que estas características podían ser aprendidas de forma \textbf{jerárquica} por medio de capas que aprendieran conjuntamente, de forma que las primeras capas podrían aprender por ejemplo, a detectar características simples como bordes, colores, o texturas. Luego, las capas más profundas aprenderían a extraer características más complejas, como ojos, orejas, hojas..., y las capas aún más profundas aprenderían a extraer objetos completos como personas ó árboles. La última capa, recibiría una información muy resumida de la imagen, y aprendería a realizar la tarea concreta a partir de esa información (por ejemplo, la tarea de clasificar la imagen).

Con esta idea, podemos pensar que las redes más profundas, tendrán la capacidad de extraer conceptos cada vez más complejos de las imágenes.

\subsection{Factores que han hecho posible esta tendencia}

\subsubsection{Más datos}
\label{mas_datos}

En general, las redes con muchas capas necesitan una cantidad enorme de datos para poder aprender representaciones complejas y superar a los enfoques clásicos. Por lo que hasta 2009, cuando el conjunto de datos ImageNet con 1 millón de imágenes fue publicado, no era posible sacar provecho de las posibles ventajas de la profundidad.

Desde ese momento, otros conjuntos de datos de gran tamaño y nuevos retos han sido publicados, impulsando en gran medida la investigación en estos modelos profundos.

\subsubsection{Mejor hardware}
\label{gpu}

El entrenamiento de un modelo de aprendizaje profundo requiere que el conjunto de datos ``pase'' a través de muchas capas de operaciones algebraicas con alta complejidad desde el punto de vista computacional. Este es otro motivo por el que las redes profundas no se desarrollaron antes.

El desarrollo de las GPU (graphics processing unit) hizo que el entrenamiento de las redes profundas pudiera realizarse en un tiempo mucho más reducido.

Aunque queda fuera del ámbito de este trabajo, daremos una ligera intuición de por qué las GPU funcionan mejor para este tipo de tareas que las CPU (central processing unit).

La idea es que las CPUs están diseñadas para cargas de trabajo más generales. Por contra, las GPUs son menos flexibles, pero están diseñadas para calcular en paralelo una misma instrucción sobre muchos datos al mismo tiempo, haciendo uso de una gran cantidad de núcleos. Las redes neuronales profundas están estructuradas de forma que en cada capa, miles de de operaciones \textit{iguales} tienen que ser realizadas. Por tanto, la estructura de una red neuronal profunda encaja bien con el tipo de computación que puede realizar una GPU.

Las desventajas que podríamos encontrar en una GPU, son la menor capacidad de memoria frente a una CPU, y la menor velocidad del reloj, que hace que en tareas secuenciales (que no es el caso de las redes neuronales) no vayan a rendir tanto como las CPU.

\textbf{En resumen}, las GPUs funcionan muy bien frente a las CPUs para esta tarea por dos motivos: (i) tienen una alta capacidad para el cálculo paralelo de operaciones y (ii) la arquitectura de las redes neuronales encaja bien con el tipo de cálculo que una GPU puede hacer de forma eficiente.


\subsection{Redes menos profundas}

Como última \textbf{aclaración} acerca de la profundidad de las redes, queremos destacar que aunque a lo largo de la (corta) historia de las redes ha habido una tendencia al incremento de la profundidad, no existe un modelo que funcione bien para todos los problemas (de nuevo, No free lunch), y por tanto \textit{no siempre nos ayudará la profundidad.}

Existen algunas arquitecturas modernas que no se centran únicamente en la profundidad, por ejemplo ResNext \cite{resnext}, introduce el concepto de \textit{cardinalidad} (que se define como el número de convoluciones paralelas que se realizan sobre una misma entrada), y defienden que incrementar la cardinalidad es una mejor forma de aumentar la capacidad del modelo, en lugar de aumentar la profundidad ó la anchura (número de filtros por capa).

\section{Transfer learning}
\label{preentrenamiento}

Ya sabemos que entrenar un modelo profundo y conseguir una buena generalización requiere una enorme cantidad de datos, en ocasiones, del orden de millones de ejemplos.

Sin embargo, en muchos problemas, etiquetar o recoger los datos puede consumir mucho tiempo, por lo que \textbf{no tendremos un conjunto de datos lo suficientemente grande}.

Una posible solución es aplicar \textit{transfer learning} (transferencia de aprendizaje), que consiste en transferir el conocimiento aprendido con un determinado conjunto de datos, a otro conjunto de datos.

Por ejemplo, supongamos que queremos aprender a clasificar 10 tipos de tortugas, pero tenemos un conjunto de datos de apenas 10000 ejemplos, y el esfuerzo de recolectar más datos y etiquetarlos no es factible. Entonces, lo que podemos hacer es partir de un modelo profundo entrenado con ImageNet (que no es un conjunto de datos de tortugas) y transferir el conocimiento útil aprendido en este conjunto de datos, a nuestro problema concreto (ahora veremos de forma aproximada cómo realiza esta transferencia).

Podríamos pensar, que por no ser un conjunto de datos de tortugas, la información aprendida en ImageNet no sería útil, sin embargo, es posible que el modelo entrenado en este conjunto de datos haya aprendido a distinguir \textbf{información genérica útil} \cite{yosinski2014transferable}, como bordes, texturas, formas geométricas, ..., que podría ser útil para clasificar tortugas.

Una primera aproximación para transferir este conocimiento sería la siguiente: \begin{enumerate}
\itemsep0em
    \item Entrenar un cierto modelo de red neuronal en un conjunto de datos mayor que el que disponemos (ImageNet por ejemplo).
    \item Eliminar la última capa de la red, que es la que se encarga de la clasificación final, y estará muy relacionada con el conjunto de datos concreto.
    \item Añadir una última capa para la clasificación de nuestro problema. En el caso de nuestro ejemplo, podríamos añadir una capa totalmente conectada de 10 neuronas para clasificar los 10 tipos de tortugas.
    \item Entrenar el modelo con nuestro conjunto de datos de tortugas. Ahora, la última capa aprendería desde cero, pero el resto de capas partirían del conocimiento previo obtenido con ImageNet, y sólo se ``refinarían'' los pesos para adaptarse al nuevo conjunto de datos.
\end{enumerate}

Por tanto, hemos visto una ligera idea de por qué es útil la técnica de transfer learning, y cómo podríamos aplicarla. En la \textbf{sección \ref{transfer_learning}}, veremos en mayor detalle esta técnica, y la aplicaremos a nuestro problema concreto.

\begin{tcolorbox}[
  colback=Green!5!white,
  colframe=Green!75!black,
  title={Recapitulación}]

Ideas clave a recordar:

\begin{itemize}
\small
\itemsep 0em 

    \item La redes neuronales convolucionales son una excelente opción para la clasificación de imágenes.
    \item Las redes convolucionales realizan una extracción progresiva de la información, desde características simples (como bordes) en las primeras capas, hasta objetos completos en las capas profundas.
    \item Las redes residuales (ResNets) nos permiten aumentar la profundidad asegurando que se mantiene la expresividad del modelo.
    \item Poseer un conjunto de datos muy grande es esencial para conseguir una buena generalización en el caso de clasificación de imágenes.
    \item En caso de tener un conjunto de datos pequeño, podemos tratar de transferir el conocimiento obtenido con otro conjunto de datos mayor.
    
\end{itemize}

\end{tcolorbox}
	
	\part{Problema a tratar}
	


\chapter{Conjunto de datos} 
\label{datos}
\begin{tcolorbox}[
  colback=SkyBlue!5!white,
  colframe=SkyBlue!75!black,
  title={Sumario}]
  
Cuando estudiamos los elementos de un problema de clasificación, uno de los principales era el conjunto de datos con el que entrenaremos el modelo de aprendizaje automático. En este capítulo hablaremos sobre los datos:
 
\begin{itemize}
\itemsep 0.1em 
    \item Explicaremos de dónde se han tomado los datos, y la selección de imágenes que el tutor de este trabajo nos facilitó.
    \item Qué son las imágenes PET y MRI, por qué su preprocesado es adecuado, y qué tipos de preprocesado realizaremos sobre ellas.
\end{itemize}

\end{tcolorbox}

\section{Alzheimer's Disease Neuroimaging Initiative}

ADNI \cite{adni} (Alzheimer’s Disease Neuroimaging Initiative) es un estudio que tiene el objetivo de descubrir, optimizar, estandarizar y validar las medidas de los ensayos clínicos y los biomarcadores utilizados en la investigación actual sobre la \acrshort{ad}. Todos los datos de ADNI se recopilan en una base de datos segura para que los científicos que estudian la enfermedad de Alzheimer puedan acceder a ellos para realizar investigaciones científicas, o para usarlos en la enseñanza \cite{adni_explicacion}.

De entre los conjuntos de datos que ADNI ofrece, nos interesan los de imágenes de resonancia magnética (\acrshort{mri}) y los de imágenes \acrshort{pet}, ya que son algunos de los tipos de imágenes más utilizados en el diagnóstico de la enfermedad \cite[p.~226]{compendio}.

\subsection*{Selección de las imágenes}
\label{pocos_datos}

Uno de los objetivos de este trabajo, era estudiar el uso simultáneo de imágenes \acrshort{pet} y \acrshort{mri} para el diagnóstico de la enfermedad. El problema es que esto requiere encontrar a aquellos pacientes a los que se les haya tomado ambos tipos de imágenes, y descargarlas. Dada la gran cantidad de datos que existe en ADNI, este último proceso puede ser realmente \textbf{lento}.

Para que pudiéramos enfocarnos en lo que realmente queremos estudiar, que es la aplicación de redes convolucionales, el director de este trabajo, \textit{Fermín Segovia Román}, nos facilitó una selección de imágenes \acrshort{mri} y \acrshort{pet} de pacientes a los que se le habían tomado ambas, ahorrándonos una enorme cantidad de tiempo.

Más concretamente, nuestro conjunto de datos está formado por un total de \textbf{249 imágenes de cada modalidad}, de las cuales:\begin{itemize}
\itemsep0em
    \item \textbf{70} están clasificadas como \acrshort{ad}.
    \item \textbf{111} están clasificadas como \acrshort{mci}.
    \item \textbf{68} están clasificadas como \acrshort{cn}.
\end{itemize}

Un último detalle a destacar sobre este conjunto de datos, es el hecho de que las imágenes cerebrales se consideran personales y están protegidas, por lo que antes del acceso a las imágenes facilitadas por el tutor, se firmó un contrato en el que nos comprometemos a \textbf{eliminar todos los datos} una vez terminado este trabajo de fin de grado.

\section{Imágenes PET}

Aunque no es el objetivo de este trabajo el estudio en profundidad de las imágenes médicas, siempre es útil tener, al menos, un conocimiento general sobre los datos que estamos tratando. Por ello, en esta sección daremos una idea general de algunos conceptos acerca de las imágenes \acrshort{pet}.

La tomografía por emisión de positrones (\acrshort{pet}) es una técnica de imagen médica que tiene por objetivo medir la actividad metabólica de las células de un determinado tejido u órgano, así como otras actividades fisiológicas, como el flujo de sangre \cite{johns_hopkings_pet}. Este tipo de imágenes que miden algún tipo de actividad se conocen como imágenes funcionales. 

Para tomar este tipo de imágenes, se requiere introducir una pequeña cantidad de sustancia radiactiva (conocida como radiotrazador) en el cuerpo del paciente, que luego puede localizarse mediante algún tipo de detector.

Las imágenes \acrshort{pet} son principalmente utilizadas por oncólogos, neurólogos, y cardiólogos \cite{johns_hopkings_pet}. En el ámbito de la neurología, un ejemplo de uso es en la enfermedad de Alzheimer: en esta enfermedad, el metabolismo de la glucosa y el oxígeno se ven reducidos, por lo que este tipo de imágenes pueden ser utilizadas para detectar esta reducción y facilitar su diagnóstico \cite{calsolaro2016alterations}.

\subsection*{¿Cómo se realiza?}

El primer paso para generar una imagen \acrshort{pet} es administrar el \textbf{radiotrazador} al paciente por vía intravenosa.

Los radiotrazadores que se usan para las imágenes PET se fabrican uniendo átomos de alguna sustancia radiactiva a algún tipo de sustancia química que el órgano o tejido que se quiere estudiar utilice de forma natural durante sus procesos metabólicos \cite{johns_hopkings_pet}.

Por ejemplo, en el caso de querer estudiar el cerebro, como éste hace un alto uso de la glucosa, un radiotrazador muy utilizado es el ${}^{18}$F-FDG.  Este radiotrazador se forma partiendo de una molécula análoga de la glucosa, la fluorodesoxiglucosa (FDG), y convirtiendo el flúor de esta molécula en flúor-18, que es un isótopo radiactivo emisor de positrones \footnote{El positrón es la antipartícula del electrón, es decir, es idéntica al electrón, pero con carga eléctrica positiva.}.

Cuando este radiotrazador es administrado, como consecuencia del metabolismo se descompone, emitiendo dos positrones, y además, se producen rayos gamma durante esta emisión.

Una vez que se ha administrado el radiotrazador al paciente y se ha esperado un cierto tiempo, el escáner \acrshort{pet} se irá desplazando lentamente sobre la parte del cuerpo a ser estudiada. Este escáner es capaz de detectar los rayos gamma producidos por la emisión de positrones, hasta que finalmente, un ordenador recoge la información sobre la llegada de los rayos gamma y la procesa para generar una imagen tridimensional del tejido u órgano \cite{shukla2006positron}.

\begin{figure}[H]
    \centering
    \includegraphics[width=4cm]{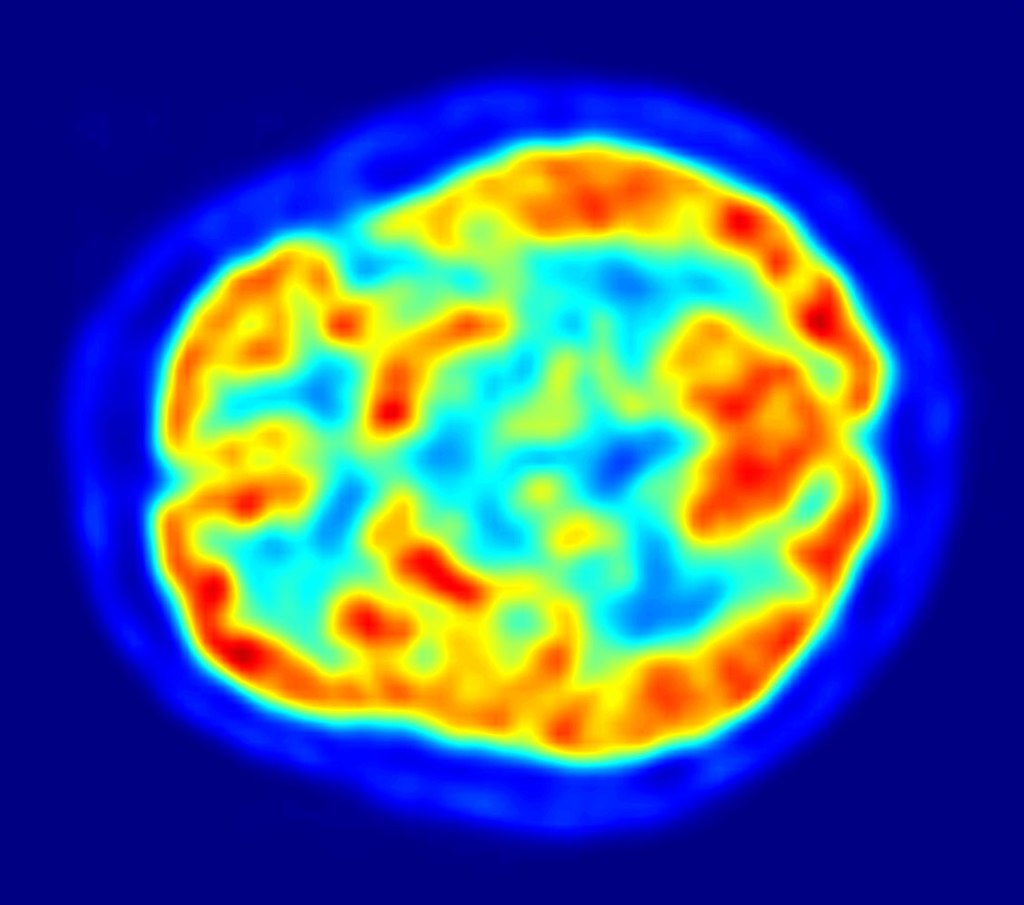}
    \caption[Ejemplo de imagen PET]{Uno de los cortes de una imagen PET \cite{imagen_pet_wikipedia} (recordemos que son imágenes 3D). En rojo aparecen las zonas con una mayor degradación del radiotrazador (mayor actividad celular), y en azul las de menor actividad.}

    \label{fig:ejemplo_pet}
\end{figure}

\section{Imágenes MRI}
\label{mri}

Al igual que hicimos con las imágenes \acrshort{pet}, a continuación vamos a dar una idea general sobre las imágenes \acrshort{mri}.

La imagen por resonancia magnética (\acrshort{mri}) es una potente herramienta de diagnóstico que utiliza fuertes imanes y ondas de radio para generar imágenes tridimensionales de los tejidos del cuerpo \cite{mri_scan}. A diferencia de las imágenes anteriormente vistas, que aportaban información sobre el metabolismo (reacciones químicas), estas aportan información sobre la anatomía (estructura).

Este tipo de imagen es enormemente utilizada en traumatología, cardiología y neurología. En la enfermedad de Alzheimer, también se trata de una técnica útil que hace posible detectar la atrofia (pérdida de células) que se produce en determinadas zonas del cerebro \cite{duara2008medial}.

\begin{figure}
    \centering
    \includegraphics[width=5cm]{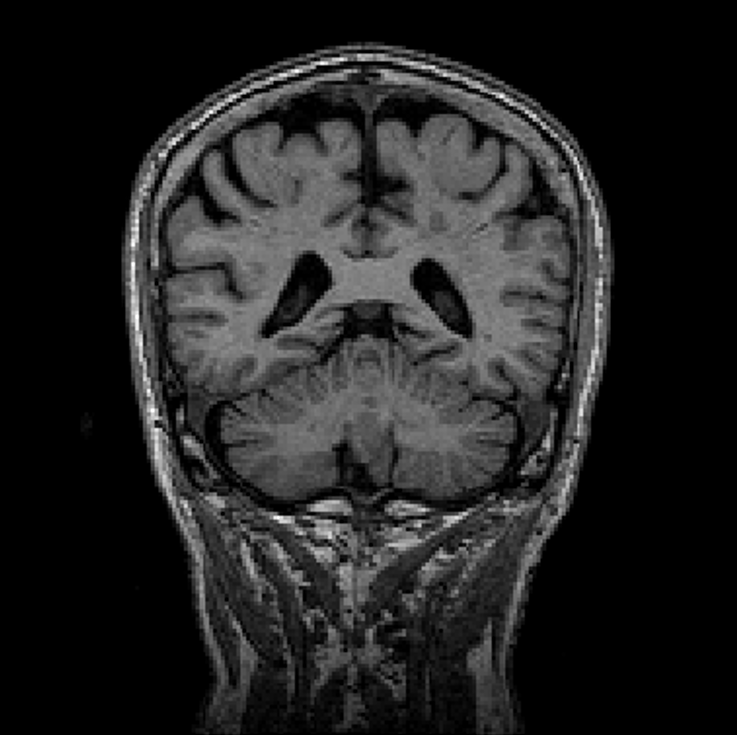}
    \caption[Ejemplo de MRI]{Ejemplo de resonancia magnética \cite{ejemplo_mri}. Vemos que este tipo de imagen aporta un gran detalle sobre la estructura del cerebro, permitiendo detectar la atrofia de determinadas zonas.}
    \label{fig:ejemplo_mri}
\end{figure}

\subsection*{¿Cómo se realiza?}

Ahora, veremos cómo los potentes imanes usados en esta técnica interactúan con el cuerpo para formar estas imágenes.

La gran parte del cuerpo humano está hecho de moléculas de agua, y cada una de estas moléculas de agua está formada por un átomo de oxígeno unido a dos de hidrógeno. En el centro de cada átomo de hidrógeno hay un protón (cargado positivamente), que se comporta como un pequeño imán orientado de forma aleatoria en condiciones normales.

Cuando el cuerpo se introduce dentro del campo magnético producido por los potentes imanes de un escáner \acrshort{mri},  los protones se alinean con este campo magnético y giran (como una peonza) a una determinada frecuencia conocida como frecuencia de resonancia, que es función de la fuerza del campo magnético \cite{talagala1991introduction}.

Aunque todos están alineados, parte de ellos están alineados en la misma dirección que el campo magnético, y otros en dirección opuesta.

Además del potente imán, el escáner contiene una antena mediante la que se envían pulsos de ondas de radio a la zona del cuerpo que se quiere estudiar. Estas ondas de radio se envían con una frecuencia igual a la frecuencia de resonancia de los protones, causando que aquellos protones alineados en la misma dirección del campo magnético, absorban la energía y se alineen en dirección opuesta. Cuando el pulso termina, los protones vuelven a su orientación original, emitiendo en este proceso una señal de radio que es recibida de vuelta por la antena del escáner.

Estas señales aportan información suficiente sobre la localización exacta de estos protones en el cuerpo. Además, permiten distinguir distintos tipos de tejidos dado que los protones en distintos tipos de tejidos se realinean a distintas velocidades y producen señales distintas \cite{mri_scan}.

Por último, estas señales emitidas por muchos protones al mismo tiempo, permiten construir finalmente una imagen (tridimensional) por medio de un ordenador, haciendo uso de algoritmos complejos \cite{van1999basic}.

\section{Preprocesado}
\label{preprocesado}

En este apartado estudiaremos el porqué es necesario preprocesar las imágenes, y cuáles son los tipos de preprocesado que utilizaremos. 

Como veremos, algunas de las técnicas son complejas y su entendimiento en profundidad se escapa del alcance de este trabajo, por lo que utilizaremos herramientas especializadas (como SPM) en lugar de implementarlas. Sin embargo, es adecuado tener una idea general del funcionamiento y conocer el efecto que estas técnicas producen en las imágenes, ya que \textit{realizar un preprocesado con total desconocimiento puede resultar en datos inservibles}.

\subsection{Necesidad}
\label{necesidad_preprocesado}

En la \textbf{sección \ref{extrayendo_caracteristicas}} expusimos el hecho de que cuando utilizamos redes convolucionales profundas, no se necesita un preprocesado manual de las imágenes como el que solía hacerse antes de 2012, ya que estas redes son capaces de aprender todas las transformaciones de las imágenes necesarias. Y en el caso de realizar algún preprocesado, suelen ser sencillos, como una simple estandarización (media 0, varianza 1).

Debemos recordar otra cosa que se dijo en la sección \ref{mas_datos}, y es que para que una red convolucional profunda tenga un buen comportamiento, se necesita una \textbf{enorme cantidad de datos} (del orden de millones de imágenes en ocasiones).

Hemos visto que tenemos tan solo 249 imágenes de cada modalidad, por lo que está muy claro que \textbf{no cumplimos el requisito de tener gran cantidad de datos}.

En este caso lo que se puede hacer es intentar ``simplificar'' el problema a la red, y la forma de simplificarlo es aplicando algún tipo de preprocesamiento ``manual''.

Vamos a aclarar un poco más la idea de por qué aplicar un procesamiento previo a las imágenes podría simplificar el trabajo a la red convolucional, y por qué podría esto ayudar con el problema de tener pocos datos. 

Para ello, vamos a suponer que tenemos la tarea de ``dada la foto de un dígito manuscrito, clasificarlo según qué dígito sea'', y vamos a suponer por ejemplo, que tenemos únicamente 10 fotos de cada letra, en total, 270 imágenes. Además, vamos a suponer que este conjunto de datos ha sido tomado por distintas personas, y tengamos en cuenta que es muy probable que cada persona haya tomado la foto desde una perspectiva distinta, con distintas condiciones de iluminación, y por qué no, incluso con la letra al revés.

Si pensamos en entrenar directamente una red con esas fotos, debemos darnos cuenta de que la red deberá aprender una función muy compleja con el objetivo de clasificar las imágenes independientemente de la perspectiva, tamaño de la letra, o brillo (intuitivamente podríamos decir que tiene que aprender a hacer muchas cosas). Y como sabemos, para representar esa función tan compleja (y desconocida), necesitaremos una red con una capacidad suficiente, es decir, con una profundidad y número de parámetros suficientes. El problema está en que una red con una capacidad alta, tiene también una alta capacidad de \textit{memorización}, y memorizaría sin ningún problema las 270 imágenes. Por tanto, siguiendo este enfoque, las predicciones ante nuevas fotos serían completamente inútiles debido al gran sobreajuste.

Se une además otro factor importante, y es que el hecho de tener imágenes muy diferentes representando una misma clase, hace que tengamos una muestra muy variable, facilitando aún más a la red memorizar esa \textit{variabilidad de los datos}, en lugar de la información subyacente.

Ahora, supongamos que ponemos la \textbf{restricción} de que aunque las letras no tienen por qué ser exactamente iguales, todas las fotos deben estar tomadas en las \textbf{mismas condiciones exactas}. Si ponemos esta restricción, ahora la red tendrá que ``fijarse'' sólo en la parte de la imagen en la que aparece la letras, y podrá centrarse en aprender sólo la información interesante. Entonces, gracias a esta restricción conseguiríamos reducir la complejidad que añade el hecho de poder clasificar letras fotografiadas desde distintas perspectivas, con lo que podríamos utilizar una red con menor capacidad, disminuyendo el sobreajuste, y mejorando en consecuencia la bondad de nuestra red. También es muy importante darse cuenta de que, para que la red funcione bien ante nuevos datos, estos tendrán que cumplir las mismas restricciones (lo que no siempre es posible).

Entendido el motivo por el que necesitamos preprocesar las imágenes, vamos a ver los dos tipos de preprocesado principales que utilizaremos: normalización en intensidad y normalización espacial.

\subsection{Normalización en intensidad (PET)}
\label{normalizacion_intensidad}

Cuando se capturan imágenes \acrshort{pet} de varios pacientes, un valor determinado de intensidad, no se corresponde con una misma cantidad de metabolismo en todas las imágenes. Esta variabilidad viene dada por varios factores como el peso, la edad, o la cantidad de radiotrazador administrado al paciente \cite{nugent2020selection, stoeckel2003outils}.

Con el objetivo de hacer estas imágenes comparables, se realiza un proceso de normalización en intensidad. Más concretamente, realizaremos una normalización a un valor máximo $I_{max}$ similar al usado en otros estudios (\cite{illan201118f}). Esta normalización consiste en calcular el promedio del 1\% de los vóxeles \footnote{Concepto análogo al de píxel, pero en una imagen de tres dimensiones.} de mayor intensidad de la imagen, y dividir el valor de intensidad de cada vóxel entre $I_{max}$.

Este tipo de normalización se fundamenta en el hecho de que existen zonas del cerebro que raramente son afectadas por la enfermedad de Alzheimer \cite{compendio, Smith4135}, y se hace la suposición de que la actividad máxima en esas zonas debe ser similar a la actividad máxima en el cerebro de sujetos sanos. 

\subsection{Normalización espacial (PET)}

El otro aspecto de las imágenes que ``complica'' nuestro problema, es que como cabía esperar, existen diferencias en la forma de los cerebros de distintos pacientes, y además, es probable que no todas las imágenes se tomen con el paciente en la misma posición exacta.

La normalización espacial trata de transformar una imagen de forma que se adapte a una plantilla común (cerebro estándar), consiguiendo así que sea cual sea la imagen, unas determinadas coordenadas siempre se refieran a la misma posición anatómica \cite[p.~52]{stoeckel2003outils}.

Esta normalización espacial comienza con un primer paso en el que se trata de reducir la diferencia cuadrática media entre la imagen fuente y la plantilla de referencia, usando para ello transformaciones afines (rotación, escalado, traslación, ó inclinado) \cite{friston1995spatial, woods2000spatial, szeliski2010computer}. 

Tras esta normalización mediante transformaciones afines, se realiza un refinamiento mediante un nuevo proceso de minimización, en este caso aplicando transformaciones no lineales (elásticas) más complejas \cite{ashburner1999nonlinear}.

\begin{figure}[t]
\centering
    \subfloat{\includegraphics[scale=0.25, angle=0]{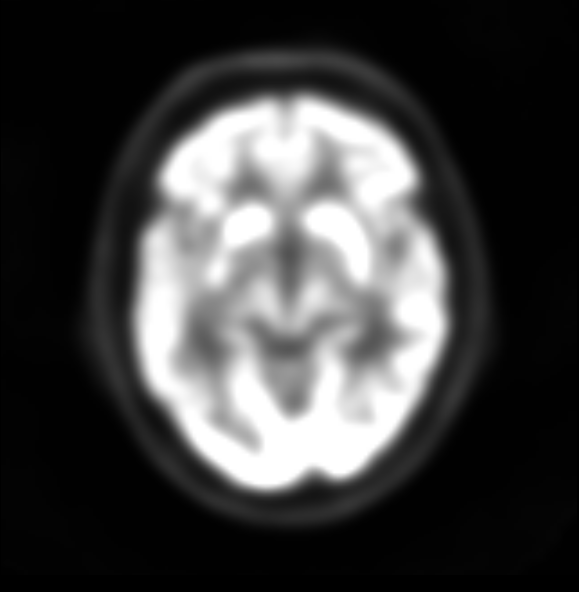}}\hfil
    \subfloat{\includegraphics[scale=0.25, angle=0]{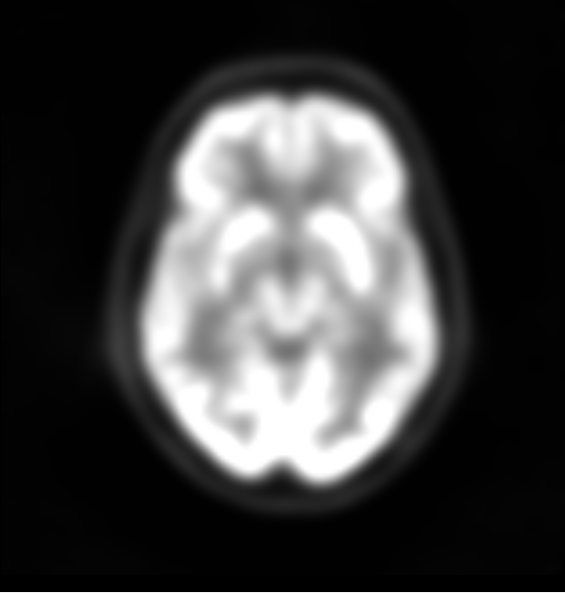}}
    
    \subfloat{\includegraphics[scale=0.3, angle=0]{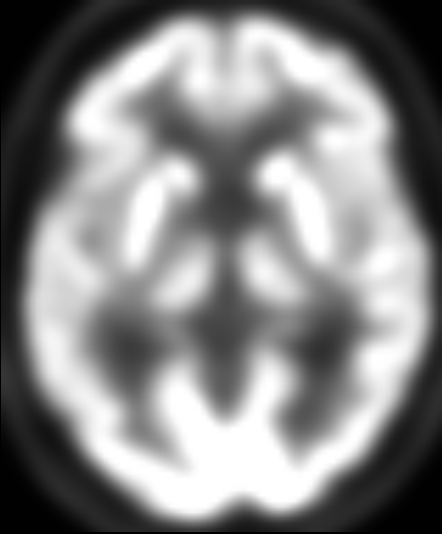}}\hfil
    \subfloat{\includegraphics[scale=0.3, angle=0]{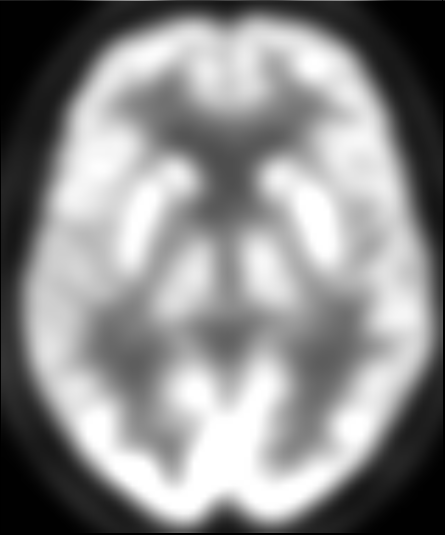}}
    \caption[Ejemplo de normalización espacial]{Ejemplo de normalización espacial. Arriba vemos dos imágenes \acrshort{pet} pertenecientes a dos pacientes distintos, en las que se aprecia la diferencia de forma entre sujetos. Tras ser normalizadas a una plantilla común mediante transformaciones no lineales, abajo vemos que sus formas coinciden en gran medida.}
    \label{fig:normalizacion_espacial}
\end{figure}

Debemos tener en cuenta, que cuando aplicamos este tipo de transformaciones, aunque nos estemos beneficiando de la simplificación que esto supone para el problema, \textit{se elimina alguna información} (como la forma y el tamaño), que podría ser importante para el estudio de la enfermedad en cuestión (por ejemplo, un corazón demasiado grande podría ser signo de una enfermedad).

\subsection{Segmentación de tejidos  (MRI)}

Como se puede apreciar en la figura \ref{fig:ejemplo_mri}, cuando se toma una imagen por resonancia magnética, \textbf{no sólo se obtiene información sobre el tejido que queremos estudiar}, sino que aparecerán tejidos inútiles. Por ejemplo, el cráneo no nos aporta ninguna información para el diagnóstico de \acrshort{ad}, mientras que en la \textbf{materia gris} del cerebro sí que es posible apreciar daños producidos por la enfermedad \cite{serra2010grey}.

El proceso de segmentación nos permitirá extraer únicamente aquellos tejidos que nos interesan (como la materia gris en nuestro caso). Se trata de un proceso complejo, que haciendo uso de mapas de probabilidad de tejido (plantillas que indican la probabilidad de que haya un tipo de tejido en un determinado vóxel), realiza la segmentación al mismo tiempo que se lleva a cabo una normalización espacial \cite{ashburner2005unified}. De nuevo, no entraremos en más detalle sobre este método, y confiaremos en implementaciones muy usadas en el ámbito médico.

\begin{tcolorbox}[
  colback=Green!5!white,
  colframe=Green!75!black,
  title={Recapitulación}]

Importante recordar:

\begin{itemize}
\small
\itemsep 0em 
    \item Las imágenes \acrshort{pet} miden el metabolismo, mientas que las \acrshort{mri} hablan de la estructura de los tejidos.
    \item El hecho de tener muy pocos datos nos obliga a \textbf{movernos del enfoque común en el deep learning}, y trabajar más en el preprocesado para ``simplificar el problema''.
    \item Usaremos principalmente tres técnicas de \textbf{preprocesado}:\begin{itemize}
        \item Normalización en \textbf{intensidad}: el objetivo es conseguir que un mismo valor de intensidad se corresponda con un mismo valor de actividad metabólica en distintos pacientes.
        \item Normalización \textbf{espacial}: conseguir que un mismo vóxel en distintas imágenes corresponda a la misma posición anatómica.
        \item Segmentación de \textbf{tejidos}: conseguir quedarnos con sólo aquellos tejidos que nos interesan, eliminando por ejemplo, el cráneo.
    \end{itemize}
    
\end{itemize}

\end{tcolorbox}

\chapter{Retos}

\begin{tcolorbox}[
  colback=SkyBlue!5!white,
  colframe=SkyBlue!75!black,
  title={Sumario}]
En este capítulo, explicaremos cómo trataría un médico el problema al que nos enfrentamos, y los principales retos que plantea para enfocarlo como un problema de aprendizaje automático.
\end{tcolorbox}

\section{Cómo actuaría un médico}

En ocasiones, antes de abarcar directamente el problema a resolver, puede resultar muy útil entender cómo un \textbf{experto} resuelve el problema, ya que así podremos entender la dificultad de este, y quizás podamos obtener alguna información extra que ayude a resolverlo.

Por tanto, supongamos nuevamente el problema de clasificar a un paciente entre una de las clases \acrshort{ad}, \acrshort{mci}, \acrshort{cn}. Pero en lugar de usar una red convolucional, vamos a suponer que es un médico el que trata de realizar esta clasificación. A continuación, vamos a ver brevemente cuál sería el proceso seguido.

Ante la llegada a la consulta de un paciente con un posible cuadro de demencia, el fin fundamental es clasificar dicha demencia. Para ello en la mayoría de los casos se siguen los siguientes pasos \cite{olazaran2011diagnosticarse}: \begin{itemize}[label=\ding{212}]
\itemsep0em
    \item \textbf{Anamnesis:} consiste en interrogar al paciente o a sus familiares para recabar toda la información sobre el progreso de la sintomatología, preguntando explícitamente sobre la pérdida de memoria y su repercusión en las actividades básicas de la vida diaria. 
    \item \textbf{Exploración física:} se hace para encontrar comorbilidades\footnote{Dos o más enfermedades que presenta una misma persona al mismo tiempo.} médicas que nos puedan orientar sobre el tipo de demencia.
    \item \textbf {Examen del estado mental:} este paso es fundamental para analizar cada una de las áreas cognitivas afectadas, que son de utilidad tanto para el diagnóstico como para la clasificación de la enfermedad. Para llevar a cabo este análisis, se usan test de valoración del estado mental estandarizados, entre ellos el de uso más extendido es el test ``Mini-mental'', que se explicará posteriormente.
\end{itemize}

Llegados a este punto, si no se tiene un diagnóstico completamente seguro de enfermedad de Alzheimer, o se sospecha cualquier otro tipo de demencia, se realizará una prueba de neuroimagen \cite{gifford2000systematic}.

\subsection{Test mini-mental (MMSE)}
El \acrshort{mmse} (Mini-Mental State Examination), es una prueba escrita que púntua como máximo 30 puntos, en la que un valor bajo hace referencia a un deterioro cognitivo mayor. Este test debe ser realizado en un lugar cómodo, para evitar cualquier tipo de estrés en la realización del mismo y es necesario que nos aseguremos de la participación voluntaria del paciente, ya que puede interferir en los resultados.

Para calcular la puntuación, este test se basa en la realización de una serie de preguntas que se agrupan en áreas como las siguientes: orientación espacio temporal, atención, memoria, o cálculo matemático \cite{llamas2015versiones}.

\subsection{¿Para qué sirven las imágenes?}
Sabemos que ante la imposibilidad de realizar un diagnóstico claro de enfermedad de Alzheimer, se recurre a las técnicas de imagen. Dentro de las más usadas encontramos las imágenes PET y MRI. El problema es que su análisis es explorador dependiente, es decir, depende en parte, de la subjetividad del profesional que las estudie \cite{mayo_diagnosed}.

En muchos casos, estas imágenes pueden ayudar en el diagnóstico diferencial de afecciones como son las hemorragias o los tumores, o bien, en la identificación de las diferentes demencias \cite{kelley2007alzheimer}. En este último ámbito, la clasificación de las demencias, encontramos un gran obstáculo y es que los mismos profesionales no dan el mismo diagnóstico ante una misma imagen, ya que por ejemplo, existen solapamientos entre lo que se podría considerar un deterioro normal por la edad, y el deterioro producido por la enfermedad de Alzheimer en etapas tempranas \cite{jack1997medial}.

Por lo tanto, las imágenes en la actualidad sólo serán útiles como apoyo para la realización de un diagnóstico diferencial y siempre habiendo realizado previamente un análisis de los hallazgos clínicos del paciente.

\subsection{Ambigüedad en el diagnóstico de MCI}
\label{mci_ambiguedad}

Hasta ahora hemos visto de forma aproximada cómo sería el proceso de diagnosticar a un paciente con la enfermedad de Alzheimer, y cómo podría diferenciarse de un paciente que presenta otro tipo de demencia, o alguna otra enfermedad. La pregunta es: ¿dónde queda la clase \acrshort{mci}?. Lo cierto es que en esta clase caerían aquellos pacientes que aunque tienen un ligero deterioro cognitivo, no presentaban dificultades para llevar a cabo las actividades de la vida diaria, y además no han podido ser diagnosticados de otra enfermedad.

El problema es que la definición actual de esta enfermedad (\acrshort{mci}) también admite un deterioro leve en las actividades de la vida diaria, lo que difumina la línea entre \acrshort{mci} y la enfermedad de Alzheimer, y además, algunos investigadores entienden el deterioro cognitivo leve como una etapa muy temprana de la \acrshort{ad} \cite{heerema_2021}. 

Asimismo, algunos de los pacientes que se diagnostican con \acrshort{mci} acaban evolucionando a \acrshort{ad}, y en ocasiones, cuando un medicamento causa un deterioro cognitivo, o cuando un paciente sufre ciertos tipos de depresión, puede ser diagnosticado erróneamente de \acrshort{mci} \cite{alz_mci}.

En definitiva, podemos decir que las líneas que separan la clase \acrshort{mci} del resto de clases es difusa, y no es raro que se comentan errores tanto en la clasificación \acrshort{mci} vs \acrshort{ad}, como en \acrshort{mci} vs \acrshort{cn}.

\section{¿Existe f?}

Cuando vimos los elementos de un problema de clasificación (sección \ref{elementos_clasificacion}), decíamos que existía una función desconocida $f : X \rightarrow Y; y^{(i)} = f(x^{(i)})$ que resolvía nuestro problema.

Siendo muy optimistas, podríamos pensar que existe esta $f$, y que además es capaz de clasificar las imágenes de forma \textbf{perfecta}. 

En el caso de muchos problemas como la detección del habla o de imágenes naturales, es fácil saber que $f$ existe, ya que los seres humanos conocemos esa $f$, y entonces podremos afirmar que hay una posibilidad de descubrirla, independientamente de que se pueda descubrir o no mediante una red convolucional.

Sin embargo, en este problema \textbf{no tenemos ninguna evidencia de que la $f$ ideal que buscamos exista}, ya que como vimos, excepto en casos concretos, un médico no sería capaz de clasificar imágenes cerebrales sin tener información extra, e incluso con esta información, las equivocaciones en el diagnóstico son frecuentes \cite{beach2012accuracy}.

Con todo esto \textbf{no queremos decir que el problema no tenga solución}, sino que no tenemos certeza de que el problema se pueda resolver con tanto éxito como el que esperamos, pero también es posible, que aunque un médico no sea capaz de aproximarse del todo a esa $f$ desconocida, una red convolucional lo pueda hacer de una forma más precisa.

\section{Etiquetas probablemente ruidosas}

Antes hemos visto que existe un solapamiento entre el deterioro del cerebro producido por la edad, y el causado por la enfermedad de Alzheimer. Esto puede llevar a discrepancias en el análisis de las imágenes por parte de distintos médicos: lo que uno puede considerar normal de la edad, otro lo puede considerar causado por la enfermedad.

Además, la línea que separa los pacientes \acrshort{cn} los \acrshort{mci}, así como de los \acrshort{mci} de \acrshort{ad} es \textbf{difusa}, y depende de la subjetividad del médico.

Por tanto, podríamos decir que la etiqueta \acrshort{mci} es ruidosa por naturaleza, por lo que podemos esperar un cierto porcentaje de imágenes ``mal'' clasificadas.

\section{Tamaño de las imágenes} 

Otro reto importante que presenta este problema es el hecho de que las imágenes son 3D. Aunque conceptualmente no sea un problema mayor, estas imágenes ocupan gran cantidad de memoria RAM al ser cargadas, lo que nos obligará a tomar determinadas decisiones en la implementación (sección \ref{tfrecords}). Además nos ``obligará'' a usar convoluciones 3D, con un alto coste computacional.

\section{Pocos datos}
 
Aunque hemos hablado repetidas veces sobre la falta de datos, sólo queremos recordar que esta falta de datos, será probablemente el mayor reto a la hora de resolver este problema.

\begin{tcolorbox}[
  colback=Green!5!white,
  colframe=Green!75!black,
  title={Recapitulación}]

Ideas importantes:
\begin{itemize}
\small
\itemsep 0em 
    \item Para poder clasificar correctamente las imágenes, \textbf{un médico necesitaría conocer la clínica} del paciente.
    \item La línea que separa la clase \acrshort{mci} del resto de clases no está claramente definida.
    \item Resolver el problema mediante aprendizaje automático plantea una serie de retos:\begin{itemize}
    \itemsep0em
        \item No tenemos \textbf{certeza} de que la $f$ desconocida y deseada exista.
        \item La subjetividad en el diagnóstico hace posible la existencia de \textbf{etiquetas incorrectas}.
        \item Las imágenes son grandes y ocupan mucho espacio en memoria RAM. Además por ser tridimensionales, su manejo será computacionalmente más costoso.
        \item Tenemos muy \textbf{pocos datos}.
    \end{itemize}
    
\end{itemize}

\end{tcolorbox}

\chapter{Estudios relacionados}
\label{estudios_relacionados}

\begin{tcolorbox}[
  colback=SkyBlue!5!white,
  colframe=SkyBlue!75!black,
  title={Sumario}]

Antes de desarrollar una propuesta, puede ser de gran utilidad revisar estudios similares al que queremos llevar a cabo, ya que de esta forma, podremos evitar caer en errores cometidos por otros, y además, es muy probable que nos ayude a tomar ciertas decisiones.\vspace{1em}

En este capítulo repasaremos brevemente algunos de los hechos más destacables encontrados en los artículos revisados.

\end{tcolorbox}

\section{Búsqueda}

Para la revisión de diferentes estudios relacionados con el nuestro, hemos realizado una búsqueda en plataformas como PubMed, ScienceDirect, Scopus ó ResearchGate de publicaciones sobre clasificación de \acrshort{ad} utilizando redes neuronales convolucionales sobre imágenes médicas --- concretamente resonancia magnética (\acrshort{mri}) y tomografía por emisión de positrones (\acrshort{pet}) ---. El resultado de esta búsqueda fue una enorme cantidad de estudios. Dada la gran cantidad de estudios, leerlos todos no era viable, por lo que sólo hemos leído con atención algunos de los que tenían más citaciones en el periodo 2017-2021, aunque por supuesto esto no quiere decir que fuesen los mejores necesariamente. 

En las secciones siguientes, vamos a especificar algunos de los detalles más relevantes que hemos visto al revisar estos estudios.

\section{Distintas tareas de clasificación}

En la búsqueda efectuada, la tarea de clasificar entre pacientes con \acrshort{ad} de sujetos cognitivamente normales (\acrshort{cn}) es la más ampliamente realizada, aunque quizás, también es la de menor interés desde el punto de vista médico (más sencilla). De los estudios revisados en detalle, los artículos \cite{backstrom2018efficient, cheng2017multi, lian2018hierarchical, gunawardena2017applying, korolev2017residual} se centran únicamente esta tarea de clasificación. 

Pero como sabemos, antes del desarrollo de la \acrshort{ad}, los pacientes pasan por una fase previa (\acrshort{mci}) durante la cual tienen algunos síntomas, aunque no suficientemente severos como para considerarse demencia. En otros estudios \cite{aderghal2018classification, senanayake2018deep}, se ha tratado además, esta tarea de distinguir entre pacientes con \acrshort{mci} y \acrshort{ad} ó entre \acrshort{mci} y \acrshort{cn} , y en una última parte de los estudios \cite{Valliani, islam, WANG2019145} se ha tomado el enfoque de distinguir directamente entre las tres clases \acrshort{mci}, \acrshort{ad}, \acrshort{cn}  (tarea más compleja) --- y enfoque que nosotros tomaremos ---.

\section{Uso de las imágenes} 
\label{uso_imagenes}

Entre los estudios revisados, hemos visto distintos enfoques a la hora de utilizar las imágenes tridimensionales: \begin{itemize}
    \item \textbf{Cortes 2D.} En  \cite{aderghal2018classification, Valliani, gunawardena2017applying, islam}, en lugar de utilizar las imágenes 3D, apuestan por extraer ciertos cortes 2D de las imágenes, lo que les da la ventaja de poder utilizar redes convolucionales ya implementadas para la clasificación de imágenes naturales, pero por otra parte pierden parte de la información espacial.
    \item \textbf{Imagen 3D completa.} En \cite{cheng2017multi, senanayake2018deep, korolev2017residual, WANG2019145}, utilizan directamente las imágenes 3D, lo que les permite mantener las relaciones entre los píxeles en todas las dimensiones, aunque a cambio será necesario usar convoluciones 3D, que tienen un coste computacional muy elevado.
    \item \textbf{Regiones de interés.} En otros estudios \cite{backstrom2018efficient, lian2018hierarchical} se ha optado por utilizar sólo pequeñas partes de las imágenes que se consideran de interés. Aparentemente es un buen enfoque, ya que se consigue aumentar el tamaño del conjunto de datos y reduce el coste computacional, sin embargo, hace necesario tener un conocimiento médico amplio para poder extraer esas regiones, así como herramientas para extraerlas correctamente.
\end{itemize}


Respecto al tipo de imágenes utilizadas, lo cierto es que la inmensa mayoría de los estudios utilizan imágenes \acrshort{mri}, mientras que es difícil encontrar estudios que hagan uso de \acrshort{pet} (quizás por existir menos imágenes de este tipo, ya que son más costosas \cite{pet_price}).

\section{Conjuntos de datos}
Respecto al tamaño de los conjuntos de datos que se utilizan en los distintos estudios, tenemos algunos que utilizan conjuntos muy pequeños con a penas 231 ejemplos \cite{korolev2017residual}, otros tienen conjuntos de entre 400 y 800 ejemplos \cite{aderghal2018classification, cheng2017multi, senanayake2018deep, lian2018hierarchical, Valliani, islam, WANG2019145}, y una pequeña parte de ellos superan (no por mucho) los 1000 ejemplos \cite{backstrom2018efficient, gunawardena2017applying}.

Por otra parte, no todos los estudios utilizan las mismas bases de datos de imágenes para crear sus conjuntos de datos, por lo que habrá \textbf{diversidad} en la resolución y calidad de las imágenes, así como en diversos factores dependientes del paciente (edad, etnia, nivel de estudios, etc...).

\section{Aumento de datos y transfer learning}

A pesar de que las técnicas de aumento de datos y transfer learning son muy utilizadas en el ámbito de la clasificación de imágenes mediante redes convolucionales, pocos de los estudios revisados las utilizan para este problema médico.

De forma muy breve, vamos a analizar cómo se han usado estas técnicas en esta minoría de los estudios:\begin{itemize}
    \item En el artículo de \textbf{Karim Aderghal y cols. \cite{aderghal2018classification}} se utiliza transferencia de aprendizaje, empleando como conjunto de preentrenamiento imágenes cerebrales de pacientes con enfermedad de Alzheimer, pero de una modalidad distinta a las imágenes que utilizan para entrenar finalmente su red (\acrshort{mri}).
    \item En \textbf{Bo Cheng y cols. \cite{cheng2017multi}}, también se utiliza transfer learning para el problema de clasificación de dos clases, y de nuevo se valen de imágenes cerebrales para el preentrenamiento, y de la misma enfermedad.
    
    \item Los artículos de \textbf{Jyoti Islam y Yanquing Zhang} y de \textbf{Ally Valliani y Ameet Soni} (\cite{islam,Valliani}) utilizan la técnica de aumento de datos. Es importante resaltar que ambos habían tomado el enfoque de usar cortes 2D de las imágenes, por lo que el uso de esta técnica es muy sencillo (está implementado en varias bibliotecas).
    
    Otro dato importante, es que ambos \textbf{reportan mejoras al utilizar esta técnica}.
    
\end{itemize}

\section{Problemas encontrados}

Si bien es cierto que la gran \textbf{mayoría} de estudios utilizan métodos de obtención de resultados \textbf{correctos en términos estrictos}, desde nuestro punto de vista, y especialmente, en conjuntos de datos pequeños, algunos de estos estudios utilizan métodos que no son del todo adecuados, ya que los resultados dados pueden estar sesgados de forma optimista.

El problema de estos resultados sesgados, es que si alguno de estos modelos se ``enfrentara'' a la realidad, es probable que su comportamiento empeorara de forma notable respecto al que se muestra en el estudio.

En nuestra opinión, sería más útil aportar resultados que informen (dentro de lo posible) del comportamiento real de nuestros modelos, por lo que veremos a continuación algunas de las formas en las que se ha podido introducir este sesgo optimista en algunos de los estudios, \textit{de forma que más tarde podamos evitarlo}.

\subsection{Fuga de datos}

\textbf{Nota:} la fuga de datos sí que es un error, los resultados de un modelo que presente fugas de datos en su entrenamiento, quedarán completamente invalidados.

La fuga de datos consiste en \textbf{utilizar información de fuera del conjunto de entrenamiento} durante el entrenamiento de un modelo. Esta información adicional puede permitir al modelo aprender algo que de otra forma no habría aprendido, y en consecuencia \textbf{invalidar por completo} el desempeño estimado del modelo en cuestión \cite{brownlee_data_leakage}.

Esta fuga de datos puede ocurrir de muchas formas distintas, algunos ejemplos simples serían el de hacer una separación deficiente de los conjunto de validación y test, o que al aplicar transfer learning, exista un solapamiento entre los datos del problema y  los usados para el preentrenamiento. Existen formas mucho más sutiles de que este error ocurra, como por ejemplo reutilizar el mismo optimizador utilizado para entrenar el modelo entre distintas iteraciones de k-fold, aunque no es nuestro objetivo entrar en más detalles.

En una pequeña parte de los estudios encontramos problemas de fugas de datos, vamos a repasarlos brevemente:\begin{itemize}
    \itemsep0em
    \item \textbf{Jyoti Islam y Yanqing Zhang} \cite{islam}: en este estudio se parte de un conjunto de datos de imágenes tomadas de un total de 416 pacientes, habiendo unas \textbf{3 ó 4 imágenes de cada uno}. Esto tiene un problema, y es que luego, cuando evalúan sus modelos mediante validación cruzada 5-fold, efectúan la partición de los datos de forma aleatoria, por lo que con una probabilidad muy alta , ``caerán'' imágenes del mismo paciente tanto en entrenamiento como en la porción de test, produciendo una fuga de datos clara.
    \item \textbf{Karl Bäckström y cols.} \cite{backstrom2018efficient}: mismo problema que en el estudio anterior. Se tiene como conjunto de datos 1198 imágenes tomadas de únicamente 340 sujetos, y se realiza una división de tipo train-validation-test de forma aleatoria. Con casi total seguridad, tanto en validación como en test habrá imágenes pertenecientes a los mismos sujetos de entrenamiento.
    \item \textbf{Hongfei Wang y cols.} \cite{WANG2019145}: tienen 833 imágenes tomadas de 624 pacientes, por lo que de nuevo, puede haber repetición de pacientes en las distintas particiones cuando realizan validación 10-fold. En el artículo se dice que se tiene en cuenta este problema, y se asegura que las imágenes de un mismo paciente no puedan ser repartidas entre ``porciones'' distintas. Por lo tanto, hasta aquí no parece haber fallos.
    
    El problema es que en el artículo, se afirma que aunque dos imágenes hayan sido tomadas de un mismo paciente, ellos consideran que han sido tomadas de distintos pacientes siempre y cuando exista un período de diferencia de tres o más años entre las tomas, ya que según los autores \textit{el cerebro cambia notablemente en 3 años}, pero lo cierto es que no aportan ninguna referencia sobre esta afirmación. Además, ninguno de los autores tiene estudios oficiales en medicina, y por nuestra parte tampoco hemos encontrado ninguna referencia que apoye este hecho.
    
\end{itemize}

\subsection{No existencia de conjunto de test}
\textbf{Nota}: la no existencia de un conjunto de test no suele considerarse un fallo desde el punto de vista técnico, aunque nosotros consideramos que es adecuado tenerlo.

Como se vio en la sección \ref{train_validation_test}, de cara a obtener una estimación \textbf{no sesgada} del error que nuestro modelo cometerá ante datos nunca antes vistos, lo que se hacía era reservar un pequeño conjunto de \textbf{test} para evaluar nuestro modelo final (y que no se usa en el proceso de selección de un modelo), además, esto era especialmente importante en el caso de tener pocos datos. 

Sin embargo, \textit{algunos} de los estudios no realizan esta partición, y consideran que la estimación del error obtenida mediante k-fold (con k=5 ó 10) les libera de la necesidad de un conjunto de test. Debemos recordar que especialmente cuando tenemos pocos datos, con las suficientes pruebas acabaremos \textbf{sesgando de forma optimista} el resultado arrojado por k-fold, por lo que sería conveniente reservar un conjunto exclusivo para test. 

También cabe destacar, que aunque es cierto que buena parte de los estudios revisados \cite{Valliani, aderghal2018classification, backstrom2018efficient, senanayake2018deep, lian2018hierarchical, gunawardena2017applying} sí que hacen uso de un conjunto de test, no todos ellos lo usan como un conjunto \textbf{independiente} (por lo que este conjunto pierde su función, y no se le debería llamar test).

Concretamente, los estudios \cite{senanayake2018deep, Valliani, gunawardena2017applying} hacen \textit{solamente} una partición del tipo train-test (al menos, eso dan a entender), de modo que para ajustar sus modelos probablemente hayan usado test (lo cual no es correcto, ya que estarían ``memorizando el conjunto de test manualmente'').

Por otro lado, en el artículo de Karl Bäckström y cols. (\cite{backstrom2018efficient}), aunque sí que realizan una partición del tipo train-validation-test, a la hora de dar su resultado final, realizan varias ejecuciones distintas sobre el conjunto de test, y reportan únicamente aquella ejecución con mejor valor. Esto en principio no es un error aberrante, pero tampoco es una buena práctica.

\subsection{Métrica no adecuada}

Por último, aunque no podemos considerarlo exactamente un error, existen estudios \cite{islam, senanayake2018deep} que utilizan una métrica no del todo adecuada. Ambos utilizan la exactitud (ejemplos correctamente clasificados dividido por el total de ejemplos) en un conjunto de datos en el que más del 80\% de los ejemplos pertenecen a una clase (clase mayoritaria). Un clasificador que simplemente clasificara todos los ejemplos como pertenecientes a la clase mayoritaria obtendría un 80\% de exactitud.

Aunque esto \textbf{no es un error}, este tipo de métricas podrían causar confusión y hacer pensar que el modelo es enormemente mejor de lo que realmente es, por lo que debemos evitar usarlas en conjuntos altamente desbalanceados.

\section{Modelos y profundidad} 

Respecto a los modelos utilizados, podemos encontrar desde modelos que hacen uso de dos capas convolucionales \cite{gunawardena2017applying, aderghal2018classification}, pasando por modelos que usan entre cinco y quince capas \cite{backstrom2018efficient, senanayake2018deep, lian2018hierarchical, WANG2019145}, hasta llegar a arquitecturas de más de 20 capas \cite{islam, Valliani, korolev2017residual}.

Sin embargo, \textbf{es imposible obtener conclusiones} sobre qué tipo de modelo (y qué profundidad) podría comportarse mejor en este tipo de problemas, ya que como hemos visto, existen grandes diferencias entre los conjuntos de datos de los distintos estudios, además de la existencia de ciertos problemas en la evaluación de algunos de ellos.

\section{Resultados que obtienen}

Dar detalles numéricos sobre los resultados obtenidos por los distintos estudios no sería de gran utilidad debido a los mismos motivos que hemos dados respecto a la profundidad de los modelos, sin embargo, podríamos decir de forma muy genérica, que para la tarea de clasificación \acrshort{ad} vs \acrshort{cn} los resultados son muy buenos (superando en general el 90\% de exactitud), mientras que para la distinción entre las clases \acrshort{cn} y \acrshort{mci}, o entre las clases \acrshort{ad} y \acrshort{mci} observamos exactitudes en torno a un 80\%. 

Por último, para el problema de tres clases, que es el que nosotros abarcaremos, los resultados suelen ser bastante inferiores, con exactitudes en torno al 60-70\%. Sin embargo, existe un estudio \cite{WANG2019145} que muestra una exactitud superior al 94\%, que aunque es posible, siempre tenemos que tener en cuenta que en los estudios también pueden cometerse errores, y conociendo la dificultad del problema y haciendo uso del sentido común, parece probable que en este estudio haya habido algún problema (por ejemplo, a la hora de transcribir los resultados finales).

\begin{tcolorbox}[
  colback=Green!5!white,
  colframe=Green!75!black,
  title={Recapitulación}]

En los estudios revisados:
\begin{itemize}
\itemsep0em 
    \item Aparecen tres posibles enfoques para usar las imágenes 3D.\begin{itemize}
        \itemsep0em
        \item Usar las imágenes enteras.
        \item Usar determinados cortes 2D.
        \item Usar regiones interesantes. Enfoque aparentemente bueno, pero requiere un mayor conocimiento médico.
    \end{itemize}
    \item Las técnicas de aumento de datos y transfer learning no son ampliamente usadas, y la de aumento de datos sólo es usada en los estudios que toman el enfoque 2D.
    \item Aunque en general, la evaluación de los estudios es correcta en términos estrictos, algunos de los estudios presentan un sesgo optimista.
    \item Se usan distintos modelos con mayor y menor profundidad, pero no es posible concluir cuáles funcionan mejor debido a la imposibilidad de comparar los distintos estudios.
    \item En general, los resultados para el problema de clasificación de dos clases (\acrshort{ad} vs \acrshort{cn}) son excelentes, pero no es así para el problema de tres clases (\acrshort{ad} vs \acrshort{mci} vs \acrshort{cn}).

\end{itemize}
\end{tcolorbox}
	
	\part{Desarrollo y experimentación}
	\chapter{Propuesta}


\begin{tcolorbox}[
  colback=SkyBlue!5!white,
  colframe=SkyBlue!75!black,
  title={Sumario}]
Hasta ahora, hemos visto los fundamentos teóricos que nos hacen falta, entendemos el problema que vamos a tratar, y hemos analizado distintos enfoques que se han tomado en varios estudios para intentar dar una solución al problema dado.\vspace{1em}

En este capítulo, vamos explicar cuál es nuestra propuesta para tratar de resolver este problema: el uso que daremos a las imágenes, el tipo de problema de clasificación que abarcaremos, y una explicación general de los experimentos que proponemos.

\end{tcolorbox}
\section{Uso de las imágenes}

En la sección \ref{uso_imagenes} se vieron los distintos enfoques usualmente tomados para utilizar las imágenes médicas tridimensionales. 

En nuestro caso, hemos decidido utilizar las imágenes 3D completas, ya que esto nos aporta las siguientes ventajas:\begin{itemize}
    \itemsep0em
    \item No necesitamos un gran conocimiento médico, ya que no necesitamos saber qué regiones del cerebro extraer.
    \item No perdemos la información sobre las posibles relaciones de los vóxeles en tres dimensiones, que sí perderíamos al usar un enfoque 2D.
    \item Quizás existe la posibilidad de que nuestra red sea capaz de encontrar información útil en zonas del cerebro que los médicos no consideran de utilidad.
\end{itemize}

A cambio de estas ventajas, necesitaremos \textbf{ampliar el concepto de convolución a las imágenes tridimensionales}, e inevitablemente aumentaremos el coste computacional.

\subsection{Convolución 3D}

En la sección \ref{convolucion_2d} estudiamos la operación de convolución sobre imágenes 2D con $c$ canales. Ahora, en nuestro problema, tendremos imágenes 3D, que serán matrices  de dimensiones $h \times w \times d \times c$, aunque concretamente tendremos un sólo canal, por lo que $c$ será igual a $1$.

La operación de convolución de un filtro de tamaño $k \times k \times k$ con una imagen 3D es casi idéntica a la que vimos en el caso de 2D, con la diferencia de que el filtro, en lugar de ``deslizarse'' en anchura y altura, se deslizará también en profundidad. Por tanto, el resultado de aplicar la convolución de un filtro de tamaño $k \times k \times k \times c $ sobre una imagen 3D de dimensiones $h \times w \times d \times c$, será una nueva imagen (matriz) 3D de dimensiones $h' \times w' \times d' \times 1$, con $h' = h - k + 1, w' = w - k + 1, d' = d - k + 1$.

\section{Número de clases}

Hemos visto que en los estudios se usaban distintos enfoques respecto a las clases tenidas en cuenta en la clasificación. Algunos clasificaban únicamente entre \acrshort{ad} y \acrshort{cn}, otros realizaban la clasificación entre \acrshort{ad}, \acrshort{cn} y \acrshort{mci}, pero dos a dos, y por último había unos pocos que tomaban el enfoque de clasificar entre las tres clases.

\textbf{Nosotros}, nos enfrentaremos al problema de tres clases.

Queremos remarcar que en un problema de tres clases, un clasificador \textbf{aleatorio} conseguiría una exactitud media del \textbf{33\%}, mientras que en un problema de dos clases (el más común en los estudios revisados), un \textbf{50\%}.

\section{Experimentos a realizar}
\label{experimentos}

De forma muy genérica, nuestra propuesta constará de los siguientes experimentos:\begin{enumerate}
    \itemsep0em
    \item Probar redes de distinta profundidad.
    \item Hacer uso del aumento de datos.
    \item Hacer uso de imágenes sin preprocesar, para estudiar así la capacidad de las \acrshort{cnn} para extraer características.
    \item Entrenar una red con datos de COVID19 y tratar de transferir el conocimiento útil a nuestro problema.
    \item Crear una red con dos entradas, que tome simultáneamente imágenes \acrshort{pet} y \acrshort{mri}.
\end{enumerate}


\chapter{Consideraciones generales}

\begin{tcolorbox}[
  colback=SkyBlue!5!white,
  colframe=SkyBlue!75!black,
  title={Sumario}]

En este capítulo, se verán ciertas consideraciones generales que se mantendrán constantes a lo largo de todos los experimentos:\begin{itemize}
\itemsep0em
    \item Estudiaremos las métricas y el proceso utilizado para evaluar los distintos modelos.
    \item Veremos un método muy genérico que seguiremos a la hora de experimentar.
    \item Analizaremos los hiperparámetros principales que tendremos que fijar para crear nuestros modelos.

\end{itemize}

\end{tcolorbox}


\section{Evaluación de modelos}

En esta sección explicaremos la forma en la que evaluaremos la bondad de un determinado modelo. Resumidamente, una evaluación correcta es importante por los siguientes motivos:\begin{itemize}
    \itemsep0em
    \item Un buen sistema de evaluación nos permitirá comparar los distintos modelos de una forma ``fiable'', de manera que el modelo que creemos que es el mejor, lo sea realmente (o al menos con alta probabilidad).
    \item Nos permitirá dar una estimación no sesgada (o al menos poco sesgada) de la bondad del modelo ante nuevos datos.
    \item Hará que el resultado (numérico) de la evaluación sea un buen indicador de si el problema a tratar se está resolviendo correctamente o no.
\end{itemize}







\subsection{Repeated k-fold}
\label{repeated_kfold}

Ya sabemos que para comparar los distintos modelos de una forma fiable, dividir el conjunto de datos en un conjunto de entrenamiento fijo y un conjunto de validación fijo puede ser problemático, especialmente si este conjunto de validación es pequeño, ya que afirmar que un modelo \textit{A} funciona mejor que uno \textit{B} será difícil debido a que el resultado de la evaluación en validación, dependerá en gran medida del conjunto concreto (y no tanto de los modelos). Es decir, si realizamos una partición del tipo train-validation-test en un conjunto de datos pequeño, es muy probable que nos equivoquemos al decir que un modelo es mejor que otro, ya que dependerá mucho de la partición concreta de los datos \cite[p.~122]{goodfellow2016deep}

En condiciones normales, una solución para evitar de forma razonable este problema es utilizar k-fold, método que se explicó en la sección \ref{k_fold}. El problema es que cuando tenemos un conjunto de datos extremadamente pequeño (como en nuestro caso), aunque es cierto que realizando k-fold reducimos este problema, tenemos un segundo problema muy similar, y es que los resultados pueden oscilar de forma notable en función de factores aleatorios (partición concreta de los datos, inicialización de los pesos de los modelos, etc...).

Está claro que esta oscilación aleatoria en los resultados, nos hace difícil comparar de una forma rigurosa los distintos modelos, y además, hace que los resultados que obtenemos no sean repetibles.

Con el objetivo de reducir esta variabilidad, hemos utilizado una variante conocida como ``repeated k-fold''. Esta variante consiste, como su nombre indica, en repetir k-fold varias veces, utilizando una división distinta de los datos en cada una de las repeticiones \cite{brownlee_repeated_kfold}. Debemos destacar que no existe (o al menos no hemos podido encontrar) una demostración formal de por qué la variabilidad podría verse reducida \cite{rodriguez2009sensitivity} usando este método, pero lo cierto es que es una técnica utilizada por distintos científicos de datos y en plataformas como \textit{Kaggle}, y en este caso concreto parece funcionar, aportándonos una gran disminución de la variabilidad.

Además de aplicar esta técnica, nos hemos asegurado de que cada una de las $k$ divisiones mantiene la misma proporción de clases, de nuevo, con el objetivo de intentar una evaluación lo más ``realista'' posible. Para concretar aún más, hemos realizado repeated k-fold con 5 repeticiones, y $k=10$. 

Por supuesto, no todo serán ventajas, y es que con este tipo de evaluación, testear un sólo modelo requerirá de $5 * 10 = 50$ entrenamientos, lo que enlentecerá de forma exagerada nuestra experimentación.

Queremos destacar el hecho de que utilizar esta técnica \textbf{no eliminará la variabilidad en la estimación por completo}, pero sí que la reducirá, permitiendo comparar los modelos de forma más robusta.

\subsection{Separación de un conjunto de test}

El método anterior nos permitía reducir la variabilidad en la evaluación de los modelos, pero tenemos un problema de otro tipo, y es que realizando las suficientes pruebas y mejoras, al final acabaremos sesgando de forma optimista la estimación, aunque sea de forma leve.

Lo ideal es reservar un conjunto de test completamente independiente que no se utilice hasta haber seleccionado el mejor de los modelos, de forma que el resultado obtenido en este conjunto no estará sesgado. En nuestro caso, reservaremos un conjunto de test con el \textbf{20\%} de los datos iniciales, tomados de forma aleatoria.

El motivo de usar 20\% es que si utilizáramos muchos más datos para test, tendríamos aún menos datos para entrenar, lo que empeoraría nuestro modelo. Por otra parte, si usáramos menos datos, la estimación dependería en gran medida de los ejemplos concretos tomados para test, con lo que no sería una estimación representativa de la realidad.

Podríamos haber decidido cualquier otra proporción, pero lo cierto es que no existe una regla ni proporción óptima. Nosotros consideramos que un 20\% es razonable en este caso.

Otro detalle a destacar, es que con el objetivo de que este 20\% de ejemplos sean lo más representativos posibles, hemos realizado la división de forma que la proporción de ejemplos de cada clase sea la misma que la existente en el conjunto de datos (lo que se conoce a veces como división estratificada).

Un último detalle a destacar es que aunque con este método no tengamos un sesgo optimista, sí que tenemos un problema, y es que cuando este conjunto de test es pequeño (50 ejemplos en nuestro caso), \textbf{el resultado será muy dependiente de los ejemplos concretos de este conjunto}.

\subsection{Métricas y matriz de confusión}
\label{metrica_explicacion}

La evaluación de los modelos que hemos visto hasta ahora nos dará como resultado un determinado valor, pero es necesario que ese valor sea un buen indicativo de cómo de bien está resolviendo nuestro modelo el problema en cuestión, para lo que tendremos que decidir una \textbf{métrica} adecuada al problema, es decir, tenemos que establecer cómo se calcula ese valor.

Por otro lado, para saber cómo de bien realiza la clasificación un clasificador, existe una herramienta muy utilizada y que utilizaremos habitualmente: la \textbf{matriz de confusión}.

\subsubsection{Matriz de confusión}

Se trata de una matriz en la que en cada columna indica el número de elementos de cada clase que ha predicho el clasificador, mientras que en cada fila hay un recuento del número de ejemplos que hay realmente en cada clase.

Por ejemplo, supongamos que tenemos sólo 12 pacientes de test, siendo 4 de cada una de nuestras tres clases (\acrshort{ad}, \acrshort{mci}, \acrshort{cn}), y pasamos estos 12 ejemplos a un supuesto clasificador, la matriz de confusión obtenida podría tener la siguiente forma:

\newcommand\items{3}   
\arrayrulecolor{white} 

\begin{center}
\begin{tabular}{cc*{\items}{|E}|}
\multicolumn{1}{c}{} &\multicolumn{1}{c}{} &\multicolumn{\items}{c}{Predicho} \\ \hhline{~*\items{|-}|}
\multicolumn{1}{c}{} & 
\multicolumn{1}{c}{} & 
\multicolumn{1}{c}{\rot{CN}} & 
\multicolumn{1}{c}{\rot{AD}} & 
\multicolumn{1}{c}{\rot{MCI}} \\ \hhline{~*\items{|-}|}
\multirow{\items}{*}{\rotatebox{90}{Real}} 
& CN  & 2   & 0  & 2   \\ \hhline{~*\items{|-}|}
& AD  & 0   & 3  & 1   \\ \hhline{~*\items{|-}|}
& MCI & 1   & 1   & 2   \\ \hhline{~*\items{|-}|}
\end{tabular}
\end{center}

Si nos fijamos por ejemplo en la primera fila, vemos que hay en total 4 ejemplos (2 + 2), lo que significa que existen 4 ejemplos cuya clase es \acrshort{cn}, aunque el clasificador ha predicho que 2 de ellos son \acrshort{cn}, y en los otros dos ha cometido un error prediciendo que son \acrshort{mci}.

Por tanto, en la diagonal principal tenemos el número de ejemplos de cada clase que el clasificador ha clasificado correctamente, mientras que en el resto de casillas, encontramos el número de \textbf{confusiones} de cada tipo que ha tenido el clasificador, como por ejemplo:\begin{itemize}
    \itemsep0em
    \item En la primera fila y segunda columna, tendríamos el número de ejemplos cuya etiqueta real es \acrshort{cn} pero el clasificador ha etiquetado como \acrshort{ad} (en este caso 0).
    \item En la tercera fila y segunda columna, vemos que existe un ejemplo de la clase \acrshort{mci} que ha sido etiquetado como \acrshort{ad}.
\end{itemize}

\subsubsection{Exactitud (accuracy)}

La métrica principal que hemos decidido utilizar para este problema es la exactitud. Se trata de una métrica muy simple, que se define como la fracción del número de ejemplos bien clasificados entre el número total de ejemplos, es decir, esta métrica mide la \textbf{proporción de ejemplos bien clasificados}.

Para que esta métrica sea realmente la adecuada, nuestro problema debe cumplir los siguientes requisitos que vamos a analizar brevemente:\begin{itemize}
    \item El conjunto de datos debe estar \textbf{razonablemente balanceado}. Este primer requisito es muy importante, ya que si por ejemplo, tuviéramos un problema de dos clases en el que el 99\% de los ejemplos pertenecen a una de ellas, con simplemente clasificar todos los ejemplos como si fueran de la clase mayoritaria, obtendríamos una exactitud del 99\%, y en absoluto sería un buen clasificador.
    
    En nuestro problema, la clase mayoritaria, \acrshort{mci}, representa un \textbf{44.5\%} de los ejemplos, por lo que no supondrá un problema utilizar la exactitud, aunque tendremos que tener en cuenta que si el clasificador consiguiera una exactitud del 44\%, es probable que solamente haya aprendido a clasificar todos los ejemplos como \acrshort{mci}.
    
    \item Todas las clases deben tener \textbf{igual importancia}. Existen casos en los que es más importante etiquetar una clase que otra, por ejemplo, si tuviéramos el problema de distinguir entre un tumor benigno o maligno, sería extremadamente importante no cometer el fallo de clasificar un tumor maligno como benigno (ya que esto podría implicar la muerte de una persona), mientras que equivocarse etiquetando un tumor benigno como maligno, no sería de tanta importancia (aunque es cierto que tampoco es deseable).
    
    En nuestro problema, es cierto que quizás la clase \acrshort{ad} es algo más importante que las demás, ya que un diagnóstico temprano podría permitir retrasar algo los síntomas, como vimos en la sección \ref{enfermedad_alzheimer}. Sin embargo, es difícil establecer un criterio sobre cuánto más importante es esta clase (no podemos decir que sea el doble de importante que otra clase, ni el triple).
    
    Para nuestros objetivos (y dado que no podemos establecer un criterio claro de la importancia de cada clase), es razonable considerar que las tres clases son igualmente importantes, y nuestro objetivo principal es que se clasifiquen correctamente el mayor número de ejemplos posibles.
    
    \item Nuestro clasificador debe dar como salida \textbf{una clase}. En este caso, nuestros clasificadores (redes neuronales convolucionales), darán como salida una clase, que será concretamente la de mayor probabilidad. 
    
\end{itemize}

Como cumplimos en buena medida los criterios anteriores, utilizaremos la exactitud, que además nos aporta tres ventajas: es \textbf{muy sencilla de calcular}, es \textbf{muy intuitiva}, y nos permite evaluar un modelo con \textbf{un único valor}.

\textbf{A cambio de estas ventajas}, tendremos que asumir que estamos dando a las tres clases exactamente la misma importancia.

\subsubsection{Sensibilidad y especificidad}

Para comparar los modelos, hemos visto que es adecuado utilizar la exactitud como métrica. Sin embargo, cuando demos los resultados finales (en test), utilizaremos algunas medidas extra que nos dirán cómo de bien se está clasificando concretamente la enfermedad de Alzheimer, que consideramos algo más importante (aunque no sabemos cuanto). Estas medidas son las siguientes:

\[Sensibilidad = \frac{VP}{ VP + FN}\]
\[Especificidad = \frac{VN}{VN + FP}\]

Siendo $VP$ los verdaderos positivos, $FN$ los falsos negativos, $VN$ los verdaderos negativos y $FP$ los falsos positivos.

La \textbf{sensibilidad} nos indica qué proporción de los positivos han sido clasificados correctamente. Una alta sensibilidad implica que el clasificador es capaz de detectar todos los casos positivos (por lo que sería adecuado tener una alta sensibilidad para la clase \acrshort{ad}).

Por otro lado, la \textbf{especificidad} nos indica qué proporción de los negativos han sido correctamente clasificados. Un clasificador con una alta sensibilidad pero una baja especificidad, sería un clasificador con un alto número de falsos positivos (por lo que idealmente, nos gustaría tener una alta especificidad).

Estas métricas se usan habitualmente en problemas de clasificación binaria, y en el ámbito de la medicina, usualmente la clase positiva es la presencia de una determinada enfermedad. Como en este caso tenemos 3 clases, diremos que la \textbf{clase positiva} es la \acrshort{ad}, mientras que la \textbf{clase negativa} es la suma de los pacientes \acrshort{cn} y \acrshort{mri}.

\section{Método general de experimentación}
\label{metodo_experimentacion}

En este trabajo tendremos que conseguir encontrar una red convolucional que resuelva lo mejor posible nuestro problema de clasificación.

Aunque tenemos conocimientos teóricos de las distintas decisiones que podemos tomar, y una ligera idea de lo que puede funcionar, no es posible (salvo que tuviéramos mucha suerte) encontrar un modelo con un buen comportamiento sin realizar experimentos.

Como vamos a tener que experimentar con distintas decisiones, es adecuado tener un método, aunque sea muy general, que podamos seguir (aunque sea aproximadamente) para guiar los distintos experimentos. Este ``método'' que seguiremos será el siguiente: \label{algoritmo_generico} \begin{enumerate}
    \itemsep0em
    \item Implementar un modelo base, rápidamente, sin perder el tiempo en intentar crear un modelo perfecto.
    \item Entrenarlo y analizar la curva de aprendizaje (figura \ref{fig:early_stopping}). Tras este análisis podremos diagnosticar que nuestro modelo sufre de:\begin{itemize}
        \itemsep0em
        \item Underfitting (en general fácil de resolver): en este caso tomaremos alguna de las siguientes decisiones\begin{enumerate}
            \itemsep0em
            \item Crear una red más grande (más profunda o con más parámetros).
            \item Entrenar por más tiempo (épocas).
            \item Ajustar el optimizador o usar otro.
        \end{enumerate}
        \item Overfitting (más difícil de resolver): en este caso tomaremos alguna de las siguientes decisiones\begin{enumerate}
            \item Buscar más datos (usualmente no es posible).
            \item Regularizar (muchas posibles formas).
            \item Buscar otro tipo de arquitectura (muchas posibles).
        \end{enumerate}
    \end{itemize}
       \item Tomar una decisión según el diagnóstico del paso 2 y volver a repetirlo.
\end{enumerate}

Queremos dejar claro que esto \textbf{no se trata de un procedimiento cerrado que se pueda seguir con total rigurosidad}, sino que es muy genérico, y cada uno de sus pasos abre una puerta a infinidad de decisiones que tomar, pero sí que sirve para dar una idea aproximada del proceso que seguiremos para llegar a obtener los modelos finales.

\section{Hiperparámetros principales}
\label{hiperparametros_principales}

En el paso 2 del algoritmo general anterior, hemos visto una serie de decisiones \textbf{muy generales} a tomar según nuestro modelo sufra underfitting u overfitting. 

Ahora, concretaremos \textit{algunas} decisiones más específicas que encajan dentro de estas decisiones generales. Estas decisiones específicas consistirán en ajustar una serie de \textbf{hiperparámetros}. 

Aunque en el capítulo \ref{capitulo_cnn} hablamos de forma breve sobre el concepto de hiperparámetro, vamos a recordar brevemente este concepto. Para ello, vamos a recordar primero el concepto de ``parámetro de un modelo'': un parámetro no es más que cada uno de los pesos que se aprenden durante el entrenamiento, es decir, cuando hablamos de parámetros, nos referimos a los parámetros \textit{entrenables}.

El resto de variables libres que  \textbf{tendremos que decidir} (no se aprenden) a la hora de entrenar un modelo, se conocen como hiperparámetros. Veamos algunos de los más importantes:\begin{enumerate}
    \itemsep0em
    \item \textbf{Learning rate} (tasa de aprendizaje): se trata de un hiperparámetro referente al \textit{optimizador} que indica la ``longitud'' de los pasos a la hora de actualizar los pesos del modelo (algoritmo \ref{alg:descenso_gradiente}). Usualmente nos referimos a él con las siglas \acrshort{lr}, o con la letra $\alpha$.
    
    \item \textbf{Número de épocas}: como su nombre indica, es el número de épocas durante el que entrenamos el modelo. Hay que ajustarlo ya que si es demasiado alto, es probable que aparezca sobreajuste, mientras que si es demasiado bajo el modelo no habrá aprendido lo necesario (underfitting).
    
    \item \textbf{Número de capas ocultas}: ya conocemos este concepto, y con este número estableceremos la profundidad de la red. Cabe destacar que \textit{uno de los objetivos principales de este trabajo es estudiar el efecto de este hiperparámetro para nuestro problema}.
    
    \item \textbf{Número de filtros}: este hiperparámetro debe ser establecido para cada una de las capas. A mayor número de filtros, mayor es el número de parámetros entrenables.
    
    \item \textbf{Función de activación}: decidir si usar la función sigmoide, ReLU, tangente hiperbólica, Leaky ReLU ...
    
    \item \textbf{Optimizador}: decidir el optimizador a usar SGD, Adam, RMSProp ...
    
    \item \textbf{Tamaño del minibatch}: se refiere al número de ejemplos que se tomarán durante el entrenamiento para realizar una actualización de los pesos (también podemos entenderlo como el número de ejemplos que se toman para calcular el gradiente). Un mayor tamaño de minibatch suele requerir un mayor \acrshort{lr}.
    
    \item \textbf{Hiperparámetros de regularización}: aquí caben muchos hiperparámetros, como la tasa de dropout (en el caso de capas dropout), el hiperparámetro $\lambda$ en la regularización L2, hiperparámetros relacionados con el aumento de datos (que veremos en la sección \ref{parametros_aumento}), etc...
    
\end{enumerate}

\subsection{¿Cómo se ajustan?}

Como hemos visto en los pasos 2 y 3 de nuestro algoritmo genérico (sección \ref{algoritmo_generico}) el proceso de tomar decisiones es cíclico, es decir, requiere realizar varias pruebas (experimentos).

Queremos enfatizar que para ajustar un hiperparámetro, podríamos darle infinidad de valores distintos. Si diéramos estos valores de forma aleatoria la experimentación sería un proceso eterno, por lo que siempre nos apoyaremos en nuestro conocimiento teórico y en la \textbf{experiencia}.

Por si quedara ambigüedad en la explicación sobre el proceso de ajustar un determinado hiperparámetro, vamos a poner un ejemplo de cómo ajustaríamos la tasa de aprendizaje (\acrshort{lr}).

En primer lugar estableceríamos un valor inicial para esta tasa de aprendizaje (teniendo en cuenta nuestra experiencia y conocimientos). Supongamos que establecemos $LR = 0.00001$. Ahora, evaluaríamos nuestro modelo conforme al método de evaluación que hayamos establecido (en nuestro caso \textbf{repeated k-fold}), y también dibujaríamos la curva de aprendizaje (también calculada con repeated k-fold).

Observando el resultado de la evaluación y la curva, podríamos ver, por ejemplo, que el aprendizaje está siendo excesivamente lento. Lo que nos haría sospechar de que estamos utilizando una tasa de aprendizaje excesivamente baja, por lo que probaríamos con una tasa de aprendizaje más alta, por ejemplo $LR = 0.001$.

Tras esta decisión, volveríamos a evaluar el modelo y tomar una nueva decisión según lo que ocurra, hasta conseguir un resultado razonable.

Debemos tener en cuenta que \textbf{la modificación de un hiperparámetro, en general, afectará a otros hiperparámetros}, y en realidad, cualquier decisión que tomemos, como cambiar el tipo de preprocesado, afectará a los hiperparámetros óptimos, por lo que este proceso de optimización de hiperparámetros se trata de una tarea muy compleja y larga.

Por último, queremos hacer saber que existen herramientas par la \textbf{optimización de hiperparámetros automática}, como OpTuna \cite{optuna_2019} ó Keras tuner \cite{omalley2019kerastuner}, pero debido al método de evaluación que hemos tomado (obligados por las particularidades de nuestro problema), con la carga computacional que esto conlleva, y unido a las limitaciones en el tiempo de utilización del hardware (sección \ref{desarrollo_nube}), no hemos podido utilizar de forma satisfactoria estas herramientas, por lo que la optimización de hiperparámetros, será un laborioso proceso manual.

\begin{tcolorbox}[
  colback=Green!5!white,
  colframe=Green!75!black,
  title={Recapitulación}]
 Los puntos más importantes a recordar son:
\begin{itemize}
    \itemsep0em
    \item Evaluaremos nuestros modelos mediante repeated k-fold, que nos dará una estimación poco sesgada del error cometido, y nos aportará un buen grado de repetibilidad.
    \item Reservaremos un conjunto de test completamente independiente para tener una estimación no sesgada del desempeño de los modelos.
    \item Como métrica, utilizaremos principalmente la exactitud, una métrica intuitiva que indica la proporción de ejemplos bien clasificados.
    \item Para crear una arquitectura y ajustar distintos hiperparámetros, hemos presentado un ``método'' cíclico basado principalmente en analizar la curva de aprendizaje en sucesivas iteraciones.
    \end{itemize}
\end{tcolorbox}

\chapter{Aspectos de implementación}
\begin{tcolorbox}[
  colback=SkyBlue!5!white,
  colframe=SkyBlue!75!black,
  title={Sumario}]
Antes de comenzar con el desarrollo y la experimentación, en este capítulo explicaremos de forma muy breve algunos de los elementos principales que hemos utilizado para la implementación y ejecución de nuestros experimentos: las bibliotecas más importantes, algunos entornos de desarrollo en la nube (y el hardware al que estos nos dan acceso), y otras herramientas útiles.\vspace{1em}

No es el objetivo explicar en detalle ninguno de estos elementos, sino dar a conocer el motivo de su uso o para qué nos sirven.

\end{tcolorbox}

\section{Bibliotecas principales}

Antes de pasar a explicar algunas de las bibliotecas más útiles para nuestro trabajo, conviene decir que el lenguaje de programación que hemos utilizado es Python. Nuestro motivo para elegirlo es que ya teníamos experiencia con él, y además se trata de un lenguaje muy popular para el desarrollo de proyectos de aprendizaje automático, lo que nos dará ventajas para encontrar rápidamente información de ayuda cuando sea necesario.

De entre las numerosas bibliotecas de código abierto existentes para este lenguaje, algunas de las que han sido \textbf{fundamentales} para el desarrollo de nuestro trabajo son las siguientes:\begin{itemize}
    \itemsep0em
    \item \textbf{Tensorflow} \cite{tensorflow2015-whitepaper}: nos permite la creación de modelos de aprendizaje automático de forma relativamente sencilla y eficiente, además de darnos acceso a numerosas funciones que nos facilitarán, entre otras cosas, la carga de datos de forma eficiente, o el entrenamiento de modelos de forma distribuida.
    \item \textbf{Keras} \cite{chollet2015keras}: actúa como una API de alto nivel a la biblioteca Tensorflow, y está diseñada de forma que nos permitirá diseñar redes neuronales profundas de una forma rápida. 
    \item \textbf{Scikit-learn} \cite{scikit-learn}: se trata de otra biblioteca de aprendizaje automático. Nos aporta gran cantidad de funciones, siendo especialmente útiles para nosotros las relacionadas con la evaluación de modelos y el particionamiento de los conjuntos de datos.
    \item \textbf{Scipy} \cite{2020SciPy-NMeth}: nos aporta funciones para el procesado de imágenes tridimensionales (filtrado, transformaciones geométricas...).
    \item \textbf{Numpy} \cite{harris2020array}: una biblioteca esencial para el cálculo numérico y manejo eficiente de vectores n-dimensionales (numpy \textit{ndarray}).
    \item \textbf{Nibabel} \cite{brettfreec84}: para la lectura (y escritura) de imágenes médicas.
    \item \textbf{Matplotlib} \cite{Hunter:2007}: para la creación de gráficas.
    
\end{itemize}

\section{Desarrollo en la nube}
\label{desarrollo_nube}

Como ya vimos (sección \ref{gpu}), el entrenamiento de redes convolucionales profundas en un tiempo razonable requiere el uso de GPU.

En nuestro caso, no tenemos acceso a ningún ordenador con una GPU dedicada, e incluso si lo tuviéramos, lo más probable es que debido al uso de imágenes 3D, una tarjeta gráfica de gama media no fuese suficiente para entrenar nuestras redes en un tiempo admisible para nuestros objetivos.

Ante este problema, hemos decidido apoyarnos en entornos de desarrollo que nos permiten programar código en Python directamente en el navegador, y que nos dan acceso gratuito (limitado) a \textbf{GPUs} con un gran rendimiento, e incluso a otro tipo de hardware concebido explícitamente para el entrenamiento y uso de redes neuronales \cite{cloud_tpu}: las \textbf{unidades de procesamiento tensorial} (TPUs), que además, están optimizadas específicamente para su uso con Tensorflow.

\subsection{Google Colaboratory}

Uno de estos entornos de desarrollo es Google Colaboratory (o Colab simplemente).

Entre sus \textbf{ventajas} principales encontramos:\begin{itemize}
    \itemsep0em
    \item Posibilidad de usar gratuitamente tarjetas gráficas de una gran potencia (Nvidia k80, T4, P4 y P100), aunque no podemos elegir cuál usar, sino que se asignan según la disponibilidad.
    \item Acceso a unidades de procesamiento tensorial, aún más rápidas que las GPUs si se usan adecuadamente, y con mayor memoria.
    \item Dispone de 16GB de memoria RAM, que en principio será suficiente para nosotros.
    \item Permite el acceso directo al almacenamiento en Google Drive, lo que nos puede ser útil para cargar los datos.
\end{itemize}

\noindent Por otro lado, encontramos las siguientes \textbf{desventajas}:\begin{itemize}
    \itemsep0em
    \item El tiempo de uso de GPU es limitado (unas 12 horas seguidas), y una vez pasado ese límite habrá que esperar unas 12 horas para volver a poder usarlas. Además, si se supera este límite frecuentemente, hemos observado que este tiempo de uso máximo se va reduciendo, y el servicio va empeorando (se producen cortes en la ejecución).
    
    Para superar este problema, hemos optado por suscribirnos a \textbf{Colab Pro}, que nos dará posibilidad de un uso continuado de las GPU durante 24 horas, además de una disponibilidad casi continua de la \textbf{Nvidia P100} (la más potente de las que ofrecen).
    
    \item A pesar de que tenemos acceso a TPUs, estas requieren que los datos se encuentren en Google Cloud Storage, un servicio no gratuito (y caro si se sobrepasan ciertos límites), por lo que para evitar el riesgo, no las usaremos.
    \item Aunque tengamos acceso a Google Drive, la cantidad de almacenamiento vendrá determinado por el plan al que estemos suscritos (con el plan gratuito sólo 15 gigabytes). Sin embargo, esta limitación la podemos evitar utilizando nuestra \textbf{cuenta de Google de la UGR}.
\end{itemize}

\subsection{Kaggle}

Otro entorno de desarrollo, que también se ejecuta directamente en el navegador, es el que nos ofrece la comunidad de aprendizaje automático y ciencias de datos Kaggle.

Entre sus \textbf{ventajas}, encontramos:\begin{itemize}
    \itemsep0em
    \item Nos permite la creación de conjuntos de datos de hasta 100GB, y un número ilimitado de conjuntos de datos, que pueden ser privados (como en nuestro caso).
    \item En nuestra experiencia, el acceso a estos conjuntos de datos es enormemente más rápido que el acceso a los datos almacenados en Google Drive.
    \item Nos da acceso al mismo hardware que Google Colaboratory.
    \item Nos permite alojar los conjuntos de datos en Google Cloud Storage de forma gratuita, por lo que podremos explotar el potencial de las TPU.
\end{itemize}

\noindent Por su parte, las principales \textbf{desventajas} son:\begin{itemize}
    \itemsep0em
    \item El acceso al hardware es aún más limitado que en el caso de Colab: 9 horas de uso continuado, y 30 horas semanales como máximo.
    \item No permite el acceso a código alojado en Github.
\end{itemize}

\subsection{Nuestra solución: Colab + Kaggle}

En un primer momento, decidimos utilizar Colab Pro, ya que en principio parecía que los recursos eran suficientes para poder ejecutar todos nuestros experimentos en un tiempo razonable, pero lo cierto es que debido al método de evaluación que utilizamos (sección \ref{repeated_kfold}), y al uso de redes 3D (con las cuales no teníamos ninguna experiencia), incluso con los recursos de Colab las ejecuciones eran excesivamente lentas.

Para aligerar estas ejecuciones, decidimos usar paralelamente las GPU, y especialmente las TPU de Kaggle, donde hemos ejecutado los experimentos especialmente pesados, consiguiendo una reducción de hasta 8 veces el tiempo de ejecución (aproximadamente).

Cabe destacar que algunos experimentos, \textbf{sólo han podido ser ejecutados gracias al uso de las TPU} (y sus 128GB de memoria), ya que en ocasiones los 16GB de memoria en las GPU no eran suficientes (en teoría deberían ser suficientes, pero cuando usamos Tensorflow y Colab, en ocasiones se van almacenando objetos en la memoria que acaban llenándola, y existe actualmente poca información sobre cómo liberarla adecuadamente).

\section{Otras herramientas}

Otras herramientas que hemos utilizado en este proyecto son \textbf{SPM12}, que se trata de un software que se ejecuta sobre el lenguaje Matlab, y que se utilizará para el preprocesado de las imágenes, y \textbf{Mricron}, útil para visualizar las imágenes y comprobar que no hemos cometido errores importantes durante el preprocesado.

\chapter{Datos: preprocesado y carga}

\begin{tcolorbox}[
  colback=SkyBlue!5!white,
  colframe=SkyBlue!75!black,
  title={Sumario}]

Aunque ya se habló de los datos y su preprocesado en el capítulo \ref{datos}, en este capítulo veremos, de forma muy breve: \begin{itemize}
    \itemsep0em
    \item Cómo hemos realizado este preprocesado con el software SPM12.
    \item Algunas de las formas en las que se puede cargar un conjunto de datos.
\end{itemize}

\end{tcolorbox}

\section{Preprocesado con SPM12}

Ya vimos que para nuestro problema, parecía adecuado realizar ciertos tipos de preprocesamiento (sección \ref{preprocesado}), siendo los más complejos la normalización espacial, y la segmentación de tejidos. Implementar estos preprocesados sería enormemente complejo, pero vimos que por suerte existen programas como SPM12 que nos permiten hacerlo de forma sencilla.

A continuación vamos a explicar de forma breve cómo se realiza el preprocesado.

\subsection{Normalización espacial (PET)}

\begin{figure}[H]
    \centering
    \includegraphics[width=10cm]{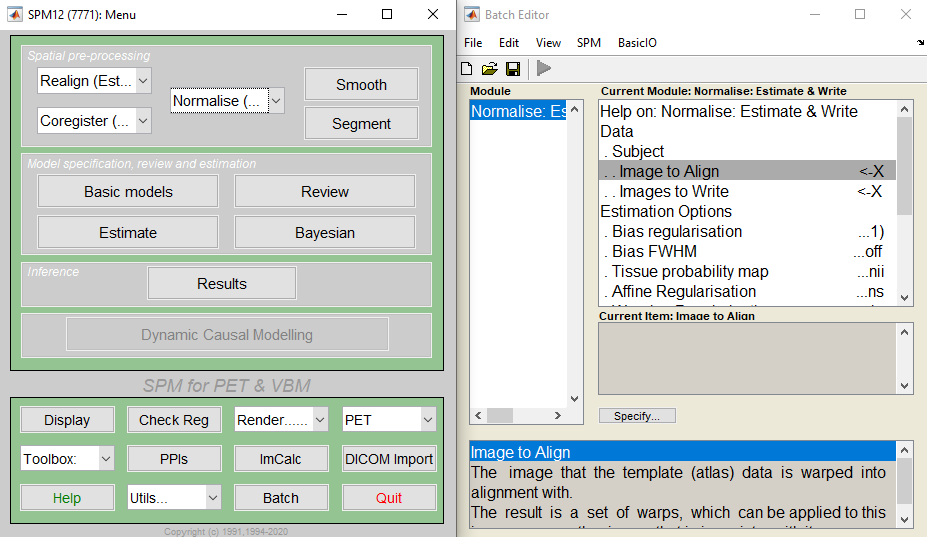}
    \caption{SPM. Interfaz gráfica para PET.}
    \label{fig:spm_pet}
\end{figure}

SPM12 tiene una interfaz gráfica (figura \ref{fig:spm_pet}) dedicada al manejo de las imágenes PET.

En esta interfaz, simplemente tendremos que dejar todos los parámetros que aparecen por defecto en la ventana izquierda de la figura \ref{fig:spm_pet}, e ir seleccionando una a una las imágenes \acrshort{pet} que queremos normalizar espacialmente en la ventana que aparece en la derecha (seleccionándola tanto en ``Image to Align'' como en ``Image to Write'').

Respecto a los parámetros (que no hemos modificado), lo cierto es que no son muy difíciles de entender y pueden consultarse en el manual gratuito de SPM \cite{ashburner2014spm12}. Sin embargo, para nuestros objetivos, es más que suficiente con que comprendamos la idea de lo que se consigue con la normalización espacial (sección \ref{fig:normalizacion_espacial}), por lo que no los explicaremos. \textbf{Únicamente queríamos mostrar que efectivamente, este preprocesado se hace de forma sencilla gracias a esta herramienta}.

Un último detalle sin mucha importancia, es que si quisiéramos evitar tener que introducir las imágenes una a una para ser preprocesadas, que puede ser un proceso largo, SPM nos permite generar el código en Matlab para preprocesar una imagen, por lo que de forma relativamente sencilla podemos programar un bucle que haga el preprocesado de todas las imágenes de una sóla vez.

\subsection{Segmentación de tejidos (MRI)}
\label{segmentacion_gris}

En el caso de la segmentación de tejidos, tenemos también una interfaz gráfica idéntica a la anterior, y el preprocesado es incluso más sencillo, ya que en este caso SPM sí nos permitirá seleccionar todas las imágenes \acrshort{mri} de una sola vez. Simplemente tendremos que seleccionar todas las imágenes, y especificar los tejidos que queramos extraer (en nuestro caso, la materia gris), sin más que seleccionarlos con un clic.

\section{Carga de datos}

Para entrenar un modelo de aprendizaje automático, este tendrá que recibir datos de los que aprender, que usualmente tendremos almacenados en el disco duro o en algún otro dispositivo de almacenamiento permanente. 

En nuestro problema concreto, lo que tendremos que hacer es ``alimentar'' a nuestro modelo con imágenes (matrices), y la etiqueta de cada una de esas imágenes (un vector de tres elementos), pero como hemos dicho, estas imágenes las tenemos en el disco duro.

En Tensorflow, existen varios métodos posibles para alimentar a nuestro modelo con estas imágenes etiquetadas. En las próximas secciones, veremos resumidamente algunas de las formas en las que esto puede hacerse.

\subsection{Datos cargados en memoria}

La forma más sencilla (y probablemente más común) de introducir los datos para el entrenamiento de un modelo, es creando un vector (Numpy array) $\boldsymbol{X}$ en el que introducimos las imágenes (cada posición contiene una imagen), y otro vector $\boldsymbol{y}$ de la misma longitud que contiene las etiquetas. Cargar todos los datos en un vector puede tener una serie de \textbf{problemas}:\begin{enumerate}
    \itemsep0em
    \item El vector será una variable de nuestro programa en Python, y como cualquier variable, se almacenará en la memoria RAM del ordenador.
    
    En nuestro caso, con el conjunto de datos de \acrshort{ad} no tendremos problema, ya que es muy pequeño (menos de 5GB). El problema nos llega cuando queremos realizar el preentrenamiento con datos de COVID19, que ocuparán más de 40GB, por lo que \textbf{esta forma de cargar los datos no nos valdrá}.
    
    \item Cada vez que reiniciemos el programa en Python, para poder entrenar nuestros modelos tendremos que volver a crear este vector, lo que implica leer las imágenes desde el disco. Esto es especialmente lento si nuestro ``disco'' es Google Drive.
\end{enumerate}

A pesar de estos problemas, tiene una \textbf{ventaja}, y es que una vez cargados los datos en memoria RAM, Tensorflow podrá utilizar los datos de una forma rápida, pudiendo aprovechar el rendimiento de la GPU o TPU.

\subsection{Generador de datos}

Hasta ahora, tenemos el problema de que $\boldsymbol{X}$ podría ser muy grande, y cargarlo en memoria no es factible. 

Tensorflow admite otra posible forma de recibir los datos, y es mediante un generador de Python \footnote{En esencia, un generador es una función que devuelve elementos uno a uno (en realidad, devuelve un iterador).}. A este generador le pasaríamos una \textbf{lista con el nombre de los archivos} de cada imagen, e iría cargando y devolviendo las imágenes, una a una. Este método \textbf{soluciona el problema de la RAM} porque sólo tendremos que almacenar en memoria la lista con el nombre de cada imagen (que obviamente ocupa mucho menos espacio)

Sin embargo, este método tiene un grave \textbf{problema}, y es que la carga de datos será enormemente lenta, tanto que no podremos ``alimentar'' la GPU ó TPU que utilicemos para conseguir su máximo rendimiento, es decir, la carga de datos será un cuello de botella que nos impedirá entrenar un modelo de forma rápida, y nos hará desaprovechar el tiempo de uso limitado de las TPU.

\subsection{TFRecords}
\label{tfrecords}

Hasta el momento hemos solucionado el problema de que los datos no quepan en memoria, pero tenemos el problema de la lentitud en la carga de los datos para ``alimentar'' el entrenamiento. 

Para resolver este problema en la mayor medida posible, existe un tipo de archivo propio de Tensorflow que nos ayuda, el formato TFRecord.

Sin entrar en muchos detalles, un archivo TFRecord almacenará uno o varios ejemplos (imágenes) junto a sus etiquetas, como una cadena de bytes, y nos dará una serie de ventajas \cite{oliveira_2021}:\begin{itemize}
    \itemsep0em
    \item Permite ser leído con múltiples operaciones de entrada/salida \textbf{en paralelo} (esto lo gestiona Tensorflow de forma transparente al usuario), que es especialmente útil para obtener un buen rendimiento de las TPU. 
    \item Los archivos TFRecord almacenarán los datos junto a sus etiquetas, por lo que nos podremos despreocupar de guardar las imágenes con distinta etiqueta en directorios distintos.
\end{itemize}

Tras estudiar las ventajas de este formato, hemos decidido usarlo para implementar nuestros experimentos. A cambio de las ventajas, tenemos una serie de \textbf{inconvenientes y dificultades}:\begin{itemize}
    \itemsep0em
    \item Tendremos que leer las imágenes y sus etiquetas, serializarlas \footnote{Convertirlas en un cadena de bytes.}, y almacenarlas en una serie de archivos TFRecord (almacenados en Google Drive). 
    
    En el caso de no tener experiencia con este formato (como era nuestro caso), realizar este proceso puede llegar a ser difícil, en gran parte, debido a que aunque existe mucha información sobre como realizar el proceso con imágenes 2D, hasta hoy \textbf{la información sobre el manejo de imágenes 3D es muy limitada}, y la compatibilidad de Tensorflow con este tipo de imágenes es en nuestra opinión, bastante reducida. De nuevo, no explicaremos en detalle este proceso, ya que son aspectos de implementación que quedan fuera de los objetivos de este trabajo.
    
    \item Ocuparemos un mayor almacenamiento, ya que tendremos que guardar nuestras imágenes en este formato, y usualmente es conveniente mantener el conjunto de datos en su formato original (por si quisiéramos hacer cambios futuros). En nuestro caso, esto no será un problema gracias al almacenamiento ilimitado en Google Drive que nos ofrece la Universidad de Granada.
    
    \item La dificultad para leer los datos en formato TFRecord también se verá incrementada (al menos si no tenemos experiencia), además de otras dificultades que añade el hecho de conseguir que esta lectura de datos sea eficiente.
    
\end{itemize}

\begin{tcolorbox}[
  colback=Green!5!white,
  colframe=Green!75!black,
  title={Recapitulación}]
\begin{itemize}
    \itemsep0em
    \item El preprocesado de las imágenes se hará de forma sencilla mediante la interfaz gráfica de SPM12.
    \item Cargar los datos en memoria RAM es eficiente para entrenar los modelos, pero no es factible cuando el conjunto de datos es grande (caso del conjunto COVID19).
    \item Para solucionar este problema sin perder mucho en eficiencia usaremos un formato especial de Tensorflow (TFRecord), aunque esto complicará notablemente la implementación.
    \end{itemize}
\end{tcolorbox}

\chapter{Experimentación}

\begin{tcolorbox}[
  colback=SkyBlue!5!white,
  colframe=SkyBlue!75!black,
  title={Sumario}]

En este capítulo explicaremos en detalle los experimentos realizados. El código que hemos utilizado para realizar todos estos experimentos, puede encontrarse en el \href{https://github.com/Angelvj/Alzheimer-disease-classification.git}{repositorio} de github para este proyecto.

\end{tcolorbox}

\section{Nota importante: experimentos no mostrados}

En las secciones \ref{metodo_experimentacion} y \ref{hiperparametros_principales} vimos en qué consistía aproximadamente el proceso que seguiremos para obtener finalmente una red convolucional que se adecúe a nuestro problema, así como algunos hiperparámetros que habrá que optimizar, y cómo es el proceso seguido para optimizarlos.

También hemos visto en la sección \ref{experimentos} las áreas de experimentación que definen el objeto de estudio de este trabajo: efecto de la profundidad, tipos de preprocesado, transfer learning, y uso simultáneo de \acrshort{mri} y \acrshort{pet}.

El problema es que cada experimento, de cada una de las áreas, requiere un ajuste específico de los hiperparámetros y el diseño de una arquitectura, proceso nada fácil.

A pesar de no ser el objetivo de estudio de este trabajo, estos ``experimentos secundarios'' son de \textit{igual} importancia, ya que sin ellos, no obtendríamos el mejor resultado (dentro de lo posible) de cada experimento principal, por lo que el estudio realizado carecería de sentido alguno. 

Queremos destacar que a pesar de que sólo explicaremos en cada momento un pequeño resumen de estos experimentos secundarios, \textit{estos suponen, probablemente, la tarea que ha consumido más tiempo en el desarrollo}.

\subsection{Un ejemplo}

Es posible que haya quedado algo de ambigüedad con la breve explicación anterior, por lo que vamos a explicar la importancia de estos experimentos secundarios con un ejemplo.

Supongamos nuestra tarea de estudiar la profundidad adecuada de las redes convolucionales para nuestro problema (esto es a lo que hemos llamado un ``área de experimentación''), y supongamos que tomamos como punto de partida un modelo con 3 capas convolucionales (esto es, un experimento general de esta área), sobre el que realizamos una optimización de hiperparámetros, regularización, etc... (estos son los experimentos secundarios), y conseguimos finalmente una cierta exactitud, un 60\% por ejemplo.

Como queremos estudiar la profundidad, decidimos añadir capas a este modelo, creando un modelo con 5 capas, otro con 7, y un último con 8 (estos son otros experimentos generales de la misma área: estudiar la profundidad).

Ahora, como ya hemos decidido los hiperparámetros buenos con el primer modelo de 3 capas, entrenamos todos estos modelos usando los mismos hiperparámetros, y encontramos que los modelos de 7 y 8 capas son los que obtienen peor comportamiento, por tanto, concluimos que es mejor un modelo de menor profundidad para nuestro problema. \textbf{Esta conclusión no es válida}.

El motivo de la no validez de esta conclusión, es que con cada cambio que hacemos sobre el modelo inicial (añadir capas en este caso), es muy probable (casi seguro) que los hiperparámetros, así como los métodos de regularización y otras decisiones de arquitectura dejen de ser óptimas.

Por tanto, si realmente queremos estudiar de forma correcta el efecto de la profundidad, tendremos que conseguir que cada profundidad ``dé lo mejor de sí'' (esto lo conseguimos con los experimentos secundarios) y de esa forma sí podremos concluir (al menos con mayor seguridad), qué profundidad es la más adecuada a nuestro problema.

\subsubsection{¿Y qué experimentos mostramos?}

Volviendo al ejemplo anterior, vamos a ver qué parte mostraríamos, y cual no: \begin{itemize}
    \item \textbf{Lo que no mostramos}: el detalle de todos los experimentos secundarios llevados a cabo en cada profundidad para solucionar todos los problemas que vayan surgiendo (overfitting, entrenamiento demasiado lento, underfitting...). Esto incluye la optimización de los hiperparámetros.
    
    \item \textbf{Lo que sí mostramos}: para cada profundidad, un resumen de los hechos relevantes encontrados en estos experimentos, y un razonamiento de los resultados obtenidos. Se dejará como anexo la arquitectura del mejor modelo obtenido para cada profundidad.
    
\end{itemize}

\section{Fase 1: estudio de la profundidad}

\begin{tcolorbox}[
  colback=Red!5!white,
  colframe=Red!75!black,
  title={Cómo actuaremos}, before upper={\parindent15pt}]

En este apartado realizaremos los experimentos necesarios para posteriormente dar una respuesta sobre el efecto de la profundidad de las redes neuronales convolucionales en nuestro problema.\vspace{1em}

Debido a las grandes diferencias entre las imágenes \acrshort{pet} y \acrshort{mri}, es posible que la profundidad no tenga los mismos efectos en ambas modalidades, por lo que la experimentación se realizará por separado para cada una de ellas.\vspace{1em}

\noindent El modo de actuar será el siguiente:\begin{enumerate}[label=\textbf{\arabic*})]
    \itemsep0em
    \item Comenzar con una red muy poco profunda.
    \item Ajustar sus hiperparámetros (experimentos no mostrados).
    \item Resumir los problemas encontrados, las soluciones tomadas, y argumentar si aumentar la profundidad podría ayudar en base a los resultados obtenidos.
    \item Crear una red más profunda y volver al paso 2.
    \item Repetir los pasos del 2 al 4 mientras consigamos una mejora.
\end{enumerate}

\end{tcolorbox}

\subsection{PET}

\textbf{Nota:} para todos los experimentos de esta fase, utilizamos normalización espacial de las imágenes, y normalización al 1\% de los vóxeles con mayor intensidad.

\subsubsection{Experimento 1: una capa convolucional}

Para este primer experimento, hemos creado una arquitectura extremadamente simple, con una única capa convolucional y una única capa densa.

La verdad es que no creíamos que este modelo fuera obtener grandes resultados, pero esperábamos que sirviera como un modelo de partida que consiguiera un comportamiento que supere al de una base aleatoria (33\% de exactitud en un problema de tres clases), demostrando de esta forma que existe información en los datos que nos ``habla'' de cómo clasificar las imágenes.

 Además, puede ser útil para detectar si hemos cometido algún error en la carga o en el  preprocesamiento de los datos, ya que si por ejemplo, hubiésemos visto que no consigue aprender absolutamente nada, es probable que hubiera algún error al cargar las etiquetas de las imágenes, o algún fallo grave en el preprocesado.

A continuación iremos resumiendo algunos de los problemas más relevantes que surgen en la creación de este primer modelo de una sola capa, así como las decisiones que hemos tomado.

\paragraph{Problema principal: número de parámetros y overfitting}

Como vimos en la sección \ref{numero_capas}, para aproximar una determinada función, una red poco profunda necesitará en general un número mayor de parámetros que los que necesitaría una red más profunda. Pero además vimos en la sección \ref{capacidad}, que una red con un número muy elevado de parámetros será muy buena ``memorizando'' los datos, es decir, sufrirá de overfitting.

\begin{figure}[H]
    \centering
    \subfloat[Exactitud]{%
        \includegraphics[width=0.5\textwidth]{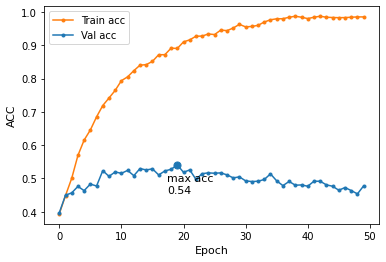}%
        }%
    \hfill%
    \subfloat[Función de pérdida]{%
        \includegraphics[width=0.5\textwidth]{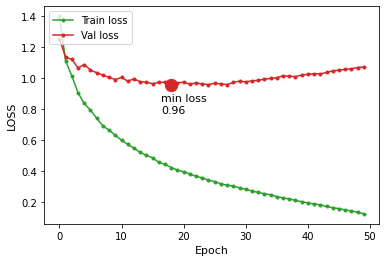}%
        }%
    \caption[Modelo inicial, ejemplo de sobreajuste]{Esta es la curva de aprendizaje de un primer modelo con un \textit{número excesivo de parámetros y que no ha sido regularizado}. Vemos como a partir de la época 20 aproximadamente, la red comienza a ``memorizar'' de forma exagerada los datos de entrenamiento (líneas naranja y verde), y el error cometido en el conjunto de validación comienza a aumentar (rojo), empeorando en consecuencia la exactitud en validación (azul).}
    \label{fig:model_0_overfitting}
\end{figure}

Este conocimiento teórico se confirma en este primer experimento (en realidad se confirmará en los siguientes experimentos), y es que para conseguir que esta primera red consiga aprender, al menos, los datos de entrenamiento (requisito necesario para que pueda comportarse bien en validación), el número de parámetros entrenables necesarios es enorme, y esto efectivamente, hace que tengamos un problema de overfitting (figura \ref{fig:model_0_overfitting}).  

\subparagraph{¿Qué solución tomamos?}

Como ya sabemos, ante un problema de overfitting debemos tratar de regularizar, o bien reducir la complejidad del modelo, y es lo que hemos intentado haciendo numerosas pruebas, de las que queremos destacar ciertas ocurrencias:\begin{itemize}
    \item A pesar de que las capas dropout y la regularización L2 \textit{suelen} aportar mejoras (especialmente dropout), no ha sido así en este caso. 
    
    Concretamente, lo que ocurre es que a pesar de que evita que el error en validación aumente más y más a lo largo de las épocas (en cierto modo regulariza), hace que el error cometido en validación no llegue a ningún mínimo aceptable (es decir, no baja del 0.96 de la figura \ref{fig:model_0_overfitting}, ni la exactitud supera en ningún momento el 54\%). Por otro lado, hace que el error y exactitud en entrenamiento empeoren, aunque eso sí era de esperar, ya que la potencia del modelo se ve reducida. Hablando formalmente, diríamos que añadir estos tipos de regularización está \textit{aumentando el sesgo, pero no disminuyendo la varianza}.
    
    Lo cierto es que no hemos conseguido saber formalmente por qué está ocurriendo esto, pero \textbf{de forma intuitiva}, \textit{creemos} que lo que está ocurriendo es lo siguiente: aunque este primer modelo tiene capacidad de sobra para memorizar los datos de entrenamiento (porque son muy pocos y los modelos poco profundos son ``buenos memorizando''), debido quizás a la escasa profundidad, no tiene una capacidad expresiva suficiente para aproximar bien a nuestra función desconocida $f$, aunque la pueda aproximar ``levemente''. En el momento que introducimos estos métodos de regularización, la capacidad del modelo se ve reducida, lo que hace que éste deje de memorizar completamente los datos de entrenamiento (y eso en principio es bueno), pero también pierde su (escasa) capacidad para aproximar a $f$, de ahí el aumento del error en validación.
    
    Nuestra intuición, nos dice que aunque ahora estas técnicas no hayan ayudado en este primer modelo, una vez que consigamos un modelo con capacidad \textit{de sobra} para aproximar a $f$, entonces sí que funcionarán, ya que harán que nuestro modelo no se ajuste tanto a las peculiaridades de los datos de entrenamiento (ruido), pero no quitarán suficiente capacidad al modelo como para que sea incapaz de aproximar a nuestra función desconocida.
    
    \item Reducir levemente el número de parámetros (disminuyendo el número de unidades en la capa densa), de forma que en lugar de conseguir una exactitud cercana al 100\% en los datos de entrenamiento, consigamos algo cercano al 90\%, consigue reducir \textit{levemente} el overfitting, mejorando los resultados en validación.
    
    \item Un entrenamiento relativamente corto, y una tasa de aprendizaje pequeña con decaimiento exponencial\footnote{El decaimiento exponencial (exponential decay), consiste en ir disminuyendo progresivamente el \acrshort{lr} a medida que avanza el entrenamiento. Existen varias formas de disminuirlo, pero nosotros hemos usado la siguiente: $lr = lr_{inicial} \cdot tasa ^ {epoca/total\_epocas}$, donde tanto $lr_{inicial}$ como $tasa$, son variables libres que tendremos que elegir.}, son las decisiones que más han ayudado. Respecto al entrenamiento corto, esto en cierto modo es una forma de aplicar la técnica de early stopping, es decir, estamos acortando el entrenamiento para evitar que la red comience a aprender de memoria (en nuestro caso concreto, hemos detenido el entrenamiento en 50 épocas, porque a partir de ahí, el comportamiento de nuestro modelo comenzaba a empeorar). En cuanto a la tasa de aprendizaje pequeña con decaimiento, la intuición es que disminuir el tamaño de esta tasa de aprendizaje hará que al principio la red aprenda relativamente rápido, pero poco a poco el entrenamiento se enlentezca, de forma que en los últimos pasos sólo se ``refine'' el valor de los pesos.
    
    \textbf{Nota:} usar una tasa de aprendizaje pequeña con decaimiento exponencial no es ningún tipo de regularización, sólo lo nombramos porque son decisiones que han aportado buenos resultados.
    
\end{itemize}

\paragraph{Resultados obtenidos}

\begin{figure}[H]
    \centering
    \subfloat[Exactitud]{%
        \includegraphics[width=0.5\textwidth]{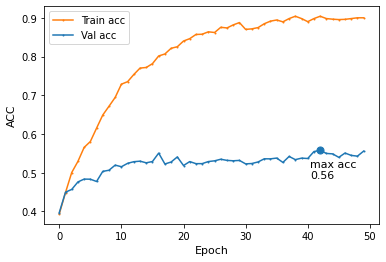}%
        }%
    \hfill%
    \subfloat[Función de pérdida]{%
        \includegraphics[width=0.5\textwidth]{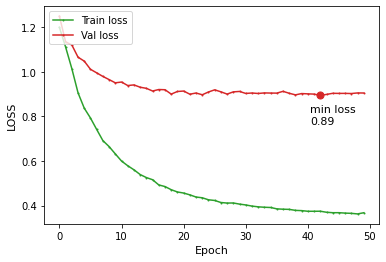}%
        }%
    \caption[Experimento 0. Modelo regularizado]{ 
     Vemos como la distancia entre las curvas de entrenamiento y validación se ha visto reducida, aunque está claro que seguimos teniendo un serio problema de sobreajuste. Notar que estas gráficas están calculadas como el promedio de 50 entrenamientos (repeated k-fold con 5 repeticiones y $k=10$).}
    \label{fig:model_0_pet}
\end{figure}

Tras regularizar y reducir la complejidad, este primer modelo base obtiene una exactitud del \textbf{55,57\%} en repeated k-fold. Su curva de aprendizaje se muestra en la figura \ref{fig:model_0_pet}, en la que se puede observar un menor efecto del sobreajuste que en la figura \ref{fig:model_0_overfitting}.

Para consultar la arquitectura e hiperparámetros concretos (que permiten reproducir este resultado), ver anexo \ref{prof_pet_exp1}. 

\paragraph{Conclusiones}

Antes de responder a la pregunta de si debemos aumentar la profundidad, vamos a ver algunas conclusiones que podemos extraer a partir de lo visto con este primer modelo y de los conocimientos teóricos que presentamos en la segunda parte de este trabajo:\begin{itemize}
    \itemsep0em
    \item El número de parámetros necesario para conseguir un ajuste suficiente en el conjunto de entrenamiento es \textit{excesivamente alto}, lo que hace que este modelo sea tendiente al sobreajuste, además de ocupar mucho espacio en memoria (unos 2GB), y nos obliga a ejecutar en TPU (consumiendo el limitado tiempo de uso).
    \item A pesar del elevado número de parámetros, es probable que debido a la escasa profundidad, nuestro modelo no tenga la representatividad suficiente para aproximar a la función desconocida que buscamos.
    \item A pesar de la regularización y la disminución de la complejidad del primer modelo inicial, sigue habiendo un \textbf{hueco muy amplio} entre la exactitud en entrenamiento y en validación, lo que nos puede estar indicando que no hemos conseguido regularizar lo suficiente, o que nuestro pequeño conjunto de datos no posee la información suficiente para aproximar bien a $f$ \cite{brownlee_2021_curve}, o que simplemente no hemos podido extraer esta información.
\end{itemize}

\paragraph{¿Tiene sentido ir más profundo?}

En nuestra opinión sí que tiene sentido, o al menos, debemos intentarlo. Los motivos para probar más profundidad ya se vieron a lo largo de los capítulos \ref{redes_neuronales} y \ref{capitulo_cnn},  pero vamos a recapitular brevemente algunos de ellos:\begin{itemize}
    \itemsep0em
    \item En la sección \ref{numero_capas} se vio que una red neuronal con más capas puede aprender funciones complejas con menos parámetros, y además, Ian Goodfellow demostró experimentalmente que las redes con más capas pueden ser más resistentes al sobreajuste (si son regularizadas adecuadamente). Nos interesa, ya que está claro que tenemos problemas de overfitting.
    \item En la sección \ref{tendencia_profundidad}, vimos que la tendencia general es hacer las redes más profundas, y que esta tendencia a la profundidad se apoyaba en cierto modo en la intuición de que las redes más profundas pueden extraer características más ricas capa a capa. Quizás estas características más complejas nos ayudan a ``exprimir'' la información subyacente en las imágenes de nuestro pequeño conjunto de datos, por lo que este argumento también apoya a nuestra decisión de aumentar la profundidad.
\end{itemize}

\subsubsection{Experimento 2: dos capas convolucionales}

Ya hemos visto las razones por las que parecía adecuado aumentar la profundidad de nuestro primer modelo. En este segundo experimento, hemos añadido una segunda capa convolucional, además de una capa de \textit{Max Pooling}, lo que nos ha permitido reducir el número de parámetros entrenables a unos \textbf{35 millones} (el modelo anterior tenía 400 millones). 

A pesar de una reducción en más de 10 veces del número de parámetros entrenables (debido principalmente a la capa de Max Pooling), este nuevo modelo consigue una exactitud (en repeated k-fold) del \textbf{57,56\%}.

Respecto a los experimentos llevados a cabo para conseguir este segundo modelo con dos capas, un detalle que nos ha llamado la atención (simplemente porque es una técnica que suele funcionar bien) es que la técnica de dropout \textit{sigue sin darnos buenos resultados} en este caso. El resto de detalles e hiperparámetros de este modelo, que hemos obtenido mediante estos experimentos, se especifican en el anexo \ref{prof_pet_exp2}.

Esta mejora de los resultados nos hace pensar que efectivamente el aumento de la profundidad ha ayudado.

\paragraph{¿Aún más profundidad?}

Hasta ahora, el aumento de la profundidad nos ha dado dos beneficios:\begin{enumerate}
\itemsep0em
    \item \textbf{Drástica reducción} del número de parámetros, lo que hace que nuestro modelo ocupe mucha menos memoria (se puede ejecutar en la GPU sin problema).
    \item \textbf{Mejora del resultado} (en exactitud y en la función de pérdida).
\end{enumerate}
Es muy difícil, incluso podríamos aventurarnos a decir que por el momento es casi imposible saber el motivo concreto por el que esta profundidad ha ayudado (quizá sea por alguno de los motivos que recordamos anteriormente, por una combinación de ellos, o por alguno que desconocemos).

A pesar de esto, es indudable que ha habido una mejora, y nada nos indica que no debamos seguir aumentando la profundidad de la red. 

\subsubsection{Experimento 3: tres capas convolucionales}

Nuevamente, al añadir una nueva capa convolucional obtenemos una mejora sobre el modelo anterior, concretamente obtenemos una exactitud del \textbf{58,12\%} (en repeated k-fold).

En los experimentos realizados para obtener este tercer modelo, no hemos encontrado detalles relevantes a destacar y que no se hayan explicado en los dos experimentos anteriores. 

De nuevo, los detalles de este modelo que serían necesarios para reproducir los resultados, se encuentran en el anexo \ref{prof_pet_exp3}.

\subsubsection{Experimento 4: seis capas convolucionales}

Dada la tendencia a mejorar, decidimos seguir añadiendo capas convolucionales una a una hasta llegar a seis capas convolucionales (obviamos estos experimentos intermedios por no aportar ninguna información relevante). 
\paragraph{Nuevos problemas.}
Al ir aumentando la profundidad, nos encontramos con los siguientes problemas: \begin{itemize}
\itemsep0em
    \item El entrenamiento (problema de minimización) de los modelos se complica \cite{pascanu2013difficulty} \cite[Capítulo~5]{nielsen2015neural}, y se hace necesario un número de épocas mucho mayor, hasta 150. Sin embargo no podemos asumir tanto tiempo de cómputo en un entrenamiento, por lo que debemos buscar otra solución.
    
    En nuestro caso, hemos decidido añadir capas de Batch Normalization (para ver la ubicación exacta de estas capas, ver anexo \ref{prof_pet_exp4}) que efectivamente han corregido el problema, acelerando enormemente el entrenamiento. 
    \item La capacidad del modelo aumenta, aumentando la tendencia al overfitting. En este caso, el uso de capas dropout \textit{ha aportado buenos resultados}, y junto a las medidas tomadas en el experimento 1 (entrenamiento corto y decaimiento del \acrshort{lr}), ha sido suficiente para evitar (dentro de lo posible) este problema.
\end{itemize}
Este modelo con seis capas convolucionales logra una exactitud del \textbf{60,86\%}.

\subsubsection{Experimento 5: ocho capas convolucionales}
\label{prof_exp_5}

\begin{figure}[H]
    \centering
    \includegraphics[width=7cm]{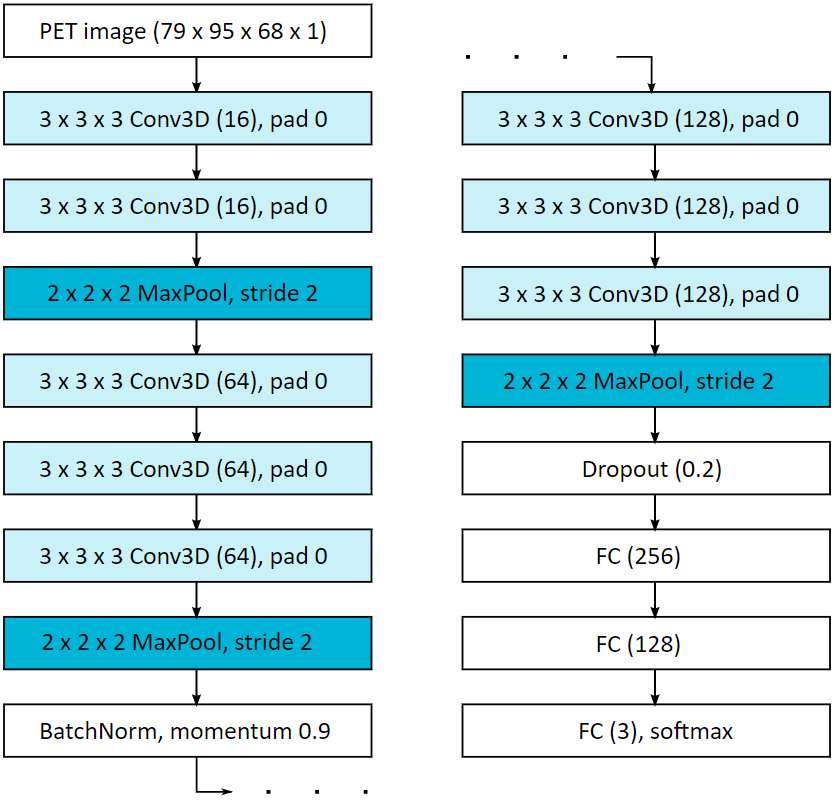}
    \caption[PET. Ocho capas convolucionales.]{Modelo con ocho capas convolucionales. Entre paréntesis, se indica el número de filtros utilizados en cada capa. Aunque no se muestre en la figura, además, en cada capa convolucional se ha usado una regularización L2, con $\lambda = 1\mathrm{e}{-5}$.}
    \label{fig:modelo4pet}
\end{figure}

Con un último modelo de ocho capas convolucionales hemos conseguido una exactitud del \textbf{63,42\%}. La arquitectura de este último modelo se muestra en la figura \ref{fig:modelo4pet}, y el resto de hiperparámetros utilizados son los siguientes: \begin{itemize}
    \itemsep0em 
    \item \textbf{Learning rate}: decaimiento exponencial. Valor inicial $1\mathrm{e}{-5}$, con tasa de decaimiento de $0.1$.
    \item \textbf{Optimizador}: Adam. 
    \item \textbf{Épocas}: 50.
    \item \textbf{Tamaño de Batch}: 4.
    \item \textbf{Activación}: ReLU.
    \item \textbf{Inicializador de pesos}: Glorot uniform \cite{glorot2010understanding}.
\end{itemize}

Un detalle a destacar es que este modelo tiene un número de parámetros entrenables inferior a \textbf{4 millones}, frente a los 400 millones del primer de los modelos. Esto lo hace mucho más ligero desde el punto de vista de almacenamiento, aunque más lento de entrenar debido a la profundidad.

\paragraph{Análisis detallado}

\begin{figure}[H]
    \centering
    \subfloat[Exactitud]{%
        \includegraphics[width=0.5\textwidth]{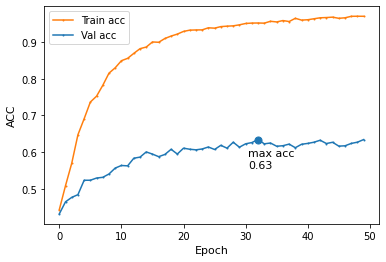}%
        }%
    \hfill%
    \subfloat[Función de pérdida]{%
        \includegraphics[width=0.5\textwidth]{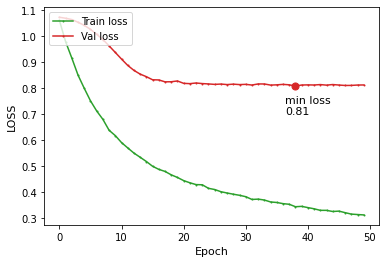}%
        }%
    \caption[Experimento 5. Curva de aprendizaje]{Curva de aprendizaje del experimento 5.}
    \label{fig:exp_5_prof_curva}
\end{figure}

Por ser el \textbf{mejor modelo} que hemos conseguido dentro de este primer bloque de experimentos, vamos a analizar en más detalles el proceso de aprendizaje y sus resultados, de los que iremos sacando algunas conclusiones. En primer lugar, vamos a ver su curva de aprendizaje (figura \ref{fig:exp_5_prof_curva}).

De esta curva de aprendizaje de la figura podemos extraer alguna información que podría ser de utilidad:\begin{itemize}
    \item[$-$] La red es capaz de aprender rápidamente los datos del conjunto de entrenamiento, como vemos en las líneas naranja y verde. Esto no nos da una información excesivamente útil, pero nos dice que el modelo tiene al menos capacidad para memorizar el conjunto de entrenamiento.
    
    \item[$-$] Sigue existiendo un \textbf{gran hueco} entre el error de training y el de validación, aunque siendo notablemente menor que en nuestro primer modelo (figura \ref{fig:model_0_pet}). Esto nos vuelve a decir que o bien no estamos siendo capaces de regularizar suficiente, o bien no existe suficiente información en los datos para conseguir una buena generalización, o por el contrario nuestro modelo no es capaz de extraer toda la información útil que existe.
    
    \item[$-$] La disminución del valor de la función de pérdida lleva consigo el aumento de la exactitud, lo que nos indica que la función de pérdida elegida (entropía cruzada) es adecuada. La verdad es que suponíamos que sería adecuada, pero nunca está de más verificarlo.
\end{itemize}

\renewenvironment{formal}{
\def\FrameCommand{{\color{Red}\vrule width 2pt}\hspace{2pt}}
\MakeFramed{\advance\hsize-\width}
\vspace{2pt}\noindent\hspace{-7pt}\vspace{3pt}
}{\vspace{3pt}\endMakeFramed}
\begin{formal}
\label{posible_duda}
{\bf Nota}

Es posible que al ver las distintas curvas de aprendizaje (figura \ref{fig:exp_5_prof_curva}, por ejemplo) surja la siguiente duda: si sigue habiendo sobreajuste y el modelo memoriza exageradamente los datos de entrenamiento, ¿por qué no reducir su capacidad más, o regularizar de una forma más ``exagerada''?.

La respuesta es que si reducimos más la capacidad del modelo, es cierto que no memorizará por completo los datos de entrenamiento, pero esta reducción de capacidad también deteriora el comportamiento en validación, por lo que hasta ahora, no nos queda otra opción que admitir este sobreajuste (ya que es lo que mejor resultado nos ha dado).

\end{formal}

\renewenvironment{formal}{
\def\FrameCommand{{\color{YellowOrange}\vrule width 2pt}\hspace{2pt}}
\MakeFramed{\advance\hsize-\width}
\vspace{2pt}\noindent\hspace{-7pt}\vspace{3pt}
}{\vspace{3pt}\endMakeFramed}

Además de la curva de aprendizaje, vamos a dar una estimación no sesgada del error que cometería nuestro modelo ante nuevos datos, usando para ello nuestro conjunto reservado para test, que \textbf{no ha sido utilizado hasta ahora}. Y queremos recalcar nuevamente, que este resultado \textbf{no} se utilizará para tomar ninguna decisión (ya que estaríamos sesgando esta estimación).

Dicho esto, nuestro modelo consigue en el conjunto de test (con 50 ejemplos) una exactitud del \textbf{62\%}, valor bastante próximo al 63.42\% que habíamos estimado mediante repeated k-fold. Para ver más en detalle la clasificación, vamos a hacer uso de la matriz de confusión:

\renewcommand\ColCell[1]{
  \pgfmathparse{#1<50?1:0}  
    \ifnum\pgfmathresult=0\relax\color{white}\fi
    \pgfmathsetmacro\compA{#1/22}   
    \pgfmathsetmacro\compB{1-#1/22} 
    \pgfmathsetmacro\compC{1}        
  \edef\x{\noexpand\centering\noexpand\cellcolor[\colorModel]{\compA,\compB,\compC}}\x #1
  } 

\begin{center}
\begin{tabular}{cc*{\items}{|E}|}
\multicolumn{1}{c}{} &\multicolumn{1}{c}{} &\multicolumn{\items}{c}{Predicho} \\ \hhline{~*\items{|-}|}
\multicolumn{1}{c}{} & 
\multicolumn{1}{c}{} & 
\multicolumn{1}{c}{\rot{CN}} & 
\multicolumn{1}{c}{\rot{AD}} & 
\multicolumn{1}{c}{\rot{MCI}} \\ \hhline{~*\items{|-}|}
\multirow{\items}{*}{\rotatebox{90}{Real}} 
& CN  & 9   & 0  & 5   \\ \hhline{~*\items{|-}|}
& AD  & 0   & 8  & 6   \\ \hhline{~*\items{|-}|}
& MCI & 4   & 4   & 14   \\ \hhline{~*\items{|-}|}
\end{tabular}
\end{center}

En la matriz de confusión podemos ver un detalle muy llamativo, y es que \textbf{no existen confusiones entre la clase \acrshort{ad} y \acrshort{cn}}. Es decir, en ningún caso nuestra red ``ha dicho'' a un paciente que padezca la enfermedad de Alzheimer que sea cognitivamente normal, ni viceversa, y por tanto, \textbf{las confusiones vienen de la clase \acrshort{mci}}: existen pacientes de las clases \acrshort{ad} y \acrshort{cn} que han sido clasificados como \acrshort{mci}, y también algunos de \acrshort{mci} que han sido clasificados como \acrshort{ad} y \acrshort{cn}. Este resultado \textbf{apoya el hecho de que existe ambigüedad} en los ``límites'' de la clase \acrshort{mci} (sección \ref{mci_ambiguedad}).

Por último, algo que dijimos es que podría ser interesante saber cómo de bueno es nuestro modelo clasificando la enfermedad de Alzheimer frente al resto de clases (si bien es cierto que en nuestro enfoque se busca una correcta clasificación general, sin dar prioridad a ninguna clase). Para ello vamos a ver la sensibilidad y especificidad para la clase \acrshort{ad}.

La \textbf{sensibilidad} para \acrshort{ad} es de un \textbf{57.14\%}, lo que nos está diciendo que nuestro modelo produce una cantidad considerable de falsos negativos, o visto de otro modo, nos está diciendo que el 57\% de los pacientes con \acrshort{ad} han sido diagnosticados correctamente (al resto se les ha diagnosticado con \acrshort{mci}). Por otro lado, tenemos una \textbf{especificidad} del \textbf{88.88\%}, lo que nos dice que nuestro clasificador tiene muy pocos falsos positivos, o que diagnostica muy bien a aquellos que no presentan la enfermedad.

De estas métricas, podríamos obtener también la idea (intuitiva) de que hemos conseguido un clasificador que clasifica bien a aquellos pacientes que presentan de forma clara la enfermedad de Alzheimer, pero clasifica como \acrshort{mci} a todos aquellos que ``no tiene claro''.

\paragraph{¿Más profundidad?}
A partir de aquí, el resto de arquitecturas de más profundidad que hemos intentado diseñar para superar a este modelo, no han tenido éxito. Concretamente, hemos intentado crear arquitecturas de 9, 10 y 11 capas, pero todas ellas han dado peores resultados que la de 8 capas.

Formulamos las dos siguientes hipótesis posibles:\begin{enumerate}
\itemsep0em
    \item Simplemente, no hemos logrado encontrar una arquitectura de más profundidad que mejore a la anterior, pero esto no nos dice que no exista.
    \item Para este problema concreto, más profundidad no aporta mejoras (no free lunch).
\end{enumerate}
Nosotros apostamos más por la primera hipótesis, ya que encontrar una arquitectura profunda y regularizarla adecuadamente requiere de muchos experimentos y en consecuencia, de mucho tiempo (del cual no disponemos), pero haciendo las suficientes pruebas, creemos que habríamos encontrado algún modelo mejor, aunque sea levemente.

Consideramos que habiendo llegado hasta aquí tenemos suficiente información para posteriormente poder extraer conclusiones útiles sobre el efecto de la profundidad. Por otro lado, creemos que llegados a este punto de estancamiento, existen otras vías para mejorar los resultados de una forma más eficiente. Estas otras vías son las que estudiaremos en los siguientes experimentos.

\subsection{MRI}

En esta sección vamos a proceder de la misma forma que con las imágenes \acrshort{pet}, es decir, partiremos de una arquitectura muy poco profunda e iremos añadiendo capas mientras seamos capaces de obtener una mejora. 

Dado que muchos de los razonamientos que hicimos a lo largo de la experimentación con imágenes \acrshort{pet} son aplicables a la experimentación con \acrshort{mri}, para evitar repetir en exceso, explicaremos ciertos detalles en menor profundidad.

\begin{formal}
{\bf Nota sobre el preprocesado}

Como sabemos, en el caso de las imágenes \acrshort{mri} hemos decidido utilizar únicamente la materia gris (sección \ref{segmentacion_gris}).

Además de esta segmentación para obtener la materia gris, hemos realizado una \textbf{estandarización} a nivel de imagen, proceso en ocasiones realizado para introducir imágenes a una red convolucional \cite{jia_2021, cross_validated_normalization} \cite[Sección~3.3.3]{sermanet2014deep}. Esta estandarización hace que la media de las intensidades de los píxeles de la imagen sea cero, y la varianza uno, lo que puede beneficiar el proceso de optimización. Se calcula como: \[ \frac{X - \Bar{X}}{\sigma(X)}\]

Siendo $X$ la matriz imagen.

\end{formal}

\begin{formal}
{\bf Nota sobre el tamaño de las imágenes}

Las imágenes \acrshort{mri} de nuestro conjunto de datos tienen unas dimensiones de $121 \times 145 \times 121$, notablemente mayores que las de las imágenes \acrshort{pet} ($79 \times 95 \times 68$).

Aunque estas dimensiones mayores pueden tener ventajas (pueden darnos más información), tenemos un problema, especialmente en los modelos poco profundos, y es que el gran tamaño de las imágenes hace que al final, el número de parámetros sea excesivamente grande, lo que además de poder darno problemas de sobreajuste, hace que la capacidad computacional deba ser aún mayor a la que tenemos, por lo que no nos podemos permitir usar este tamaño de imagen en un principio.

Una solución razonable para tratar este primer problema, es simplemente redimensionar las imágenes a un tamaño más pequeño, y es lo que hemos hecho. Concretamente hemos reducido su tamaño (manteniendo las proporciones todo lo posible) a $75 \times 90 \times 75$. Para realizar esta reducción hemos utilizado una función de la biblioteca \textit{skimage}, que lo que hace aproximadamente es ``emborronar'' la imagen con un filtrado gausiano (para evitar el aliasing), y luego eliminar ciertas filas y columnas.

En las arquitecturas más profundas, sí que utilizaremos las imágenes \acrshort{mri} en su formato original, ya que capa a capa, las dimensiones espaciales de la imagen se irán reduciendo lo suficiente para ser ``manejables''.

\end{formal}

\subsubsection{Experimento 1: una capa convolucional}

Al igual que hicimos con las imágenes \acrshort{pet}, partiremos de un primer experimento usando una red con una única capa convolucional. De nuevo, esta primera red nos servirá como punto de partida, y para verificar que no hemos cometido fallos con los datos.

\paragraph{Problema: overfitting (de nuevo)}

Como era de esperar, tenemos el mismo problema que cuando usamos una arquitectura tan poco profunda con \acrshort{pet}: necesitamos un gran número de parámetros para lograr aprender de una forma razonable los datos de entrenamiento (necesario), y debido a este gran número de parámetros, nuestra red memoriza los datos de entrenamiento, pero sin ``entender'' el problema, por lo que no generaliza bien.

Algo que nos ha llamado la atención, es que con este tipo de imágenes, \textbf{el problema de overfitting se ha acentuado aún más}. No podemos saber exactamente a qué se debe, pero nosotros apostamos por el siguiente motivo: como ya vimos en la sección \ref{mri}, estas imágenes son estructurales, es decir, nos hablan de la estructura del cerebro de cada paciente, y al igual que en el resto del cuerpo, existen diferencias entre las estructuras cerebrales de distintos pacientes. Sin embargo, estas diferencia en las estructuras no nos dicen nada sobre la presencia de la enfermedad de Alzheimer, y son zonas muy determinadas las que nos aportan información útil. Nuestra creencia es que debido a los pocos datos, la red va a aprender muy fácilmente (de memoria) estas diferencias estructurales de los pacientes, que en cierto modo no son más que ruido para nuestro problema,  pero debido a la falta de datos, será muy difícil que la red aprenda a ``fijarse'' en aquellas zonas precisas en las que la enfermedad de Alzheimer causa un daño. En resumen, podríamos decir que las peculiaridades anatómicas de cada paciente están actuando como si tuviéramos mucho \textbf{ruido} en los datos.

\subparagraph{¿Y por qué teníamos menos problemas con \acrshort{pet}?}
 De nuevo, no es una explicación formal, pero nosotros creemos que se debe a que las imágenes \acrshort{pet} son funcionales, es decir, dan información sobre la actividad del cerebro (incluso podríamos decir que a alto nivel, ya que tienen poca resolución), y es probable que la actividad del cerebro sea más similar entre todos los pacientes, por lo que la red no las va a poder memorizar tan fácilmente fijándose en las particularidades de cada paciente (ya que es posible que no haya tantas).
 
 \subparagraph{¿Qué solución damos?}
 
 Si tomamos como referencia los experimentos realizados con \acrshort{pet}, lo más probable es que la mejor idea para conseguir mejores resultados sea aumentar la profundidad de la red. Sin embargo, para ser rigurosos en nuestro estudio de la profundidad necesaria, hemos tratado de regularizar este primer modelo, así como buscar una capacidad ``óptima'' que reduzca el problema de sobreajuste.
 
 Un detalle a destacar es que en esta ocasión, una baja tasa de \textit{dropout nos ha dado ciertas mejoras} en este modelo poco profundo, y nuevamente, utilizar entrenamiento relativamente corto (es como si usáramos early stopping), y reducir levemente el número de parámetros respecto a lo que teníamos en \acrshort{pet}, han sido las decisiones que más han ayudado para combatir el sobreajuste. Como siempre, los detalles exactos sobre la arquitectura e hiperparámetros se dan en el anexo \ref{prof_1_mri}.

\subparagraph{Resultados obtenidos}

\begin{figure}[H]
    \centering
    \subfloat[Exactitud]{%
        \includegraphics[width=0.5\textwidth]{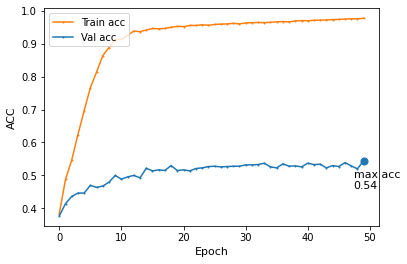}%
        }%
    \hfill%
    \subfloat[Función de pérdida]{%
        \includegraphics[width=0.5\textwidth]{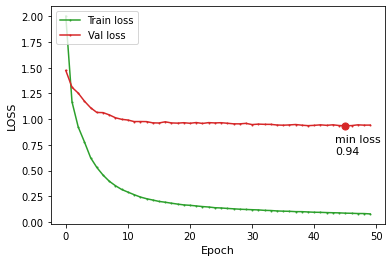}%
        }%
    \caption[MRI. Modelo base]{Si observamos detalladamente la curva, es posible apreciar que en este primer modelo de una sola capa convolucional el problema de sobreajuste es algo más acentuado que el que teníamos en el caso de \acrshort{pet} (figura \ref{fig:model_0_pet}). En apenas 10 épocas memoriza el conjunto de datos de entrenamiento por completo, mientras que en validación no consigue aprender demasiado. De nuevo, notar que no regularizamos de forma más drástica por el motivo que vimos en la sección \ref{posible_duda} (nota en color rojo).}
    \label{fig:model_0_dropout_mri}
\end{figure}

Tras refinar todo lo posible esta arquitectura de una sola capa, nuestro modelo ha logrado una exactitud del \textbf{54.49\%} (ver curva de aprendizaje en la figura \ref{fig:model_0_dropout_mri}), resultado levemente peor a los conseguidos con \acrshort{pet} (55,57\%) con un modelo de igual profundidad.

\subparagraph{¿Más profundidad?}

Los motivos para aumentar la profundidad se mantienen idénticos a los que vimos en \acrshort{pet}, por lo que debemos intentar seguir añadiendo capas.

\subsubsection{Experimento 2: hasta seis capas}

Otra vez, al igual que hicimos en nuestros experimentos con las imágenes \acrshort{pet}, intentaremos crear una serie de arquitecturas en orden creciente de profundidad, siempre y cuando consigamos una mejora en los resultados.

No queremos repetir en exceso los razonamientos que hicimos en el caso de las imágenes \acrshort{pet}, por lo que haremos un resumen breve de los hechos más interesantes encontrados durante estos experimentos:\begin{itemize}
    \itemsep0em
    \item Aumentar la profundidad de las redes ha ayudado, aunque no hemos conseguido resultados tan satisfactorios como con las imágenes \acrshort{pet}.
    \item En todos los casos, añadir dropout en las capas densas nos ha aportado beneficios para combatir levemente el sobreajuste.
    \item A partir de seis capas convolucionales, volvemos a ver que el uso de Batch Normalization nos ayuda en el entrenamiento, aunque no es así en el caso de menos profundidad.
\end{itemize}

En la tabla siguiente mostramos un resumen de algunas de las arquitecturas que nos han aportado mejoras, junto al resultado obtenido en repeated k-fold, y el enlace a la página en la que se especifican la arquitectura e hiperparámetros utilizados:

\arrayrulecolor{black}
\begin{table}[H]
\begin{tabular}{|c|l|l|}
\hline
\# Capas convolucionales & Exactitud & Arquitectura\\ \hline
3   & 55.36\%   & Anexo  \ref{prof_3_mri}      \\ \hline
4   & 55.74\%   & Anexo   \ref{prof_4_mri}     \\ \hline
6   & 57.66\%   & Anexo   \ref{prof_6_mri}     \\ \hline
\end{tabular}
\end{table}

\subsubsection{Experimento 3: nueve capas convolucionales}

Dada la (leve) mejora obtenida aumentando la profundidad, continuamos en búsqueda de una red más profunda que obtuviera mejores resultados, llegando finalmente a una red con 9 capas convolucionales que mejora los resultados anteriores.

En este caso, el uso de un número más elevado de capas nos ha permitido utilizar las imágenes \acrshort{mri} en sus \textbf{dimensiones originales} ($121 \times 145 \times 121$), con lo que quizás podamos recuperar alguna información importante que quizás hayamos perdido al reducir el tamaño de las imágenes.

\begin{figure}[H]
    \centering
    \subfloat[Exactitud]{%
        \includegraphics[width=0.5\textwidth]{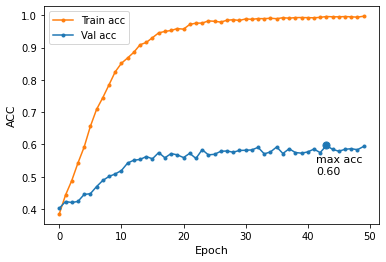}%
        }%
    \hfill%
    \subfloat[Función de pérdida]{%
        \includegraphics[width=0.5\textwidth]{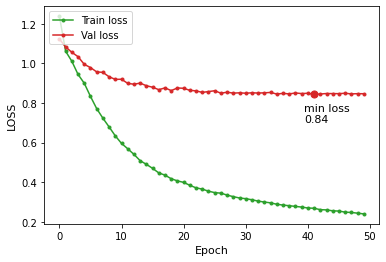}%
        }%
    \caption[Experimento 3. Curva de aprendizaje.]{Al igual que ocurría en \acrshort{pet}, la curva de aprendizaje nos muestra claramente que sigue habiendo un problema de overfitting, aunque es un resultado notablemente mejor al resultado del primer modelo que hemos visto (figura \ref{fig:model_0_dropout_mri}).}
    \label{fig:model_4_mri}
\end{figure}

\begin{figure}[H]
    \centering
    \includegraphics[width=8cm]{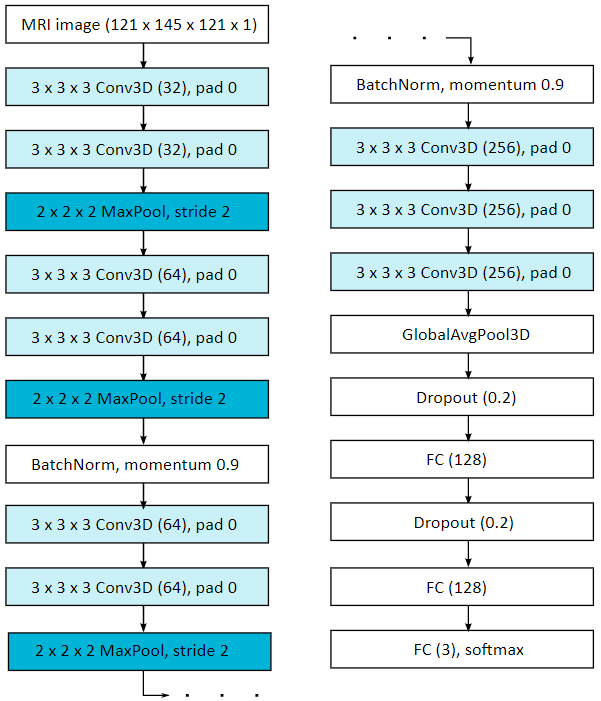}
    \caption[MRI. Nueve capas convolucionales]{Modelo con nueve capas convolucionales. Vemos que con el objetivo de reducir el sobreajuste, en este modelo usamos dos capas dropout, además de una última capa de ``media global'' (Global average pooling), que hace que el número de parámetros se reduzca enormemente.}
    \label{fig:mrimodel4}
\end{figure}

Como se puede apreciar (aproximadamente) en la curva de aprendizaje (figura \ref{fig:model_4_mri}), este modelo consigue una exactitud del \textbf{60.1\%}. Su arquitectura se muestra en la figura \ref{fig:mrimodel4} y otros de los hiperparámetros más importantes son los siguientes (gran parte de ellos coinciden con el modelo de \acrshort{pet}): \begin{itemize}
    \itemsep0em
    \item \textbf{Learning rate}: decaimiento exponencial. Valor inicial $1\mathrm{e}{-6}$, con tasa de decaimiento de $0.1$.
    \item \textbf{Optimizador}: Adam. 
    \item \textbf{Épocas}: 50.
    \item \textbf{Tamaño de Batch}: 4.
    \item \textbf{Activación}: ReLU.
    \item \textbf{Inicializador de pesos}: Glorot uniform.
\end{itemize}

Al igual que hicimos en el caso del último modelo de \acrshort{pet}, vamos a analizar brevemente los resultados que obtiene este modelo en el conjunto de \textbf{test}. 

Respecto a la exactitud, este modelo logra un \textbf{58\%}, habiendo clasificado erróneamente dos pacientes más que el modelo de \acrshort{pet}.

\arrayrulecolor{white}
\begin{center}
\begin{tabular}{cc*{\items}{|E}|}
\multicolumn{1}{c}{} &\multicolumn{1}{c}{} &\multicolumn{\items}{c}{Predicho} \\ \hhline{~*\items{|-}|}
\multicolumn{1}{c}{} & 
\multicolumn{1}{c}{} & 
\multicolumn{1}{c}{\rot{CN}} & 
\multicolumn{1}{c}{\rot{AD}} & 
\multicolumn{1}{c}{\rot{MCI}} \\ \hhline{~*\items{|-}|}
\multirow{\items}{*}{\rotatebox{90}{Real}} 
& CN  & 8   & 1  & 5   \\ \hhline{~*\items{|-}|}
& AD  & 0   & 7  & 7   \\ \hhline{~*\items{|-}|}
& MCI & 4   & 4   & 14   \\ \hhline{~*\items{|-}|}
\end{tabular}
\end{center}

En la matriz de confusión, podemos ver que los errores en la clasificación son casi idénticos a los que cometió el modelo de \acrshort{pet}, aunque en este caso sí que ha habido un fallo más ``preocupante'', y es que existe un paciente que siendo cognitivamente normal, ha sido diagnosticado con la enfermedad de Alzheimer.

Respecto a las métricas de sensibilidad y especificidad para la enfermedad de Alzheimer, tenemos un \textbf{50\%} y \textbf{86.1\%} respectivamente, lo que nos dice que este clasificador produce más falsos negativos, y también más falsos positivos.

En general, aunque no por un margen muy amplio, podríamos decir que este modelo es \textbf{peor} que el modelo que conseguimos obtener con las imágenes \acrshort{pet}, y estos peores resultados se deben, probablemente, a la mayor dificultad para extraer la información de las imágenes \acrshort{mri}, o quizás, a que las imágenes \acrshort{mri} realmente contengan menos información útil para realizar la clasificación.

\subsection{Algunas conclusiones}

\begin{tcolorbox}[
  colback=Green!5!white,
  colframe=Green!75!black,
  title={Algunas conclusiones de esta fase}]
  
De los experimentos de esta fase queremos recalcar algunas conclusiones importantes que hemos extraído: \begin{itemize}
    \itemsep0em
    \item Se verifica que nuestro pequeño conjunto de datos hace que exista mucho overfitting, que convierte a este problema en un reto muy difícil de resolver.
    \item En ambas modalidades, aumentar la profundidad (hasta ocho o nueve capas) ha ayudado, aunque aún no tenemos suficiente información para concluir del todo si todavía más profundidad ayudaría.
    \item Aunque regularizar ayude, no nos va a dar ninguna información extra que no se encuentre en los datos, por lo que es imposible que regularizar nos solucione por completo el problema de tener pocos datos.
    \item Es más difícil extraer información útil de las imágenes \acrshort{mri} que de las imágenes \acrshort{pet}.
\end{itemize}
\end{tcolorbox}

\section{Fase 2: aumento de datos}

En todos los experimentos realizados hasta ahora, la mayoría de nuestros intentos para combatir el sobreajuste se han llevado a cabo con las técnicas de regularización \textit{early stopping, dropout y regularización L2}, y mediante el diseño de arquitecturas que no tengan una capacidad excesiva. Sin embargo, existe una técnica de regularización muy usada en este ámbito, el aumento de datos (sección \ref{data_augmentation}), que no hemos utilizado hasta ahora.

Aunque realmente la técnica de aumento de datos no es más que una técnica de regularización, y como tal podría haber formado parte de los experimentos anteriores, se trata de una técnica computacionalmente costosa, y haber hecho pruebas en todos los modelos anteriores usando esta técnica no habría sido viable.

\begin{tcolorbox}[
  colback=Red!5!white,
  colframe=Red!75!black,
  title={Cómo actuaremos}, before upper={\parindent15pt}]

En esta fase de la experimentación utilizaremos la técnica de aumento de datos únicamente sobre los \textbf{mejores} modelos de la fase anterior. Concretamente actuaremos de la siguiente forma (sobre \acrshort{pet} y \acrshort{mri} por separado):

\begin{enumerate}[label=\textbf{\arabic*})]
    \itemsep0em
    \item Seleccionar un tipo de aumento de datos.
    \item Entrenar el modelo. \begin{itemize}
        \itemsep0em
        \item Si hemos conseguido una mejora notable: \textbf{refinar} los hiperparámetros del modelo.
        \item En caso contrario: volver al paso 1.
    \end{itemize}
\end{enumerate}

Por supuesto, si hemos probado varios tipos de aumento de datos y no conseguimos mejora alguna, simplemente concluiremos que la técnica no parece dar buenos resultados.

\end{tcolorbox}

\subsection{Implementación}
\label{parametros_aumento}

En condiciones normales, hacer uso de esta técnica es muy sencillo (y relativamente eficiente) en Tensorflow, sin más que usar funciones ya implementadas con los parámetros adecuados. El problema es que en el caso de imágenes 3D, \textbf{nada de esto está implementado}, por lo que hemos tenido que implementarlo nosotros.

No aportaría información de interés explicar los detalles finos de esta implementación, pero sí que consideramos interesante describir brevemente qué tipos de aumento de datos hemos implementado, y dar una ligera idea de cómo se ha realizado.

En esencia, lo que hemos hecho es crear una función (en realidad son varias) por la que pasarán los datos de entrada a la red convolucional, y les realizará una serie de modificaciones (aumentos) antes de ser pasados finalmente a la red. Más concretamente, hemos implementado los siguientes tipos de aumento de datos:\begin{itemize}
    \itemsep0em
    \item \textbf{Rotaciones aleatorias}: consiste en realizar una rotación de un ángulo aleatorio entre unos límites superior e inferior, y sobre un eje también aleatorio. Para realizar estas rotaciones aleatorias, hemos podido ayudarnos de la biblioteca \textit{scipy}, que permite rotar matrices tridimensionales sin más que indicarle un ángulo y eje de giro.
    
    \item \textbf{Ampliado aleatorio}: un aumento aleatorio, de nuevo, entre unos límites superior e inferior. En este caso no hemos encontrado funciones ya implementadas, pero basta con transformar la imagen con una matriz de escalado de la forma: \[ \begin{bmatrix}
zoom_x & 0 & 0 & 0 \\
0 & zoom_y & 0 & 0 \\
0 & 0 & zoom_z & 0 \\
0 & 0 & 0 & 1 
\end{bmatrix}\]
Para aplicar la transformación afín a la imagen, se usa la función \textit{affine\_transform} de scipy.

\item \textbf{Volteo aleatorio}: voltear la imagen sobre alguno de sus ejes aleatoriamente. En realidad, podría implementarse mediante una rotación de 180\textdegree, pero lo contamos como un caso distinto, ya que se puede implementar de forma más eficiente sin más que invertir los elementos de una determinada dimensión.

\item \textbf{Desplazamiento aleatorio}: desplazar la imagen sobre algún eje aleatorio con un desplazamiento de píxeles máximo, especificándose este máximo como una determinada proporción de las dimensiones de la imagen. Para ello realizamos la transformación de la imagen con una matriz de la forma: \[ \begin{bmatrix}
1 & 0 & 0 & shift_x \\
0 & 1 & 0 & shift_y \\
0 & 0 & 1 & shift_z \\
0 & 0 & 0 & 1 
\end{bmatrix}\]

\end{itemize}

El problema de esta implementación, es que al no ser propia de Tensorflow, su utilización \textbf{ralentizará de forma muy notable el entrenamiento de los modelos}.

\subsection{PET}

\begin{figure}[H]
    \centering
    \subfloat[Exactitud]{%
        \includegraphics[width=0.5\textwidth]{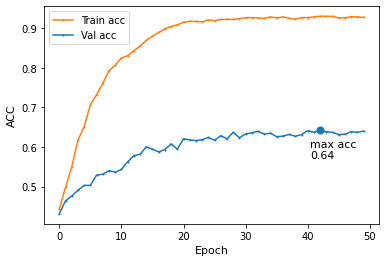}%
        }%
    \hfill%
    \subfloat[Función de pérdida]{%
        \includegraphics[width=0.5\textwidth]{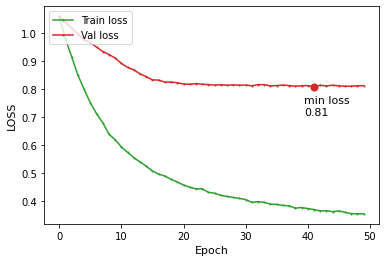}%
        }%
    \caption[PET. Aumento de datos.]{Si comparamos con la curva de aprendizaje del modelo sin aumento de datos (figura \ref{fig:exp_5_prof_curva}), podemos ver que al introducir el aumento de datos, el aprendizaje del conjunto de entrenamiento se hace más lento (no es tan fácil memorizarlo), y conseguimos una mejora aunque sea muy leve en validación.}
    \label{fig:model_4_pet_augmentation}
\end{figure}

En el caso de nuestro modelo de 8 capas para las imágenes \acrshort{pet}, hemos encontrado los siguientes detalles a destacar cuando aplicamos aumento de datos:\begin{itemize}
    \itemsep0em
    \item El volteo aleatorio daña de forma muy notable el rendimiento, lo cual tiene sentido, ya que es como si estuviéramos obligando a la red a aprender la clasificación en situaciones muy distintas.
    \item Todos los tipos de aumento de datos, al hacerse de una forma ``exagerada'', también dañan el rendimiento, probablemente por el mismo motivo anterior.
    \item El uso de la rotación aleatoria con ángulos muy pequeños (0.5 grados como máximo), y desplazamientos aleatorios muy pequeños (2 \% máximo), logra reducir el sobreajuste (ver figura \ref{fig:model_4_pet_augmentation}), consiguiendo una leve mejora de los resultados.
    
    Esta mejora se debe, probablemente, a que estos cambios muy pequeños son capaces de ``simular'' las pequeñas diferencias que existen entre distintas imágenes (a pesar de que hayan sido normalizadas espacialmente), por lo que tienen el efecto de ``aumentar'' el conjunto de datos.
\end{itemize}

Con el uso de aumento de datos y el reajuste de ciertos hiperparámetros (detalles en el anexo \ref{pet_aumento_arquitectura}) hemos conseguido una exactitud del \textbf{64.1\%}, que mejora levemente el resultado que obtuvimos sin el uso del aumento de datos.

En el conjunto de test, se ha dado la ``casualidad'' de que ha obtenido \textbf{exactamente los mismos resultados} (exactamente la misma matriz de confusión) que el modelo que no hacía uso de aumento de datos. Esto nos podría estar diciendo que aunque es posible que este modelo sea algo mejor, la nueva información aprendida no haya ayudado en nuestro conjunto de test concreto.

\subsection{MRI}

Al aplicar la técnica de aumento de datos sobre nuestro mejor modelo de \acrshort{mri}, hacemos las siguientes observaciones: \begin{itemize}
    \itemsep0em
    \item De nuevo, aplicar un aumento de datos muy ``agresivo'' daña el rendimiento de nuestro modelo.
    \item Por contra, usar un aumento de datos suave (se detalla en el anexo \ref{mri_aumento_arquitectura}) aporta notables mejoras (más que en el caso de \acrshort{pet}).
\end{itemize}

En repeated k-fold, hemos conseguido una exactitud del \textbf{63.06\%, casi un 3\% de mejora respecto al modelo que no hace uso de aumento de datos}. El motivo de esta enorme mejora de los resultados, se debe a la gran reducción del sobreajuste respecto al modelo que no hacía uso de aumento de datos, como podemos ver en la figura \ref{fig:model_4_mri_augmentation}:

\begin{figure}[H]
    \centering
    \subfloat[Exactitud]{%
        \includegraphics[width=0.5\textwidth]{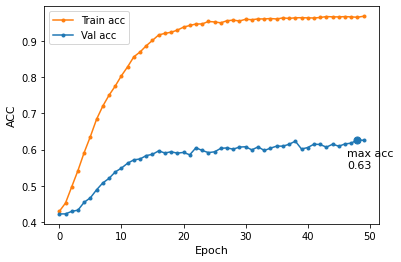}%
        }%
    \hfill%
    \subfloat[Función de pérdida]{%
        \includegraphics[width=0.5\textwidth]{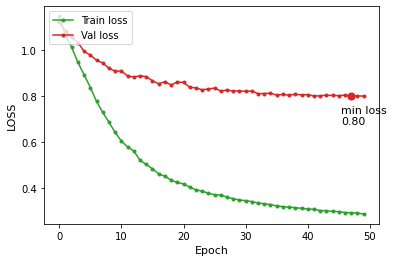}%
        }%
    \caption[Aumento de datos. MRI]{Si observamos atentamente la curva de aprendizaje de la figura \ref{fig:model_4_mri} (antes del aumento de datos), podemos apreciar la reducción del sobreajuste (separación entre las curvas de train y validación).}
    \label{fig:model_4_mri_augmentation}
\end{figure}

Dada la gran mejora que hemos logrado, vamos a ver si los resultados que obtiene este modelo en \textbf{test} mejoran a los del modelo sin aumentos (recordar de nuevo que estos resultados en test no se usan para tomar decisiones sobre los modelos, sólo lo usamos como una comprobación).

\arrayrulecolor{white}
\begin{center}
\begin{tabular}{cc*{\items}{|E}|}
\multicolumn{1}{c}{} &\multicolumn{1}{c}{} &\multicolumn{\items}{c}{Predicho} \\ \hhline{~*\items{|-}|}
\multicolumn{1}{c}{} & 
\multicolumn{1}{c}{} & 
\multicolumn{1}{c}{\rot{CN}} & 
\multicolumn{1}{c}{\rot{AD}} & 
\multicolumn{1}{c}{\rot{MCI}} \\ \hhline{~*\items{|-}|}
\multirow{\items}{*}{\rotatebox{90}{Real}} 
& CN  & 8   & 0  & 6   \\ \hhline{~*\items{|-}|}
& AD  & 0   & 10  & 4   \\ \hhline{~*\items{|-}|}
& MCI & 3   & 7   & 13   \\ \hhline{~*\items{|-}|}
\end{tabular}
\end{center}
\arrayrulecolor{black}

Vemos que este clasificador consigue diagnosticar correctamente a más pacientes con \acrshort{ad} que los anteriores, pero comete más errores clasificando los pacientes con \acrshort{mci}. 

Respecto a la exactitud en test, tenemos un \textbf{60\%}, algo peor de lo que habíamos estimado con 10-fold. Por otro lado, para la clase \acrshort{ad} tenemos unos valores de \textbf{71.43\%} y \textbf{80.56\%} de sensibilidad y especificidad respectivamente. El aumento de la sensibilidad nos dice que nuestra red es capaz de clasificar correctamente más pacientes con \acrshort{ad} que los modelos anteriores, pero a cambio, ha cometido más falsos positivos, de ahí la disminución en la especificidad.

\begin{tcolorbox}[
  colback=Green!5!white,
  colframe=Green!75!black,
  title={Algunas conclusiones de esta fase}]
  
De los experimentos anteriores podemos extraer las siguientes conclusiones:\begin{itemize}
    \itemsep0em
    \item Un aumento de datos \textit{excesivamente agresivo} empeora el comportamiento de nuestros modelos.
    \item Un aumento de datos ``suave'' es capaz de mejorar el comportamiento, a cambio de aumentar notablemente el coste computacional del entrenamiento. Debemos distinguir dos casos:\begin{itemize}
        \itemsep0em
        \item Imágenes \textbf{PET}: la mejora obtenida es muy sutil, y por tanto, sólo sería recomendable usar esta técnica en caso de que no nos importe un entrenamiento más costoso, a cambio de una mejora muy leve.
        \item Imágenes \textbf{MRI}: la mejora obtenida es realmente notable, y creemos que es muy interesante aplicar esta técnica, a pesar del mayor coste computacional.
    \end{itemize}
\end{itemize}

\end{tcolorbox}

\section{Fase 3: imágenes crudas (experimento fallido)}
Como vimos en la sección \ref{extrayendo_caracteristicas}, las \acrshort{cnn} suficientemente profundas eran capaces de extraer características complejas de las imágenes, capa a capa, lo que en teoría, podría ahorrarnos la necesidad de realizar cualquier tipo de preprocesado ``manual''. 

Hasta este momento, hemos conseguido crear arquitecturas de una profundidad notable (8 ó 9 capas convolucionales). Nuestro pensamiento es que estas arquitecturas deberían ser capaces de tomar como entrada las imágenes sin preprocesar, es decir, tal y como se han obtenido del escáner \footnote{En realidad, las imágenes que tenemos no son del todo ``crudas'', sino que presentan un preprocesado más suave que el que nosotros hemos realizado.}, y aprender a hacer todas las transformaciones que necesite para realizar la clasificación de forma correcta.

\begin{tcolorbox}[
  colback=Red!5!white,
  colframe=Red!75!black,
  title={Cómo actuaremos}, before upper={\parindent15pt}]
En esta fase, la idea es utilizar nuestros mejores modelos hasta el momento y entrenarlos, en lugar de usando las imágenes con todo su preprocesado, usando las imágenes crudas (en realidad no del todo crudas). 
\end{tcolorbox}
\renewenvironment{formal}{
\def\FrameCommand{{\color{Red}\vrule width 2pt}\hspace{2pt}}
\MakeFramed{\advance\hsize-\width}
\vspace{2pt}\noindent\hspace{-7pt}\vspace{3pt}
}{\vspace{3pt}\endMakeFramed}

\begin{formal}
\label{posible_duda_2}
{\bf ¿Tiene sentido este experimento?}
Tras haber leído la sección \ref{necesidad_preprocesado}, en la que razonamos la necesidad del preprocesado en este problema concreto, es cierto que este experimento, teniendo tan pocos datos, carece de sentido alguno.

Sin embargo, estos experimentos se llevaron a cabo antes de conseguir llegar a ese razonamiento, por lo que simplemente daremos un resumen de los resultados que se obtuvieron con las imágenes \acrshort{pet} sin preprocesar, que nos servirá además para verificar la corrección del razonamiento.

Con las imágenes \acrshort{mri}, lo cierto es que no lo probamos, ya que fue tras realizar el experimento con \acrshort{pet} cuando llegamos definitivamente a esta conclusión, y hacer las pruebas con \acrshort{mri} no habría sido más que una pérdida de tiempo (entrenar un modelo cuesta más de ocho horas).
\end{formal}

Aunque usáramos imágenes crudas, lo común es realizar algunos tipos de preprocesado muy ``suaves'' antes de introducir las imágenes en una red convolucional con el fin de facilitar el proceso de minimización. Algunos de los prepreprocesados que utilizamos son los siguientes:\begin{itemize}
    \itemsep0em
    \item \textbf{Minmax}: escalar los valores de los vóxeles en el rango $[0,1]$.
    \item \textbf{Estandarizadas}: hacer que la desviación típica de los  valores de todos los vóxeles de una imagen sea uno, y la media cero.
    \item \textbf{Minmax + estandarizadas}: aplicar las dos técnicas anteriores, una detrás de otra.
    \item \textbf{0.1 Máx}: normalización a la intensidad máxima, que se explicó en la sección \ref{normalizacion_intensidad}. 
    \item \textbf{Crudas}: imágenes tal y como las descargamos.
\end{itemize}

A continuación, mostramos una tabla con el resumen de los resultados:
\begin{table}[H]
\begin{tabular}{|l|l|l|l|l|l|}
\hline
Preprocesadas & Minmax  & Estandarizadas & Minmax + estandar. & 0.1 Máx. & Crudas  \\ \hline
63.42\%       & 51.37\% & 52.2\%         & 53.74\%            & 53.33\%  & 51.19\% \\ \hline
\end{tabular}
\caption{Resultados con imágenes crudas.}
\label{resultado_crudas}
\end{table}

Tras ver los resultados, queda claro que es una buena idea preprocesar las imágenes, al menos mientras tengamos tan pocos datos. Por tanto, durante lo que queda de este trabajo seguiremos usando las imágenes preprocesadas.

\section{Fase 4: transfer learning}

\label{transfer_learning}

Hasta este momento, hemos tratado de crear nuestras propias arquitecturas para resolver el problema de clasificación, una tarea difícil y que requiere hacer muchas pruebas, y en consecuencia mucho tiempo.

\textbf{En muchas ocasiones, no es una buena idea crear una arquitectura propia}, y en su lugar podríamos reutilizar arquitecturas diseñadas por prestigiosos investigadores (muy probablemente con más conocimiento y experiencia que nosotros), y que han demostrado excelentes resultados en muchos problemas (ResNet, DenseNet, GoogLeNet, ...).

Estas arquitecturas tienen un ``problema'', y es que en general están diseñadas para ser entrenadas con una enorme cantidad de datos (en muchas ocasiones con millones de ejemplos), una cantidad de datos de la que en general, no dispondremos.

Para solucionar esta carencia de datos, lo que se hace usualmente es partir de estas arquitecturas ya entrenadas en algún conjunto de datos muy grande (Keras nos las da directamente), y reutilizar para nuestro problema el conocimiento común entre ese conjunto de datos y el nuestro (transfer learning).

\subsection{Problemas y soluciones}

Dado que nuestro conjunto de datos es muy pequeño, y sería imposible entrenar correctamente una arquitectura conocida desde cero, parece claro que la técnica de transfer learning nos interesa, pero en este caso, nos surgen los siguientes problemas (algunos no previstos):\begin{itemize}
    \itemsep0em
    \item No hemos encontrado ninguna arquitectura conocida, en versión 3D, y preentrenada con un conjunto de datos de gran tamaño.
    \item Ni siquiera hemos encontrado ninguna implementación ``de confianza'' de una versión 3D de una arquitectura conocida. 
    \item Las arquitecturas más conocidas para clasificación de imágenes están diseñadas para funcionar bien con imágenes bidimensionales. No sabemos si funcionarán bien en el caso de imágenes volumétricas (3D), pero \textbf{confiaremos en que sí}.
\end{itemize}

Ante los problemas encontrados, tomaremos las siguientes soluciones genéricas:\begin{itemize}
    \itemsep0em
    \item Implementaremos nosotros mismos una arquitectura, adaptando alguna implementación fiable de una red 2D a 3D.
    \item Preentrenaremos la red 3D en algún conjunto de datos de mayor tamaño que el nuestro.
\end{itemize} 

\subsection{Formas de aplicar transfer learning}

Para que la técnica de transfer learning nos dé todo su potencial, lo ideal es que el conjunto de datos grande en el que ya ha sido entrenada (preentrenada) sea lo más \textbf{similar} posible a nuestro conjunto de datos \cite{yosinski2014transferable}, ya que de esa forma la información que la red aprendió a extraer en el conjunto de datos grande, será útil para nuestro conjunto de datos de menor tamaño.

Sin embargo, no siempre es posible encontrar una red preentrenada en un conjunto de datos que tenga grandes similitudes con el nuestro. Dependiendo del grado de similitud de este conjunto de datos, y del tamaño de nuestro conjunto, podemos distinguir varias situaciones que requieren distintas formas de aplicar la transferencia de aprendizaje \cite{cs231n_transfer}:\begin{enumerate}
    \itemsep0em
    \item \textit{Nuestro conjunto de datos es pequeño  y similar al conjunto de datos de preentrenamiento}. Como las características aprendidas con los datos de preentrenamiento serán similares a las que nos hacen falta, bastará con entrenar una capa densa que tome como entrada la salida de la última capa de la red preentrenada. No sería una buena idea ajustar los pesos de toda la red debido al overfitting.
    \item \textit{Nuestro conjunto de datos es grande y similar al conjunto de datos de preentrenamiento}. En este caso, como tenemos muchos datos, ajustar los pesos de toda la red sería una buena idea, ya que refinaríamos las características aprendidas sobre el conjunto de preentrenamiento, y en este caso, al tener un conjunto de datos grande, no tendríamos problemas de overfitting (si lo hacemos con cuidado).
    \item \textit{Nuestro conjunto de datos es pequeño y muy diferente al de preentrenamiento}. Como el conjunto de datos es pequeño, es mejor entrenar únicamente una capa densa (para evitar el overfitting), pero como el conjunto de datos es muy diferente, sería mejor no entrenar esta capa densa sobre la salida de la última capa de la red, ya que las características extraídas estarán muy ``pegadas'' al conjunto de datos original. Por tanto, la mejor idea sería entrenar la capa densa, pero sobre la salida de alguna capa intermedia de la red (que nos dé características más genéricas).
    \item \textit{Nuestro conjunto de datos es grande y muy diferente al de preentrenamiento}. Podríamos entrenar la red desde el principio, sin usar una red preentrenada, aunque se ha visto que usar los pesos de una red ya entrenada en otro conjunto de datos puede ser beneficioso (aunque el problema sea muy distinto). Por tanto, en este caso, entrenaríamos la red completa.
\end{enumerate}

\subsection{Conjunto de datos COVID-19}

En nuestro caso, como no hemos encontrado ninguna red 3D preentrenada con un conjunto de datos grande (o al menos, más grande que el nuestro), necesitamos encontrar un conjunto de datos para realizar el preentrenamiento de una red nosotros mismos.

Hemos decidido usar un conjunto de datos de pacientes con COVID-19 (todos los detalles de este conjunto se encuentran en el artículo \cite{covid_dataset}), que consta de 1109 ejemplos. 

No es nuestra intención entrar en muchos detalles de este conjunto de datos, pero conviene conocer al menos los siguientes detalles esenciales:\begin{itemize}
    \itemsep0em
    \item Hay una imagen 3D de los pulmones de cada uno de los 1109 pacientes.
    \item Son imágenes de tipo \acrshort{ct} (Computed Tomography). De nuevo, aunque no entraremos en detalle, estas imágenes, al igual que las \acrshort{mri}, son imágenes de tipo estructural, y aunque tienen parecido, existen diferencias tanto en la forma de tomarlas, como en la información que se puede observar en ellas.
    \item Existen 5 clases diferentes según el grado de afectación de la enfermedad, con las siguientes proporciones: CT-0, 254 (22.8\%); CT-1, 684 (61.6\%); CT-2, 125 (11.3\%); CT-3, 45 (4.1\%); and CT-4, 2 (0.2\%).
\end{itemize}

\subsection{Desarrollo del experimento}

\begin{tcolorbox}[
  colback=Red!5!white,
  colframe=Red!75!black,
  title={Cómo actuaremos}, before upper={\parindent15pt}]
Ahora que ya tenemos toda la información necesaria para comprender este experimento, vamos a ver cuáles son los pasos que seguiremos:\begin{enumerate}[label=\textbf{\arabic*})]
    \itemsep0em
    \item Crear una arquitectura conocida adapatada a 3D.
    \item Preentrenar esta red con los datos de COVID-19.
    \item Tomar una decisión sobre la forma más adecuada de aplicar transfer learning.
    \item Realizar el experimento y obtener los resultados.
\end{enumerate}
\end{tcolorbox}
\renewenvironment{formal}{
\def\FrameCommand{{\color{Red}\vrule width 2pt}\hspace{2pt}}
\MakeFramed{\advance\hsize-\width}
\vspace{2pt}\noindent\hspace{-7pt}\vspace{3pt}
}{\vspace{3pt}\endMakeFramed}

\subsubsection{ResNet-18}
\label{RESNET18}

\begin{figure}[H]
    \centering
    \includegraphics[width=8cm]{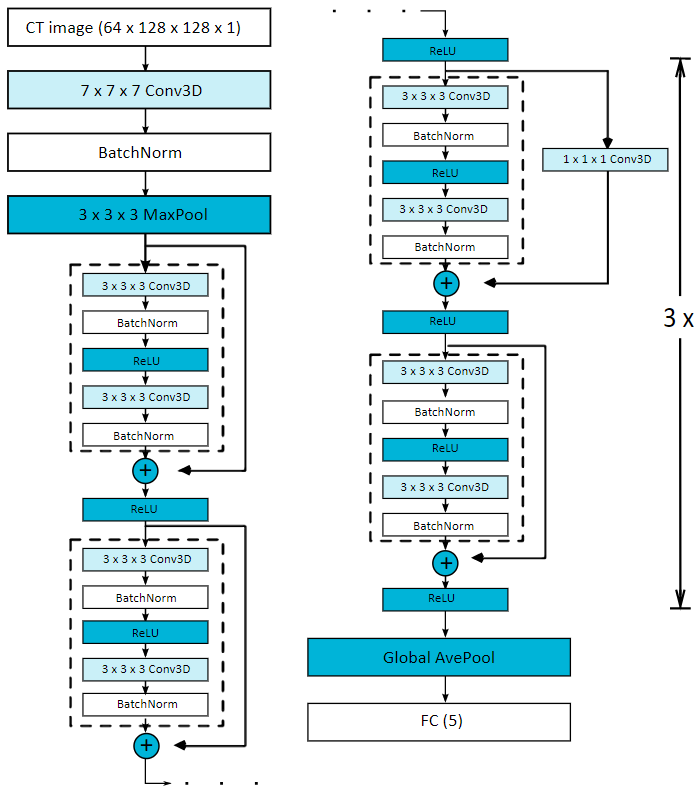}
    \caption[ResNet-18 adapatada a 3D]{ResNet-18 adaptada a 3D. Como vemos, se trata de una red residual común, pero haciendo uso de la versión 3D de las operaciones.}
    \label{fig:resnet18_3d}
\end{figure}

Como red a preentrenar, hemos decidido adaptar una arquitectura ResNet-18 (18 capas) para su aplicación a imágenes tridimensionales. La arquitectura de nuestra red se muestra en la figura \ref{fig:resnet18_3d}.

El motivo de esta elección es principalmente, que la arquitectura ResNet ha dado excelentes resultados en muchos problemas de clasificación (gracias a algunas características que estudiamos en la sección \ref{ResNets}). Por otro lado, el motivo de elegir la de 18 capas y no una aún más profunda (34, 50...), es que nuestro conjunto de preentrenamiento (COVID-19) no es demasiado grande, por lo que no parece viable entrenar una arquitectura tan profunda.

\subsubsection{Preentrenamiento}

Una vez que hemos creado la arquitectura, el siguiente paso es entrenarla en nuestro problema de COVID-19.

Al igual que en todos los experimentos que hemos hecho hasta ahora, este entrenamiento requiere fijar una serie de hiperparámetros, además de algún tipo de preprocesado de las imágenes. Sin embargo, consideramos que estos detalles no son en absoluto necesarios para la explicación del experimento llevado a cabo, y nos basta con saber que esta red ha sido entrenada en el problema de clasificación de COVID-19, logrando una exactitud de aproximadamente el \textbf{85\%}, que demuestra que ha habido aprendizaje (en este caso no se ha aplicado k-fold, ya que no tenemos ningún interés en dar una estimación exacta, además de que el conjunto de datos es más grande).

Para seguir siendo rigurosos, los detalles sobre el preprocesamiento de las imágenes y los hiperparámetros del clasificador se muestran en el anexo \ref{preprocesado_preentrenamiento}.

\subsubsection{Modo de transfer learning}

\begin{figure}[H]
    \centering
    \subfloat[Imagen \acrshort{ct}]{%
        \includegraphics[width=0.33\textwidth]{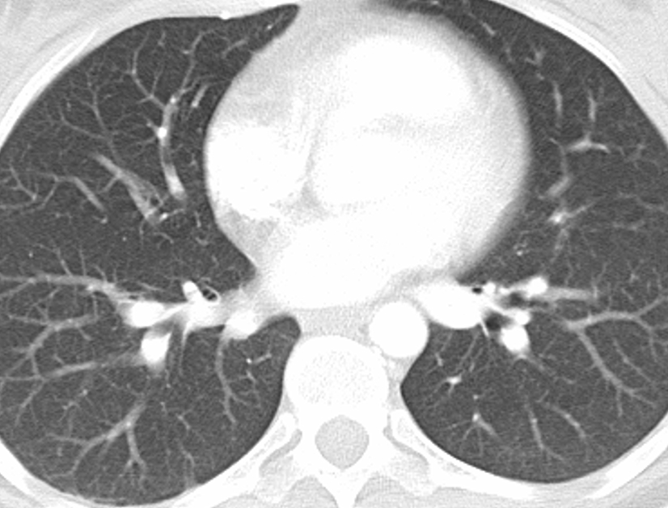}%
        }%
    \subfloat[Imagen \acrshort{pet}]{%
    \includegraphics[width=0.33\textwidth]{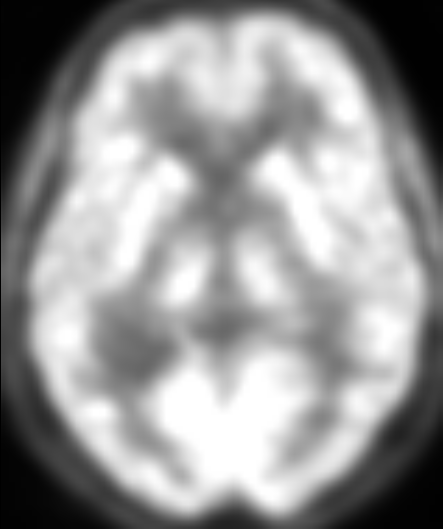}%
    }%
    \subfloat[Imagen \acrshort{mri}]{%
        \includegraphics[width=0.33\textwidth]{imagenes/ejemplo_mri.png}%
        }%
    \hfill%
    \caption{CT vs PET vs MRI}
    \label{fig:ahora_se_pone}
\end{figure}

\paragraph{PET}
\label{RESNET18_PET}
Observando la imagen de la figura \ref{fig:ahora_se_pone}, al menos intuitivamente, está claro que en el caso de \acrshort{pet}, estamos en el caso en el que tenemos \textit{un conjunto de datos pequeño y muy diferente al de preentrenamiento}. Esto nos dice que el enfoque más adecuado, probablemente sea el de utilizar la salida de alguna capa intermedia de ResNet-18 como extractor de características, y añadir una capa totalmente conectada para realizar finalmente la clasificación, y por supuesto, nuestro pequeño conjunto de datos deja totalmente fuera la posibilidad de ajustar todos los pesos de la red (tendríamos un enorme sobreajuste).

Concretamente, para aplicar transfer learning con las imágenes \acrshort{pet}
haremos lo siguiente:\begin{enumerate}[label=\textbf{\arabic*})]
    \itemsep0em
    \item Eliminar del modelo preentrenado (figura \ref{fig:resnet18_3d}) las dos últimas capas (clasificador), además de los dos últimos bloques residuales.
    \item Añadir al final una capa Global Average Pooling, y una capa totalmente conectada con tres unidades.
    \item Congelar (fijar los pesos) y poner en modo inferencia todas las capas de tipo Batch Normalization (para que no actualicen sus estadísticas internas), a excepción de las dos capas que hemos añadido. 
    \item Entrenar el modelo. Como el modelo base fue congelado, sólo se entrenarán las capas que hemos añadido.
\end{enumerate}

\paragraph{MRI} 
\label{RESNET18_MRI}
En este caso, aunque no hay una similitud ``perfecta'', sí que podemos ver que existe un buen parecido entre las imágenes \acrshort{mri} y las \acrshort{ct}: ya que ambas presentan muchos \textbf{detalles finos} de la estructura del cuerpo. Por tanto, consideraremos que estamos en el caso en el que tenemos \textit{un conjunto de datos pequeño y similar al conjunto de datos de preentranamiento}, por lo que utilizaremos un clasificador (capa totalmente conectada) situado a la salida de la última capa convolucional de ResNet-18.

En este caso, para aplicar transfer learning: \begin{enumerate}[label=\textbf{\arabic*})]
    \itemsep0em
    \item Eliminar del modelo preentrenado (figura \ref{fig:resnet18_3d}) únicamente las dos últimas capas.
    \item A partir de aquí, proceso idéntico al que realizamos con \acrshort{pet}.
\end{enumerate}

\subsubsection{Entrenamiento final y resultados}

Antes de pasar a ver los resultados obtenidos, queremos destacar algunas cuestiones sobre el entrenamiento final con nuestro conjunto de datos (el pequeño): \begin{itemize}
    \itemsep0em
    \item Debido a que el preprocesamiento realizado sobre las imágenes \acrshort{ct} hace que nuestra red espere datos acotados entre cero y uno, además del preprocesado que hemos hecho hasta ahora, hemos escalado las imágenes de nuestro conjunto de datos en el mismo rango.
    \item Ya que el aumento de datos dio buenos resultados, lo seguiremos utilizando en este caso.
    \item Gracias al preentrenamiento, el aprendizaje ha sido más rápido, lo que nos ha permitido entrenar durante un número menor de épocas.
\end{itemize}

\paragraph{MRI}

\begin{figure}[H]
    \centering
    \subfloat[Exactitud]{%
        \includegraphics[width=0.5\textwidth]{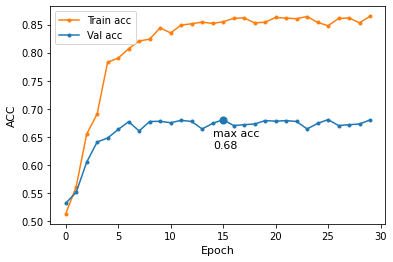}%
        }%
    \hfill%
    \subfloat[Función de pérdida]{%
        \includegraphics[width=0.5\textwidth]{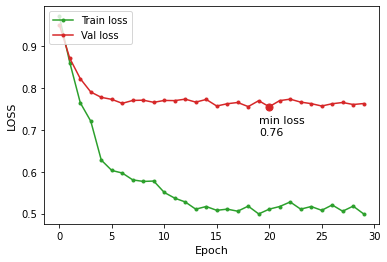}%
        }%
    \caption[Curva de aprendizaje. ResNet-18. MRI]{Curva de aprendizaje. ResNet-18. MRI.}
    \label{fig:resnet18_mri}
\end{figure}

En la figura \ref{fig:resnet18_mri}, podemos ver que ResNet-18 entrenada con \acrshort{mri} \footnote{Nótese que aunque decimos ``entrenada'' con \acrshort{mri}, en realidad esta red ha sido entrenada con COVID-19 y lo único que se ha entrenado con \acrshort{mri} es el clasificador final.} es, con diferencia, el modelo que presenta menos overfitting de todos los que hemos creado hasta ahora. Un primer detalle a destacar es que en este caso el modelo no consigue memorizar completamente el conjunto de entrenamiento (consigue aproximadamente un 85\% de exactitud), y posiblemente esto se deba a la pequeña cantidad de parámetros entrenables que hemos utilizado (menos de dos mil). Otro detalle a destacar es que a pesar de que hemos utilizado repeated k-fold, las gráficas del entrenamiento son algo más \textbf{inestables}. Quizás esto se deba a que hemos usado una tasa de aprendizaje más alta que en todos los modelos anteriores (además se acentúa visualmente porque ahora la escala es más pequeña), pero verdaderamente no lo sabemos.

Respecto al resultado final en repeated k-fold, este modelo consigue una exactitud del \textbf{68.02\%}, que supera ampliamente a la conseguida por el mejor modelo que teníamos con \acrshort{pet}. 

En test, tenemos una exactitud del \textbf{70\%}, y la clasificación detallada se muestra en la matriz de confusión:

\arrayrulecolor{white}
\begin{center}
\begin{tabular}{cc*{\items}{|E}|}
\multicolumn{1}{c}{} &\multicolumn{1}{c}{} &\multicolumn{\items}{c}{Predicho} \\ \hhline{~*\items{|-}|}
\multicolumn{1}{c}{} & 
\multicolumn{1}{c}{} & 
\multicolumn{1}{c}{\rot{CN}} & 
\multicolumn{1}{c}{\rot{AD}} & 
\multicolumn{1}{c}{\rot{MCI}} \\ \hhline{~*\items{|-}|}
\multirow{\items}{*}{\rotatebox{90}{Real}} 
& CN  & 9   & 0  & 5   \\ \hhline{~*\items{|-}|}
& AD  & 0   & 10  & 4   \\ \hhline{~*\items{|-}|}
& MCI & 1   & 5   & 16   \\ \hhline{~*\items{|-}|}
\end{tabular}
\end{center}

Como vemos, sigue sin haber confusiones entre las clases \acrshort{ad} y \acrshort{cn}, y en general el clasificador ha mejorado en la tarea de de clasificar todas las clases.

Respecto a la clase \acrshort{ad}, tenemos una sensibilidad del \textbf{71.42\%} y una especificidad del \textbf{86.11\%}: de los pacientes que no presentan la enfermedad, sólo clasificamos erróneamente a un 14\%, y por otro lado, detectamos el 71.4\% de los pacientes que presentan la enfermedad.

\paragraph{PET}

\begin{figure}[H]
    \centering
    \subfloat[Exactitud]{%
        \includegraphics[width=0.5\textwidth]{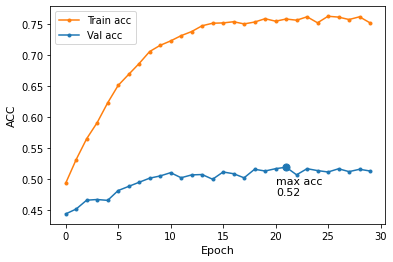}%
        }%
    \hfill%
    \subfloat[Función de pérdida]{%
        \includegraphics[width=0.5\textwidth]{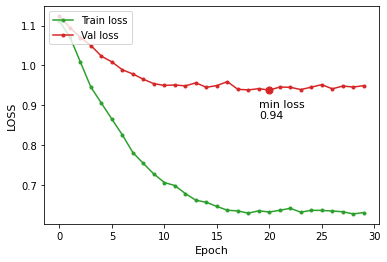}%
        }%
    \caption[Curva de aprendizaje. ResNet-18. PET]{Curva de aprendizaje. ResNet-18. PET.}
    \label{fig:resnet18_pet}
\end{figure}

Respecto al modelo entrenado con \acrshort{pet}, aunque pensábamos que las características generales extraídas por ResNet-18 serían de utilidad para nuestro problema, observando el proceso de aprendizaje (figura \ref{fig:resnet18_pet}) podemos ver que no ha sido así: nuestro modelo no ha sido capaz de aprender apenas información útil, y si nos fijamos, el resultado en validación (cercano a un 50\%) nos puede estar indicando, que prácticamente lo único que ha aprendido esta red es a clasificar a casi todos los pacientes como \acrshort{mci}, ya que es la clase mayoritaria (un 44\% de los ejemplos).

En este caso, no mostraremos los resultados en test, ya que dado el mal comportamiento, no sería de mucho interés.

Probablemente, el motivo de este mal comportamiento es que el conjunto de datos COVID-19 es excesivamente distinto a las imágenes \acrshort{pet} (que son imágenes más difusas). Para tener un mejor comportamiento, lo ideal habría sido que nuestro conjunto de datos fuera algo más grande, ya que de esta forma podríamos haber reajustado los pesos de toda la red a nuestro problema concreto (sin sufrir de mucho sobreajuste).

\section{Fase 5: dos entradas (experimento fallido)}

Hasta este momento, hemos conseguido dos modelos interesantes: \begin{itemize}
    \itemsep0em
    \item \textbf{PET:} un modelo de ocho capas convolucionales, que haciendo uso de aumento de datos, consigue una exactitud del \textbf{64.1\%}.
    \item \textbf{MRI:} una ResNet de 18 capas preentrenada con datos de COVID-19, que también hace uso de aumento de datos, y que consigue una exactitud del \textbf{68.02\%}
\end{itemize}

\begin{formal}
\label{idea_dos_entradas}
{\bf Idea}

Podemos crear una red, que en lugar de tomar como entrada una imagen para cada paciente, tome \textbf{dos entradas}, una imagen \acrshort{pet}, y otra \acrshort{mri}. Es posible que la información de las imágenes de una modalidad complemente la información que se obtiene de la otra, dando lugar a un modelo con aún mejores resultados.

\end{formal}

\subsection{Arquitectura propuesta}

\begin{figure}[H]
    \centering
    \includegraphics[width=7cm]{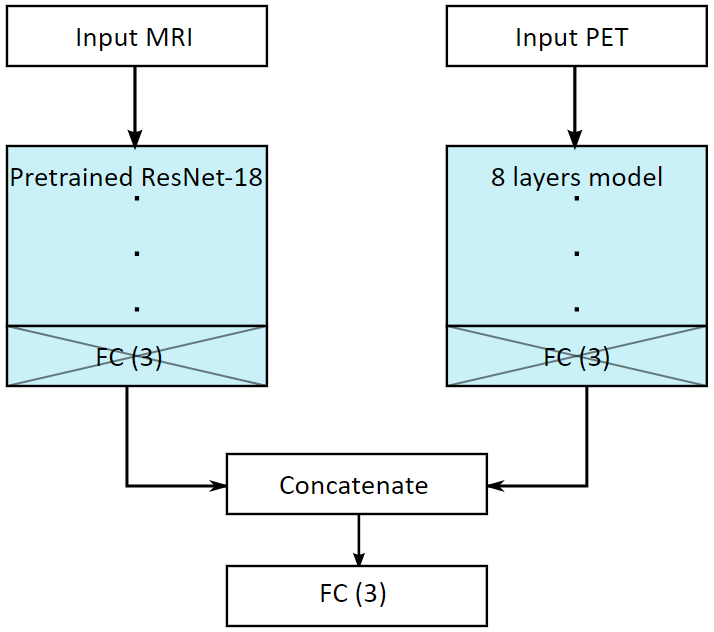}
    \caption[Modelo con dos entradas. Arquitectura]{Arquitectura del modelo con dos entradas. Fusión de los modelos de las figuras \ref{fig:resnet18_3d} y \ref{fig:modelo4pet}. La capa \textit{concatenate} no realiza más que la concatenación de las salidas de las dos arquitecturas, a las que se le ha eliminado el clasificador final.}
    \label{fig:two_input_model}
\end{figure}

Dado que tenemos dos modelos capaces de extraer información útil para cada una de las modalidades de imagen, nuestra arquitectura con dos entradas consistirá en una nueva arquitectura que reúne a nuestros dos mejores modelos (como vemos en la figura \ref{fig:two_input_model}).

\subsection{Cómo entrenarla}

Respecto al entrenamiento de esta red barajamos dos posibilidades para realizar su entrenamiento (de las que sólo hemos podido probar la primera):\begin{enumerate}
    \itemsep0em
    \item \textbf{Entrenar todo el modelo de una vez}. Esta opción consiste en entrenar el modelo de principio a fin, es decir, en cada momento recibirá dos imágenes al mismo tiempo, junto a su etiqueta, y el error cometido se propagará desde la capa totalmente conectada, hasta llegar a ambas capas de entrada.
    \item \textbf{Entrenar cada arquitectura por separado}. Esta opción consistiría en entrenar cada una de las dos ramas por separado (como hemos hecho en todos los modelos hasta ahora), y luego, entrenar una capa totalmente conectada con la concatenación de las salidas de cada arquitectura.
\end{enumerate}

Lo cierto es que no sabemos qué opción es la mejor (y debido a algunos problemas con la disponibilidad del hardware no nos podemos permitir probar ambas). Nosotros creemos que quizás la primera opción es más interesante, ya que al actualizar ambas ramas a la vez, es posible que los pesos de cada rama se aprendan ``teniendo en cuenta'' los pesos de la otra rama. Dicho de otra forma, creemos que siguiendo esta opción, la red aprenderá a obtener la información que más necesite de cada imagen (aunque es muy difícil obtener un razonamiento claro, y es posible que nuestra intuición sea errónea).

\subsubsection{Un detalle de implementación}

Un problema de esta arquitectura, es que tenemos que cuidar que para cada paciente, sus dos imágenes estén siempre emparejadas, y no podemos cometer errores de mezclar imágenes de distintos pacientes cuando realizamos barajados de los datos.

Para resolver este problema, nuestra solución ha sido crear \textbf{un archivo TFRecord por paciente}, de modo que cada archivo contiene sus dos imágenes, y la etiqueta. De este modo, como todas las manipulaciones del conjunto de datos se harán a nivel de archivo TFRecord, no corremos ningún peligro de mezclar imágenes incorrectamente.

\subsubsection{Un detalle sobre los hiperparámetros}

Otro problema que surge en esta arquitectura, es que cada una de las ramas ha demostrado un mejor comportamiento cuando se entrena usando unos hiperparámetros concretos.

Aunque quizás sería una buena opción utilizar estos hiperparámetros concretos para cada una de las ramas por separado (por ejemplo, para conseguir que la rama izquierda actualice sus pesos con un \acrshort{lr} mayor que la derecha), no tenemos constancia de que esto pueda hacerse, al menos de forma relativamente sencilla, en Tensorflow. Y por otro lado, aunque parezca una buena idea, no tenemos claro que esto sea así.

Por tanto nuestra opción ha sido la de realizar un proceso de búsqueda de los mejores hiperparámetros posibles para esta nueva arquitectura.

\begin{formal}
{\bf Problemas}

Lamentablemente, el entrenamiento de esta arquitectura con dos entradas es enormemente pesado para nuestras GPU, por lo que necesitamos utilizar obligatoriamente las TPU.

El segundo problema es que desde julio, las TPU de kaggle han tenido una gran acogida: \textit{``TPUs are popular right now. You are \#10 in the queue. You can wait, try connecting again later, or use another accelerator''}, lo que ha hecho que para utilizarlas, tengamos que esperar a que otros usuarios terminen de usarlas, y conseguir una TPU libre ha sido difícil en muchas ocasiones (días).

Este segundo problema ha hecho que no hayamos podido realizar una buena selección de hiperparámetros (con los experimentos que esto requiere), por lo que el mal comportamiento de este modelo (que ahora veremos), puede deberse, en parte, a no haber conseguido unos buenos hiperparámetros. 

\end{formal}

\subsection{Resultados}

\begin{figure}[H]
    \centering
    \subfloat[Exactitud]{%
        \includegraphics[width=0.5\textwidth]{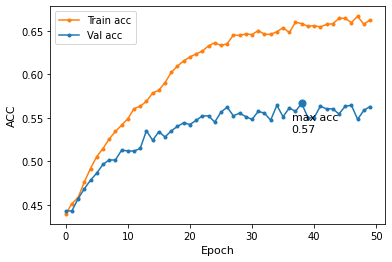}%
        }%
    \hfill%
    \subfloat[Función de pérdida]{%
        \includegraphics[width=0.5\textwidth]{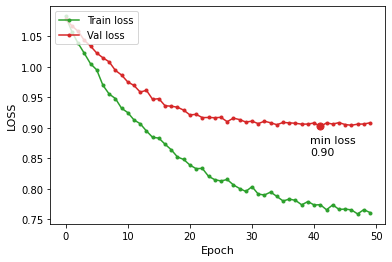}%
        }%
    \caption[Arquitectura con dos entradas. Curva de aprendizaje]{El modelo muestra un aprendizaje excesivamente lento (curvas naranja y verde), lo que nos hace sospechar de algún problema en la configuración del optimizador.}
    \label{fig:ensemble_curve}
\end{figure}

A pesar de que esperábamos conseguir el mejor resultado con este modelo, lamentablemente no ha sido así. Si vemos su curva de aprendizaje (figura \ref{fig:ensemble_curve}), es evidente que este modelo presenta dificultades en la optimización (no logra alcanzar ni el 70\% en entrenamiento).

Respecto a su estimación del error, este modelo sólo consigue un \textbf{56.3\%} en repeated k-fold, resultado que queda muy lejos de los mejores que hemos conseguido.

\subsection{¿Por qué no ha dado los resultados esperados?}

Dado que era el experimento del que esperábamos mejores resultados, hemos pensado  algunos de los problemas que podrían haber causado este mal comportamiento:\begin{enumerate}
    \item \textbf{Fallo de implementación:} cuando se comenten fallos en la implementación, lo más común es que los modelos directamente no aprendan nada (clasificación aleatoria), o que ni siquiera funcionen. En este caso, aunque aprende poco, lo cierto es que algo aprende, por lo que no parece que tengamos problemas de implementación.
    \item \textbf{Hiperparámetros (probablemente aquí esté el problema):} dado que el modelo \textbf{no consigue ni siquiera memorizar el conjunto de entrenamiento}, todo apunta a que existe algún tipo de problema con los hiperparámetros del optimizador (como quizás, un \acrshort{lr} demasiado pequeño, o una inicialización de los pesos no adecuada para esta arquitectura).
    \item \textbf{¿Información contradictoria en las imágenes?:} es una opción mucho menos probable, pero hemos pensado que quizás, existen pacientes para los que su imagen \acrshort{pet} presenta información similar a los pacientes de una determinada clase (por ejemplo, \acrshort{ad}), mientras que su imagen \acrshort{mri} presenta información más similar a los pacientes de otra clase, y esta contradicción podría empeorar la clasificación. 
\end{enumerate}

\section{Recopilación de resultados de interés}
\begin{tcolorbox}[
  colback=Green!5!white,
  colframe=Green!75!black,
  title={Sumario}, before upper={\parindent15pt}]

A lo largo de este capítulo hemos ido viendo los distintos experimentos que hemos realizado, junto a los resultados que estos han arrojado, así como una análisis detallado de los mismos.\vspace{1em}

En este último apartado recopilaremos todos aquellos resultados que consideramos de mayor interés para nuestros objetivos, de forma que se puedan consultar fácilmente de un sólo vistazo.

\end{tcolorbox}

\subsection{PET}

\begin{figure}[H]
    \centering
    \includegraphics[width=12cm]{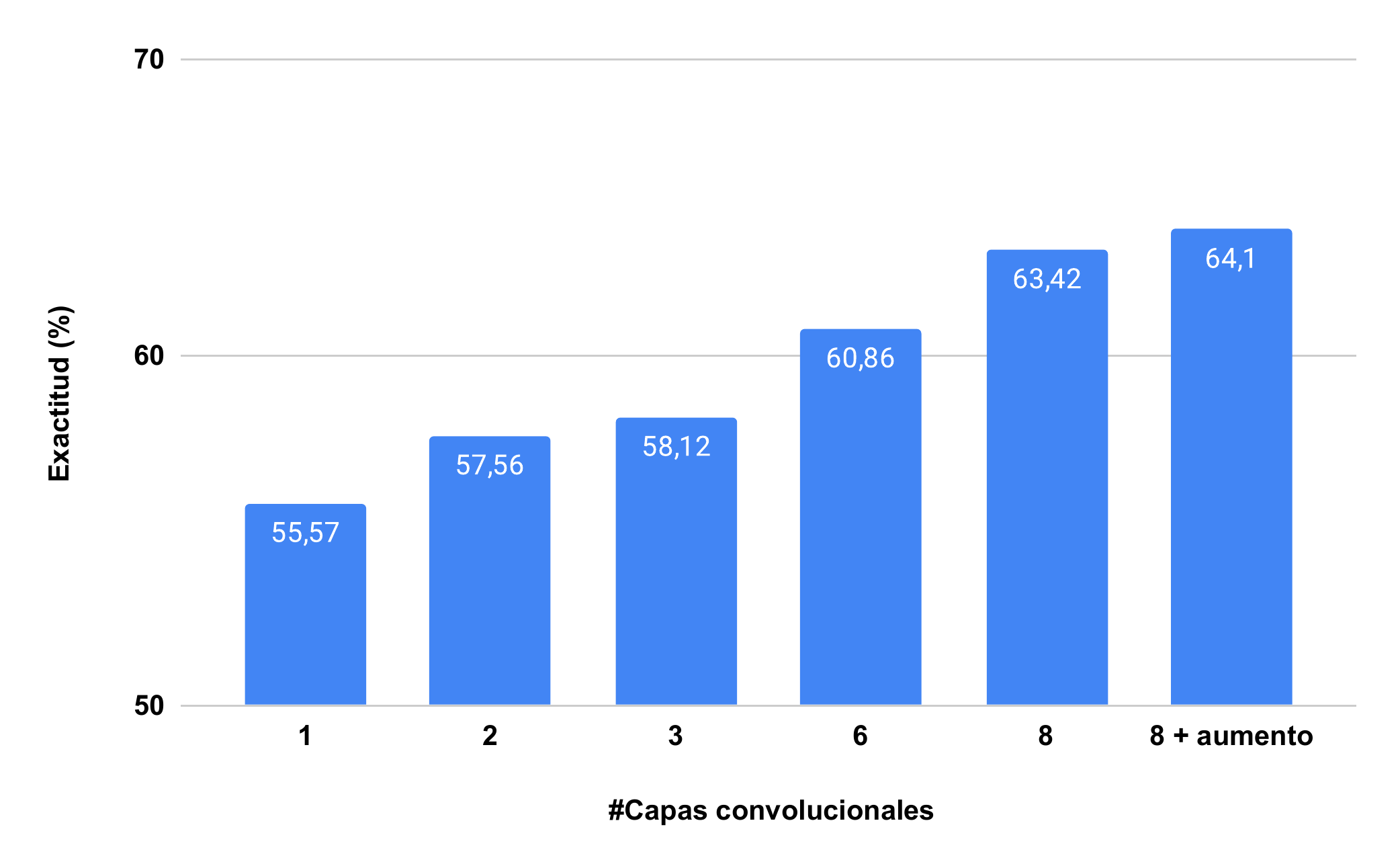}
    \caption[PET. Resumen gráfico de los experimentos.]{PET. Resumen gráfico de los experimentos. Se muestra la exactitud de cada modelo obtenida mediante repeated k-fold.}
    \label{fig:efecto_profundidad_pet}
\end{figure}

Los experimentos nos han mostrado que un aumento de la profundidad de las redes desde una capa convolucional hasta ocho capas, es capaz de aportarnos beneficios muy notables, progresando desde un 55.57\% de exactitud, hasta un 63.42\%.

Además, hemos podido observar que aplicar la técnica de aumento de datos ha aportado una pequeña mejora llegando al 64.1\%.

\newpage

\subsection{MRI}
\begin{figure}[H]
    \centering
    \includegraphics[width=12cm]{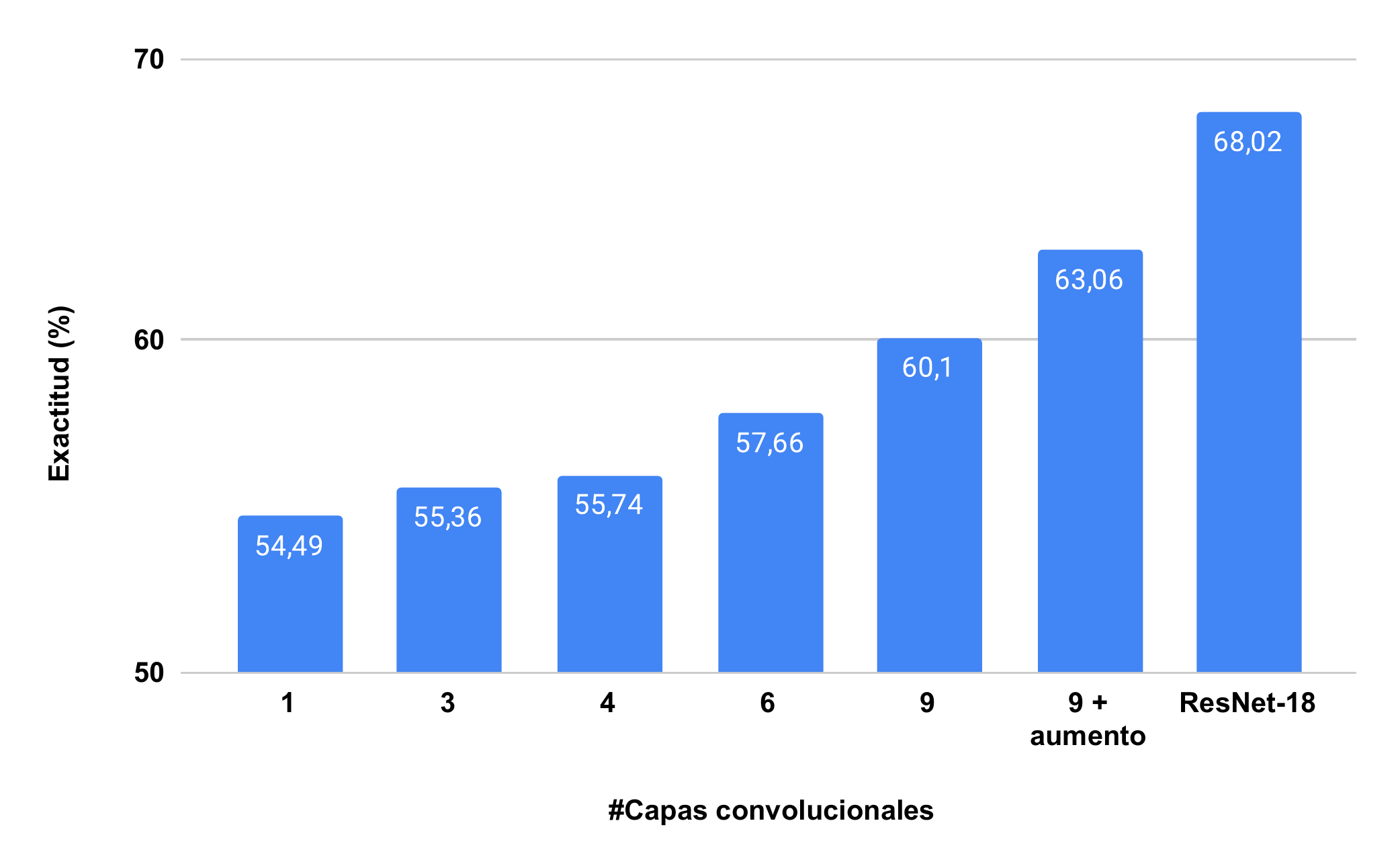}
    \caption[MRI. Resumen gráfico de los experimentos.]{MRI. Resumen gráfico de los experimentos. Se muestra la exactitud de cada modelo obtenida mediante repeated k-fold.}
    \label{fig:efecto_profundidad_mri}
\end{figure}

En el caso de \acrshort{mri}, el aumento de la profundidad de las redes también ha ayudado, aunque de una forma menos notoria que en el caso de \acrshort{pet} (desde un 54.49\% hasta un 60.1\%). 

Sin embargo aquí, el uso de aumento de datos sí que ha aportado mejoras muy notables, alcanzando un 63.06\%.

Por último, el uso de una red de 18 capas preentrenada con datos de COVID-19 ha logrado una exactitud del 68.02\%, un resultado realmente sorprendente.

\subsection{Mejor modelo final}

Por último, cabe destacar que el mejor modelo final que hemos logrado obtener, ha sido el que consiste un una arquitectura ResNet-18, preentrenada con datos de COVID-19 y que hace uso de transfer learning. Este modelo ha logrado un 68.02\% de exactitud en repeated k-fold, y del 70\% en el conjunto de test.

	\part{Conclusiones y trabajos futuros} 

    \chapter{Objetivos logrados}

Al comienzo de este trabajo enumeramos una serie de objetivos que queríamos lograr. Antes de exponer las conclusiones finales, queremos volver a enumerar estos objetivos, pero en esta ocasión, para ver si hemos conseguido cumplirlos:
    
\begin{enumerate}
    \item \textbf{Implementar una red convolucional para la clasificación de imágenes cerebrales entre las clases \acrshort{mci}, \acrshort{ad} y \acrshort{cn}}.
    
    \textbf{Cumplido}. Hemos desarrollado varias redes convolucionales, llegando finalmente a implementar una que obtiene resultados muy satisfactorios.
    
    \item \textbf{Estudiar cuál es la \textbf{profundidad} adecuada de las redes neuronales convolucionales para resolver este tipo de problemas.}
    
    \textbf{Cumplido}. Hemos realizado un estudio exhaustivo del uso con redes de distintas profundidades, tanto para \acrshort{pet}, como para \acrshort{mri}.
    
    \item \textbf{Utilizar conjuntos de datos de otras enfermedades para la aplicación de la técnica de transferencia de aprendizaje.}
    
    \textbf{Cumplido}. Hemos utilizado datos de COVID-19 para preentrenar una arquitectura ResNet-18 y transferir el aprendizaje a nuestro problema.
    
    \item \textbf{Estudiar el uso de la técnica de aumento de datos sobre imágenes 3D para la clasificación de imágenes médicas.}
    
    \textbf{Cumplido}. Hemos implementado las funciones necesarias para realizar transformaciones de las imágenes 3D en tiempo de entrenamiento, y hemos aplicado satisfactoriamente esta técnica.

    \item \textbf{Estudiar el uso simultáneo de imágenes cerebrales de dos modalidades, MRI y PET, para la mejora del diagnóstico.}
    
    \textbf{Parcialmente cumplido}. Hemos llegado a implementar una red convolucional que toma entradas de ambas modalidades simultáneamente, pero debido a la falta de disponibilidad del hardware, no hemos podido refinar suficientemente esta red.
    
    \item \textbf{Realizar una fase de experimentación y evaluación de modelos que asegure, en la medida de lo posible, la reproducibilidad de los resultados, y que estime el comportamiento de los modelos en condiciones reales.}
    
    \textbf{Cumplido}. Por medio del uso de la técnica repeated k-fold y la separación de un conjunto independiente para test.
    
    \item \textbf{Valorar distintas métricas de error para la evaluación de los modelos yseleccionar la más adecuada al problema dado.}
    
    \textbf{Cumplido}. Hemos visto que el uso de la exactitud, junto al estudio de la matriz de confusión, nos permiten evaluar adecuadamente nuestros modelos. Además hemos visto que las métricas de sensibilidad y especificidad son interesantes en el caso de considerar la clase \acrshort{ad} de mayor importancia.
    
    \item \textbf{Estudiar la necesidad del preprocesado de las imágenes médicas para la aplicación de técnicas de aprendizaje automático.}
    
    \textbf{Cumplido}. Y además hemos obtenido una respuesta muy clara, que detallaremos en la sección de conclusiones finales.
    
    \item \textbf{Realizar una implementación eficiente y escalable, por medio del uso de funciones avanzadas de Tensorflow y unidades de procesamiento tensorial (TPU)}.
    
    \textbf{Cumplido}. Hemos utilizado las TPU de Kaggle de forma eficiente, haciendo uso de archivos TFRecord y funciones de Tensorflow para el entrenamiento distribuido de los modelos.

\end{enumerate}

\chapter{Conclusiones}

\begin{tcolorbox}[
  colback=SkyBlue!5!white,
  colframe=SkyBlue!75!black,
  title={Sumario}]

A lo largo de los experimentos, y gracias al análisis que se ha ido haciendo, es posible que conozcamos algunas de la respuestas a las incógnitas que existían al comienzo de este trabajo, y que pretendíamos poder responder tras haber cumplido los objetivos. En este capítulo, daremos una \textbf{respuesta} lo más clara posible a estas preguntas. Debemos tener en cuenta, que aunque trataremos de responder de forma clara, esto no quiere decir que siempre vayamos a poder dar una respuesta totalmente cerrada, y es que cerrar por completo algunas preguntas, requeriría probablemente, años de estudio y experimentación.

\end{tcolorbox}

\section{Conclusiones: punto por punto}

\subsection{¿Es necesario el preprocesado de las imágenes?}

Mientras no dispongamos de un conjunto de datos de gran tamaño, al menos de un tamaño similar a los conjuntos de datos que se tienen para problemas con imágenes naturales, no nos cabe ninguna duda sobre la respuesta a esta pregunta: \textbf{sí, necesitamos realizar un preprocesado adecuado para simplificar nuestro problema.} 

\subsection{¿Son las técnicas de data augmentation y transfer learning útiles en este problema?}

Respecto a data augmentation, la respuesta es que esta técnica sí que puede ayudar, pero debe hacerse modificando las imágenes levemente, de forma que los cambios introducidos sean verosímiles.

Sobre la técnica de transfer learning haciendo uso de datos de otras enfermedades, podemos concluir que sí es posible obtener beneficios, al menos cuando los datos de otras enfermedades sean similares a los de nuestro problema.

\subsection{¿Es posible mejorar el diagnóstico usando distintas modalidades de imagen simultáneamente?}

Lamentablemente, debido a los problemas de hardware que ya conocemos, no podemos dar una respuesta a esta pregunta.

\subsection{¿Cuál es la profundidad adecuada de las redes en este problema?}

Dar una respuesta fija a este pregunta (2 capas, por ejemplo) no es posible, ya que que como sabemos, en el diseño de una red influyen infinidad de factores, y no sólo el número de capas.

Sin embargo, sí podemos responder a una pregunta muy relacionada, y probablemente de mayor utilidad: \textbf{¿cómo afecta la profundidad de las redes en este problema?}.

Al igual que en otros muchos problemas que se han tratado con redes neuronales convoluciones, lo cierto es que todo apunta a que la profundidad ayuda: en el caso de las imágenes \acrshort{pet}, hemos conseguido una mejora sustancial aumentando desde una sola capa, hasta ocho, y en el caso de \acrshort{mri}, hemos conseguido obtener beneficios hasta alcanzar una profundidad de dieciocho capas (ResNet-18 preentrenada).

A partir de esta misma pregunta surge otra: \textbf{¿hasta qué profundidad podríamos llegar?}. La respuesta es que probablemente, hasta más profundidad, pero no sabemos cuánto. Nuestra respuesta se fundamenta en los siguientes dos puntos:\begin{itemize}
    \itemsep0em
    \item Aunque hemos realizado numerosos experimentos, es muy probable que con más tiempo y un hardware aún más potente, pudiéramos haber encontrado alguna arquitectura aún más profunda, aunque no sabemos cuanto más.
    \item Cuando se tienen muchos datos, los problemas de sobreajuste se reducen enormemente, lo que permite crear arquitecturas muy profundas con más éxito, ya que se reduce la importancia relativa de las técnicas de regularización. Por tanto, es altamente probable que si tuviéramos más datos, pudiéramos conseguir con éxito redes más profundas.

\end{itemize}

\section{Aportaciones sobre el estado del arte}

Las aportaciones que hacen este trabajo sobre lo que ya existía, están muy relacionadas con las preguntas respondidas anteriormente, por lo que nos parece un buen momento para enumerar estas aportaciones:\begin{enumerate}
    \itemsep0em
    
    \item En los estudios revisados, existía una gran incertidumbre acerca de la profundidad: unos usaban redes muy profundas, otros utilizaban redes con apenas dos capas, y todos obtenían resultados buenos. Sin embargo, debido a las grandes diferencias entre estudios, no era posible obtener conclusiones claras sobre cómo es el efecto de la profundidad.
    
    En este trabajo hemos podido \textbf{comparar en condiciones de igualdad numerosas arquitecturas de distintas profundidades}, lo que nos ha permitido concluir que el aumento de la profundidad de las redes en general ayuda, aunque no sea posible dar una cota superior sobre la profundidad adecuada.
    
    \item Algunos de los estudios utilizaban la técnica de aumento de datos obteniendo mejoras, pero sólo con imágenes 2D. Nosotros, \textbf{hemos ampliado esta técnica a imágenes 3D}, y también hemos obtenido mejoras.
    
    \item También se había usado la técnica de transfer learning en algunos estudios revisados, pero todos ellos usaban imágenes cerebrales, y en ocasiones de la misma enfermedad.
    
    Aquí, \textbf{hemos utilizado datos de una enfermedad totalmente distinta} para realizar esta técnica, y los resultados han sido muy favorables.
    
    \item Hemos experimentado con el uso de imágenes sin preprocesar, y hemos dado razones contundentes por la que actualmente es necesario el preprocesado en problemas médicos, para los que se disponen de muy pocos datos.

\end{enumerate}

\section{Conclusión final}

En este trabajo hemos visto, que al igual que en otras muchas tareas, en el diagnóstico de la enfermedad de Alzheimer, el aprendizaje automático, y concretamente las \acrshort{cnn} han demostrado tener un gran potencial.

Hemos demostrado empíricamente, que al igual que en otros problemas, la profundidad ayuda, y hemos presentado algunas técnicas novedosas en este ámbito que han ayudado a afrontar este problema de clasificación. Disponiendo de una cantidad ínfima de datos, estas técnicas nos han permitido crear un modelo que es capaz de realizar la clasificación con aproximadamente un 70\% de exactitud, un resultado muy satisfactorio teniendo en cuenta la complejidad del problema.

Es posible que aún sea pronto para que una red convolucional sea capaz de ayudar a un médico en este ámbito, pero estos resultados son esperanzadores, y es de esperar que en un futuro próximo, el aprendizaje automático ayude a mejorar la vida de pacientes que desgraciadamente sufren la enfermedad de Alzheimer.

\chapter{Trabajos futuros}

Partiendo del trabajo realizado, consideramos que existen varios aspectos que serían interesantes de tratar en el futuro:\begin{itemize}

    \item Recolectar más datos. La falta de datos ha sido claramente el factor que hace más difícil nuestro problema. Conseguir recolectar un conjunto de datos de mayor tamaño sería imprescindible para conseguir resultados realmente buenos. 
    
    \item Con más tiempo, refinar la arquitectura que hace uso de imágenes de dos modalidades distintas, ya que aunque no hayamos conseguido resultados demasiado satisfactorios, podríamos haberlos obtenido con un ajuste adecuado.
    
    \item Avanzar un paso más con el preprocesado, y con el apoyo del conocimiento aportado por médicos, conseguir extraer solamente aquellas áreas del cerebro que se vean más afectadas por la enfermedad, lo que simplificaría nuestro problema.
    
    \item Probar un enfoque que haga uso de cortes 2D de las imágenes, ya que de esta forma, podríamos aprovecharnos de numerosas arquitecturas ya implementadas y preentrenadas en conjuntos de datos de enorme tamaño.

\end{itemize}

    \backmatter
	
	\newpage
	
	\printbibliography[heading=bibintoc,title={Bibliografía}]
	
	\mainmatter
	\begin{appendices}

\chapter{Arquitecturas e hiperparámetros}

En este anexo se ofrece un resumen de las arquitecturas de cada experimento (con una notación basada en la de Keras) junto a los hiperparámetros utilizados para entrenarlos.

\noindent\textbf{Nota:} en el caso de hiperparámetros no especificados, se supone el valor por defecto que asigna Tensorflow (versión 2.5).

\section{Fase 1: estudio de la profundidad}

\subsection{PET}

\subsubsection{Una capa convolucional}
\label{prof_pet_exp1}

\arrayrulecolor{black}

\begin{table}[H] 
\begin{tabular}{lll} 
Modelo: pet\_1 \\ \hline 
Capa                   & Dimensiones de salida               & \# Parámetros   \\ \hline \hline 
Input           & [(None, 79, 95, 68, 1)]     & 0          \\ \hline 
5 x 5 x 5 Conv3D (32), pad 0             & (None, 75, 91, 64, 32)      & 4032       \\ \hline 
2 x 2 x 2 MaxPooling3D, stride 2 & (None, 37, 45, 32, 32)      & 0          \\ \hline 
Flatten           & (None, 1704960)             & 0          \\ \hline 
FC (256)               & (None, 256)                 & 436470016  \\ \hline 
FC (3), softmax                & (None, 3)                   & 771        \\ \hline \hline 
Total de parámetros: 436,474,819 \\ 
Parámetros entrenables: 436,474,819 \\ 
Parámetros no entrenables: 0 \\ \hline

\end{tabular} 
\caption{PET. Una capa convolucional. Arquitectura.} 
\end{table}

\begin{table}[H]
\begin{tabular}{|l|l|l|l|l|l|}
\hline
LR & Optimizador & Épocas & Batch size & Activación & Inicializador \\ \hline
$1\mathrm{e}{-5}$, exp\_decay 0.1 &    Adam         &    50    &      8      &        ReLU   &       Glorot        \\ \hline
\end{tabular}
\caption{PET. Una capa convolucional. Otros hiperparámetros.}
\end{table}

\subsubsection{Dos capas convolucionales}
\label{prof_pet_exp2}
\begin{table}[H] 
\begin{tabular}{lll} 
Modelo: pet\_2 \\ \hline 
Capa                  & Dimensiones de salida                & \# Parámetros    \\ \hline \hline 
Input         & [(None, 79, 95, 68, 1)]     & 0          \\ \hline 
5 x 5 x 5 Conv3D (32), pad 0             & (None, 75, 91, 64, 32)      & 4032        \\ \hline 
2 x 2 x 2 MaxPooling3D, stride 2 & (None, 37, 45, 32, 32)      & 0          \\ \hline 
5 x 5 x 5 Conv3D (32), pad 0            & (None, 33, 41, 38, 32)      & 128032      \\ \hline 
2 x 2 x 2 MaxPooling3D, stride 2 & (None, 16, 20, 14, 32)      & 0          \\ \hline 
Flatten               & (None, 143360)              & 0          \\ \hline 
FC (256)                  & (None, 256)                 & 36700416   \\ \hline 
FC (3), softmax                   & (None, 3)                   & 771        \\ \hline \hline 
Total de parámetros: 36,833,251 \\ 
Parámetros entrenables: 36,833,251 \\ 
Parámetros no entrenables: 0 \\ \hline 
\end{tabular} 
\caption{PET. Dos capas convolucionales. Arquitectura.} 
\end{table}

\begin{table}[H]
\begin{tabular}{|l|l|l|l|l|l|}
\hline
LR & Optimizador & Épocas & Batch size & Activación & Inicializador \\ \hline
$1\mathrm{e}{-5}$, exp\_decay 0.1  &    Adam         &    50    &      4      &        ReLU   &       Glorot        \\ \hline
\end{tabular}
\caption{PET. Dos capas convolucionales. Otros hiperparámetros.}
\end{table}

\subsubsection{Tres capas convolucionales}
\label{prof_pet_exp3}

\begin{table}[H] 
\begin{tabular}{lll} 
Modelo: pet\_3 \\ \hline 
Capa                   & Dimensiones de salida                 & \# Parámetros    \\ \hline \hline 
Input           & [(None, 79, 95, 68, 1)]     & 0          \\ \hline 
5 x 5 x 5 Conv3D (16), pad 0                & (None, 75, 91, 64, 16)      & 2016       \\ \hline 
2 x 2 x 2 MaxPooling3D, stride 2 & (None, 37, 45, 32, 16)      & 0          \\ \hline 
5 x 5 x 5 Conv3D (64), pad 0              & (None, 33, 41, 28, 64)      & 128064     \\ \hline 
2 x 2 x 2 MaxPooling3D, stride 2 & (None, 16, 20, 14, 64)      & 0          \\ \hline  
5 x 5 x 5 Conv3D (128), pad 0              & (None, 12, 16, 10, 128)     & 1024128    \\ \hline 
2 x 2 x 2 MaxPooling3D, stride 2 & (None, 6, 8, 5, 128)        & 0          \\ \hline 
Flatten            & (None, 30720)               & 0          \\ \hline 
FC (256)                & (None, 256)                 & 7864576    \\ \hline 
FC (3), softmax                & (None, 3)                   & 771        \\ \hline \hline 
Total de parámetros: 9,019,555 \\ 
Parámetros entrenables: 9,019,555 \\ 
Parámetros no entrenables: 0 \\ \hline 
\end{tabular} 
\caption{PET. Tres capas convolucionales. Arquitectura.} 
\end{table}

\begin{table}[H]
\begin{tabular}{|l|l|l|l|l|l|}
\hline
LR & Optimizador & Épocas & Batch size & Activación & Inicializador \\ \hline
$1\mathrm{e}{-4}$, exp\_decay 0.01   &    Adam         &    50    &      8      &        ReLU   &       Glorot        \\ \hline
\end{tabular}
\caption{PET. Tres capas convolucionales. Otros hiperparámetros.} 
\end{table}

\subsubsection{Seis capas convolucionales}
\label{prof_pet_exp4}

\begin{table}[H] 
\begin{tabular}{lll} 
Modelo: pet\_6 \\ \hline 
Capa                  & Dimensiones de salida                 & \# Parámetros    \\ \hline \hline 
Input          & [(None, 79, 95, 68, 1)]     & 0          \\ \hline 
3 x 3 x 3 Conv3D (16), pad 0               & (None, 77, 93, 66, 16)      & 448        \\ \hline 
3 x 3 x 3 Conv3D (16), pad 0               & (None, 75, 91, 64, 16)      & 6928       \\ \hline 
2 x 2 x 2 MaxPooling3D, stride 2 & (None, 37, 45, 32, 16)      & 0          \\ \hline 
3 x 3 x 3 Conv3D (64), pad 0              & (None, 35, 43, 30, 64)      & 27712      \\ \hline 
3 x 3 x 3 Conv3D (64), pad 0              & (None, 33, 41, 28, 64)      & 110656     \\ \hline 
2 x 2 x 2 MaxPooling3D, stride 2 & (None, 16, 20, 14, 64)      & 0          \\ \hline 
BatchNorm, momentum 0.99 & (None, 16, 20, 14, 64)      & 256        \\ \hline 
3 x 3 x 3 Conv3D (128), pad 0            & (None, 14, 18, 12, 128)     & 221312     \\ \hline 
3 x 3 x 3 Conv3D (128), pad 0              & (None, 12, 16, 10, 128)     & 442496     \\ \hline 
2 x 2 x 2 MaxPooling3D, stride 2 & (None, 6, 8, 5, 128)        & 0          \\ \hline 
Flatten            & (None, 30720)               & 0          \\ \hline 
Dropout (0.1)             & (None, 30720)               & 0          \\ \hline FC (256)              & (None, 256)                 & 7864576    \\ \hline 
FC (3), softmax               & (None, 3)                   & 771        \\ \hline \hline 
Total de parámetros: 8,675,155 \\ 
Parámetros entrenables: 8,675,027 \\ 
Parámetros no entrenables: 128 \\ \hline 
\end{tabular} 
\caption{PET. Seis capas convolucionales. Arquitectura.} 
\end{table}

\begin{table}[H]
\begin{tabular}{|l|l|l|l|l|l|}
\hline
LR & Optimizador & Épocas & Batch size & Activación & Inicializador \\ \hline
$1\mathrm{e}{-6}$, exp\_decay 0.1   &    Adam         &    40    &      4      &        ReLU   &       Glorot        \\ \hline
\end{tabular}
\caption{PET. Seis capas convolucionales. Otros hiperparámetros.}
\end{table}

\subsubsection{Ocho capas convolucionales}
\label{prof_pet_exp5}

\begin{table}[H] 
\begin{tabular}{lll} 
Modelo: pet\_8 \\ \hline 
Capa                   & Dimensiones de salida                & \# Parámetros    \\ \hline \hline 
Input           & [(None, 79, 95, 68, 1)]     & 0          \\ \hline 
3 x 3 x 3 Conv3D (16), pad 0            & (None, 77, 93, 66, 16)      & 448        \\ \hline 
3 x 3 x 3 Conv3D (16), pad 0           & (None, 75, 91, 64, 16)      & 6928       \\ \hline 
2 x 2 x 2 MaxPooling3D, stride 2 & (None, 37, 45, 32, 16)      & 0          \\ \hline 
3 x 3 x 3 Conv3D (64), pad 0           & (None, 35, 43, 30, 64)      & 27712      \\ \hline 
3 x 3 x 3 Conv3D (64), pad 0             & (None, 33, 41, 28, 64)      & 110656     \\ \hline 
3 x 3 x 3 Conv3D (64), pad 0            & (None, 31, 39, 26, 64)      & 110656     \\ \hline 
2 x 2 x 2 MaxPooling3D, stride 2 & (None, 15, 19, 13, 64)      & 0          \\ \hline 
BatchNorm, momentum 0.9 & (None, 15, 19, 13, 64)      & 256        \\ \hline 
3 x 3 x 3 Conv3D (128), pad 0             & (None, 13, 17, 11, 128)     & 221312     \\ \hline 
3 x 3 x 3 Conv3D (128), pad 0            & (None, 11, 15, 9, 128)      & 442496     \\ \hline 
3 x 3 x 3 Conv3D (128), pad 0            & (None, 9, 13, 7, 128)       & 442496     \\ \hline 
2 x 2 x 2 MaxPooling3D, stride 2  & (None, 4, 6, 3, 128)        & 0          \\ \hline 
Flatten            & (None, 9216)                & 0          \\ \hline 
Dropout (0.2)            & (None, 9216)                & 0          \\ \hline 
FC (256)               & (None, 256)                 & 2359552    \\ \hline 
FC (128)               & (None, 128)                 & 32896      \\ \hline 
FC (3), softmax             & (None, 3)                   & 387        \\ \hline \hline 
Total de parámetros: 3,755,795 \\ 
Parámetros entrenables : 3,755,667 \\ 
Parámetros no entrenables: 128 \\ \hline 
\end{tabular} 
\caption{PET. Ocho capas convolucionales. Arquitectura.}
\end{table}

En cada capa convolucional se hace uso de regularización L2, con $\lambda = 1\mathrm{e}{-5}$

\begin{table}[H]
\begin{tabular}{|l|l|l|l|l|l|}
\hline
LR & Optimizador & Épocas & Batch size & Activación & Inicializador \\ \hline
$1\mathrm{e}{-5}$, exp\_decay 0.1   &    Adam         &    50    &      4      &        ReLU   &       Glorot        \\ \hline
\end{tabular}
\caption{PET. Ocho capas convolucionales. Otros hiperparámetros.}
\end{table}

\subsection{MRI}
\subsubsection{Una capa convolucional}
\label{prof_1_mri}

\begin{table}[H] 
\begin{tabular}{lll} 
Model: mri\_1 \\ \hline 
Capa                  & Dimensiones de salida                & \#  Parámetros  \\ \hline \hline 
Input          & [(None, 75, 90, 75, 1)]     & 0          \\ \hline 
5 x 5 x 5 Conv3D (32), pad 0           & (None, 71, 86, 71, 32)      & 4032       \\ \hline 
2 x 2 x 2 MaxPooling3D, stride 2 & (None, 35, 43, 35, 32)      & 0          \\ \hline 
Flatten           & (None, 1685600)             & 0          \\ \hline 
FC (200)                & (None, 200)                 & 337120200  \\ \hline 
Dropout (0.1)            & (None, 200)                 & 0          \\ \hline 
FC (3), softmax                & (None, 3)                   & 603        \\ \hline \hline 
Total de parámetros: 337,124,835 \\ 
Parámetros entrenables: 337,124,835 \\ 
Parámetros no entrenables: 0 \\ \hline 
\end{tabular} 
\caption{MRI. Una capa convolucional. Arquitectura.} 
\end{table}

\begin{table}[H]
\begin{tabular}{|l|l|l|l|l|l|}
\hline
LR & Optimizador & Épocas & Batch size & Activación & Inicializador \\ \hline
$1\mathrm{e}{-6}$, exp\_decay 0.1   &    Adam         &    50    &      4      &        ReLU   &       Glorot        \\ \hline
\end{tabular}
\caption{MRI. Una capa convolucional. Otros hiperparámetros.}
\end{table}

\subsubsection{Tres capas convolucionales}
\label{prof_3_mri}

\begin{table}[H] 
\begin{tabular}{lll} 
Model: mri\_3 \\ \hline 
Capa                   & Dimensiones de salida              & Parámetros \#    \\ \hline \hline 
Input           & [(None, 75, 90, 75, 1)]     & 0          \\ \hline 
5 x 5 x 5 Conv3D (16), pad 0     & (None, 71, 86, 71, 16)      & 2016       \\ \hline 
2 x 2 x 2 MaxPooling3D, stride 2 & (None, 35, 43, 35, 16)      & 0          \\ \hline 
5 x 5 x 5 Conv3D (32), pad 0     & (None, 31, 39, 31, 32)      & 64032      \\ \hline 
2 x 2 x 2 MaxPooling3D, stride 2 & (None, 15, 19, 15, 32)      & 0          \\ \hline 
5 x 5 x 5 Conv3D (64), pad 0     & (None, 11, 15, 11, 64)      & 256064 \\ \hline
2 x 2 x 2 MaxPooling3D, stride 2 & (None, 5, 7, 5, 64)         & 0          \\ \hline 
Flatten                          & (None, 11200)               & 0          \\ \hline 
Dropout (0.2)                    & (None, 11200)               & 0          \\ \hline 
FC (256)                         & (None, 256)                 & 2867456    \\ \hline 
FC (3), softmax                           & (None, 3)                   & 771        \\ \hline \hline 
Total de parámetros: 3,190,339 \\ 
Parámetros entrenables: 3,190,339 \\ 
Parámetros no entrenables: 0 \\ \hline 
\end{tabular} 
\caption{MRI. Tres capas convolucionales. Arquitectura.} 
\end{table}

\begin{table}[H]
\begin{tabular}{|l|l|l|l|l|l|}
\hline
LR & Optimizador & Épocas & Batch size & Activación & Inicializador \\ \hline
$1\mathrm{e}{-5}$, exp\_decay 0.1   &    Adam         &    70    &      4      &        ReLU   &       Glorot        \\ \hline
\end{tabular}
\caption{MRI. Tres capas convolucionales. Otros hiperparámetros.}
\end{table}

\subsubsection{Cuatro capas convolucionales}
\label{prof_4_mri}

\begin{table}[H] 
\begin{tabular}{lll} 
Model: mri\_4 \\ \hline 
Capa                 & Dimensiones de salida              & Parámetros \#    \\ \hline \hline 
Input           & [(None, 75, 90, 75, 1)]     & 0          \\ \hline 
5 x 5 x 5 Conv3D (16), pad 0            & (None, 71, 86, 71, 16)      & 2016       \\ \hline 
2 x 2 x 2 MaxPooling3D, stride 2 & (None, 35, 43, 35, 16)      & 0          \\ \hline 
5 x 5 x 5 Conv3D (32), pad 0             & (None, 31, 39, 31, 32)      & 64032      \\ \hline 
2 x 2 x 2 MaxPooling3D, stride 2 & (None, 15, 19, 15, 32)      & 0          \\ \hline 
3 x 3 x 3 Conv3D (64), pad 0          & (None, 13, 17, 13, 64)      & 55360      \\ \hline 
3 x 3 x 3 Conv3D (64), pad 0            & (None, 11, 15, 11, 64)      & 110656     \\ \hline 
2 x 2 x 2 MaxPooling3D, stride 2 & (None, 5, 7, 5, 64)         & 0          \\ \hline 
Flatten            & (None, 11200)               & 0          \\ \hline 
Dropout (0.25)            & (None, 11200)               & 0          \\ \hline 
FC (256)               & (None, 256)                 & 2867456    \\ \hline 
FC (3), softmax               & (None, 3)                   & 771        \\ \hline \hline 
Total de parámetros: 3,100,291 \\ 
Parámetros entrenables: 3,100,291 \\ 
Parámetros no entrenables: 0 \\ \hline 
\end{tabular} 
\caption{MRI. Cuatro capas convolucionales. Arquitectura.} 
\end{table}

\begin{table}[H]
\begin{tabular}{|l|l|l|l|l|l|}
\hline
LR & Optimizador & Épocas & Batch size & Activación & Inicializador \\ \hline
$1\mathrm{e}{-6}$, exp\_decay 0.1   &    Adam         &    50    &      4      &        ReLU   &       Glorot        \\ \hline
\end{tabular}
\caption{MRI. Cuatro capas convolucionales. Otros hiperparámetros.}
\end{table}

\subsubsection{Seis capas convolucionales}
\label{prof_6_mri}

\begin{table}[H] 
\begin{tabular}{lll} 
Model: mri\_6 \\ \hline 
Capa                   & Dimensiones de salida           & Parámetros \#    \\ \hline \hline 
Input          & [(None, 75, 90, 75, 1)]     & 0          \\ \hline 
3 x 3 x 3 Conv3D (16), pad 0           & (None, 73, 88, 73, 16)      & 448        \\ \hline 
3 x 3 x 3 Conv3D (16), pad 0             & (None, 71, 86, 71, 16)      & 6928       \\ \hline 
2 x 2 x 2 MaxPooling3D, stride 2 & (None, 35, 43, 35, 16)      & 0          \\ \hline 
3 x 3 x 3 Conv3D (32), pad 0            & (None, 33, 41, 33, 32)      & 13856      \\ \hline 
3 x 3 x 3 Conv3D (32), pad 0             & (None, 31, 39, 31, 32)      & 27680      \\ \hline 
2 x 2 x 2 MaxPooling3D, stride 2 & (None, 15, 19, 15, 32)      & 0          \\ \hline 
BatchNorm, momentum 0.9 & (None, 15, 19, 15, 32)      & 128        \\ \hline 
3 x 3 x 3 Conv3D (64), pad 0             & (None, 13, 17, 13, 64)      & 55360      \\ \hline 
3 x 3 x 3 Conv3D (64), pad 0             & (None, 11, 15, 11, 64)      & 110656     \\ \hline 
2 x 2 x 2 MaxPooling3D, stride 2 & (None, 5, 7, 5, 64)         & 0          \\ \hline 
Flatten            & (None, 11200)               & 0          \\ \hline 
Dropout (0.2)            & (None, 11200)               & 0          \\ \hline 
FC (256)               & (None, 256)                 & 2867456    \\ \hline 
Dropout (0.1)             & (None, 256)                 & 0          \\ \hline 
FC (128)               & (None, 128)                 & 32896      \\ \hline 
FC (3), softmax              & (None, 3)                   & 387        \\ \hline \hline 
Total de parámetros: 3,115,795 \\ 
Parámetros entrenables: 3,115,731 \\ 
Parámetros no entrenables: 64 \\ \hline 
\end{tabular} 
\caption{MRI. Seis capas convolucionales. Arquitectura.} 
\end{table}

\begin{table}[h]
\begin{tabular}{|l|l|l|l|l|l|}
\hline
LR & Optimizador & Épocas & Batch size & Activación & Inicializador \\ \hline
$1\mathrm{e}{-6}$, exp\_decay 0.1   &    Adam         &    50    &      4      &        ReLU   &       Glorot        \\ \hline
\end{tabular}
\caption{MRI. Seis capas convolucionales. Otros hiperparámetros.}
\end{table}

\subsubsection{Nueve capas}
\label{prof_9_mri}

\begin{table}[H] 
\begin{tabular}{lll} 
Model: mri\_9 \\ \hline 
Capa                   & Dimensiones de salida               & Parámetros \#    \\ \hline \hline 
Input          & [(None, 121, 145, 121, 1)]  & 0          \\ \hline 
3 x 3 x 3 Conv3D (32), pad 0            & (None, 119, 143, 119, 32)   & 896        \\ \hline 
3 x 3 x 3 Conv3D (32), pad 0             & (None, 117, 141, 117, 32)   & 27680      \\ \hline 
2 x 2 x 2 MaxPooling3d, stride 2 & (None, 58, 70, 58, 32)      & 0          \\ \hline 
3 x 3 x 3 Conv3D (64), pad 0             & (None, 56, 68, 56, 64)      & 55360      \\ \hline 
3 x 3 x 3 Conv3D (64), pad 0            & (None, 54, 66, 54, 64)      & 110656     \\ \hline 
2 x 2 x 2 MaxPooling3D, stride 2 & (None, 27, 33, 27, 64)      & 0          \\ \hline 
BatchNorm, momentum 0.9 & (None, 27, 33, 27, 64)      & 256        \\ \hline 
3 x 3 x 3 Conv3D (128), pad 0             & (None, 25, 31, 25, 128)     & 221312     \\ \hline 
3 x 3 x 3 Conv3D (128), pad 0             & (None, 23, 29, 23, 128)     & 442496     \\ \hline 
2 x 2 x 2 MaxPooling3D, stride 2 & (None, 11, 14, 11, 128)     & 0          \\ \hline 
BatchNorm, momentum 0.9 & (None, 11, 14, 11, 128)     & 512        \\ \hline 
3 x 3 x 3 Conv3D (256), pad 0             & (None, 9, 12, 9, 256)       & 884992     \\ \hline 
3 x 3 x 3 Conv3D (256), pad 0            & (None, 7, 10, 7, 256)       & 1769728    \\ \hline 
3 x 3 x 3 Conv3D (256), pad 0              & (None, 5, 8, 5, 256)        & 1769728    \\ \hline 
GlobalAveragePooling3D & (None, 256)                 & 0          \\ \hline 
Dropout (0.2)           & (None, 256)                 & 0          \\ \hline 
FC (128)               & (None, 128)                 & 32896      \\ \hline 
Dropout (0.2)           & (None, 128)                 & 0          \\ \hline 
FC (128)               & (None, 128)                 & 16512      \\ \hline 
FC (3), softmax              & (None, 3)                   & 387        \\ \hline \hline 
Total de parámetros: 5,333,411 \\ 
Parámetros entrenables: 5,333,027 \\ 
Parámetros no entrenables: 384 \\ \hline 
\end{tabular} 
\caption{MRI. Nueve capas convolucionales. Arquitectura.} 
\end{table}

\begin{table}[H]
\begin{tabular}{|l|l|l|l|l|l|}
\hline
LR & Optimizador & Épocas & Batch size & Activación & Inicializador \\ \hline
$1\mathrm{e}{-6}$, exp\_decay 0.1   &    Adam         &    50    &      4      &        ReLU   &       Glorot        \\ \hline
\end{tabular}
\caption{MRI. Nueve capas convolucionales. Otros hiperparámetros.}
\end{table}

\section{Fase 2: aumento de datos}

\subsection{PET}
\label{pet_aumento_arquitectura}
\subsubsection{Arquitectura}
Idéntica a la del modelo del anexo \ref{prof_pet_exp5}, salvo por los siguientes detalles:\begin{itemize}
    \itemsep0em
    \item En dropout: la tasa de dropout pasa a ser 0.1
    \item En batch normalization: momentum pasa a ser 0.95
\end{itemize}
\subsubsection{Algunos hiperparámetros importantes}

\begin{table}[H]
\begin{tabular}{|l|l|l|l|l|l|}
\hline
LR & Optimizador & Épocas & Batch size & Activación & Inicializador \\ \hline
$1\mathrm{e}{-4}$, exp\_decay 0.1   &    Adam         &    50    &      8      &        ReLU   &       Glorot        \\ \hline
\end{tabular}
\caption{PET. Aumento de datos. Hiperparámetros importantes.}
\end{table}

\subsubsection{Hiperparámetros del aumento de datos}

\begin{table}[H]
\begin{tabular}{|l|l|l|l|}
\hline
Rotación & Ampliación & Volteo & Desplazamiento \\ \hline
Máx. 0.5\textdegree & NO & NO & Máx. 2\%   \\ \hline
\end{tabular}
\caption{PET. Hiperparámetros del aumento de datos.}
\end{table}

\subsection{MRI}
\label{mri_aumento_arquitectura}
\subsubsection{Arquitectura e hiperparámetros}
Todo queda exactamente igual que en el modelo sin aumento de datos (anexo \ref{prof_9_mri}). 

\subsubsection{Hiperparámetros del aumento de datos}
\begin{table}[H]
\begin{tabular}{|l|l|l|l|}
\hline
Rotación & Ampliación & Volteo & Desplazamiento \\ \hline
Máx. 0.5\textdegree & Min. 0.95, Máx 1.05 & NO & Máx. 2\%   \\ \hline
\end{tabular}
\caption{MRI. Hiperparámetros del aumento de datos.}
\end{table}

\section{Fase 4: transfer learning}

Las arquitecturas usadas en el preentrenamiento, para \acrshort{pet}, y para \acrshort{mri} quedan descritas de forma exacta en el apartado \ref{RESNET18}. En este caso, no aportamos la arquitectura con la notación de tabla, ya que en este caso no es aclaratoria (demasiado grande y arquitectura no secuencial).

En todas las capas de Batch Normalization, hemos usado un valor de 0.99 para el parámetro \textit{momentum}.

\subsection{Preprocesado para el preentrenamiento}

Las imágenes del conjunto COVID-19 han sido preprocesadas de la siguiente forma:\begin{enumerate}
\itemsep0em
    \item Se acotan los valores en el rango $[min=-1000, max=400]$, sustituyendo cualquier valor que escape de ese rango, por el valor máximo o mínimo.
    \item Se redimensionan a $64\times128\times128$
    \item Se aplica normalización del tipo $minmax$: $X = \frac{X - min(X)}{max(X) - min(X)}$
\end{enumerate}

\noindent\textbf{Nota:} recordar que la red preentrenada esperará valores entre 0 y 1, por lo que para un comportamiento correcto, es necesario aplicar este tipo de normalización tanto a las imágenes \acrshort{pet} como a las \acrshort{mri}, además de todo su preprocesado correspondiente.

\label{preprocesado_preentrenamiento}

\subsection{Hiperparámetros del preentrenamiento}

\begin{table}[H]
\begin{tabular}{|l|l|l|l|l|l|}
\hline
LR & Optimizador & Épocas & Batch size & Activación & Inicializador \\ \hline
$1\mathrm{e}{-5}$ &    Adam         &    120    &      16      &        ReLU   &       Glorot        \\ \hline
\end{tabular}
\caption{ResNet-18. Preentrenamiento. Hiperparámetros importantes.}
\end{table}

\begin{table}[H]
\begin{tabular}{|l|l|l|l|}
\hline
Rotación & Ampliación & Volteo & Desplazamiento \\ \hline
Máx. 2\textdegree & Min. 0.9, Máx 1.1 & NO & Máx. 4\%   \\ \hline
\end{tabular}
\caption{ResNet-18. Preentrenamiento. Hiperparámetros del aumento de datos.}
\end{table}

\subsection{Hiperparámetros para PET}

\begin{table}[H]
\begin{tabular}{|l|l|l|l|l|l|}
\hline
LR & Optimizador & Épocas & Batch size & Activación & Inicializador \\ \hline
$1\mathrm{e}{-4}$ &    Adam         &    30    &      4      &        ReLU   &       Glorot        \\ \hline
\end{tabular}
\caption{PET. Transfer learning. Hiperparámetros importantes.}
\end{table}

\begin{table}[H]
\begin{tabular}{|l|l|l|l|}
\hline
Rotación & Ampliación & Volteo & Desplazamiento \\ \hline
Máx. 0.5\textdegree & NO & NO & Máx. 2\%   \\ \hline
\end{tabular}
\caption{PET. Transfer learning. Hiperparámetros del aumento de datos.}
\end{table}

\subsection{Hiperparámetros para MRI}

\begin{table}[H]
\begin{tabular}{|l|l|l|l|l|l|}
\hline
LR & Optimizador & Épocas & Batch size & Activación & Inicializador \\ \hline
$1\mathrm{e}{-4}$ &    Adam         &    30    &      4      &        ReLU   &       Glorot        \\ \hline
\end{tabular}
\caption{MRI. Transfer learning. Hiperparámetros importantes.}
\end{table}

\begin{table}[H]
\begin{tabular}{|l|l|l|l|}
\hline
Rotación & Ampliación & Volteo & Desplazamiento \\ \hline
Máx. 0.5\textdegree & Min. 0.95, Máx 1.05 & NO & Máx. 2\%   \\ \hline
\end{tabular}
\caption{MRI. Transfer learning. Hiperparámetros del aumento de datos.}
\end{table}

\end{appendices}

\end{document}